\documentclass[10pt]{asme2ej}

\usepackage{amsfonts}
\usepackage{hyperref}
\usepackage{amsmath}
\usepackage{bm}
\usepackage{amsthm}
\usepackage{algorithmic, algorithm}
\usepackage[pdftex]{graphicx}
\usepackage{subfigure}
\usepackage{caption}
\usepackage{makecell}
\usepackage{float}
\usepackage{enumitem}
\usepackage{epsfig} 
\usepackage{color}
\usepackage{pdfsync}
\usepackage{nomencl}

\newcommand{\bc}{\mathbf{c}}
\newcommand{\bg}{\mathbf{g}}
\newcommand{\bh}{\mathbf{h}}

\newcommand{\bmm}{\mathbf{m}}
\newcommand{\bt}{\mathbf{t}}
\newcommand{\bw}{\mathbf{w}}
\newcommand{\bs}{\mathbf{s}}
\newcommand{\bx}{\mathbf{x}}
\newcommand{\bz}{\mathbf{z}}
\newcommand{\bv}{\mathbf{v}}
\newcommand{\by}{\mathbf{y}}
\newcommand{\bq}{\mathbf{q}}
\newcommand{\bC}{\mathbf{C}}
\newcommand{\bG}{\mathbf{G}}
\newcommand{\bK}{\mathbf{K}}
\newcommand{\bI}{\mathbf{I}}
\newcommand{\bX}{\mathbf{X}}
\newcommand{\bA}{\mathbf{A}}

\newcommand{\bV}{\mathbf{V}}
\newcommand{\bZ}{\mathbf{Z}}
\newcommand{\bQ}{\mathbf{Q}}

\newcommand{\bW}{\mathbf{W}}
\newcommand{\balpha}{\bm{\alpha}}
\newcommand{\bbeta}{\bm{\beta}}
\newcommand{\bGamma}{\bm{\Gamma}}
\newcommand{\btheta}{\bm{\theta}}
\newcommand{\bTheta}{\bm{\Theta}}
\newcommand{\bLambda}{\bm{\Lambda}}
\newcommand{\bphi}{\bm{\phi}}
\newcommand{\bPhi}{\bm{\Phi}}
\newcommand{\R}{\mathbb{R}}

\newcommand{\calD}{\mathcal{D}}
\newcommand{\calN}{\mathcal{N}}

\newcommand{\calL}{\mathcal{L}}
\newcommand{\calU}{\mathcal{U}}
\newcommand{\qref}[1]{Eq.~(\ref{eqn:#1})}
\newcommand{\sref}[1]{Sec.~\ref{sec:#1}}
\newcommand{\fref}[1]{Fig.~\ref{fig:#1}}
\newcommand{\GP}{\operatorname{GP}}
\newcommand{\tr}{\operatorname{tr}}

\newcommand{\diag}{\operatorname{diag}}
\newcommand{\BIC}{\operatorname{BIC}}

\title{Gaussian processes with built-in dimensionality reduction: Applications in
high-dimensional uncertainty propagation}

\author{Rohit Tripathy
    \affiliation{ 
    		    School of Mechanical Engineering\\
		    Purdue University\\
		    585 Purdue Mall\\
		    West Lafayette IN 47906, USA\\
                    \href{mailto:rtripath@purdue.edu}{rtripath@purdue.edu}
    }
}

\author{Ilias Bilionis
    \affiliation{
                     School of Mechanical Engineering\\
		     Purdue University\\
		     585 Purdue Mall\\
		     West Lafayette IN 47906, USA\\
                     \href{mailto:ibilion@purdue.edu}{ibilion@purdue.edu} (Corr. author)
    }	
}

\author{Marcial Gonzalez
    \affiliation{School of Mechanical Engineering\\
		     Purdue University\\
		     Mechanical Engineering Room 1069\\
		     585 Purdue Mall\\
		     West Lafayette IN 47906, USA\\
		     \href{marcial-gonzalez@purdue.edu}{marcial-gonzalez@purdue.edu}        
    }
}

\makeglossary

\begin{document}

\maketitle    

\begin{abstract} \label{sec:abstract}
Uncertainty quantification (UQ) tasks, such as model calibration, uncertainty
propagation, and optimization under uncertainty, typically require several
thousand evaluations of the underlying computer codes.  To cope with the cost
of simulations, one replaces the real response surface with a cheap
surrogate based, e.g., on polynomial chaos expansions, neural networks, support
vector machines, or Gaussian processes (GP).  However, the
number of simulations required to learn a generic multivariate response
grows exponentially as the input dimension increases.  This curse of
dimensionality can only be addressed, if the response exhibits some special
structure that can be discovered and exploited.  A wide range of physical
responses exhibit a special structure known as an active subspace (AS). An AS
is a linear manifold of the stochastic space characterized by maximal response
variation.  The idea is that one should first identify this low dimensional
manifold, project the high-dimensional input onto it, and then link the projection to the output.  
If the dimensionality of the AS is low enough, then learning the link function 
is a much easier problem than the original problem
of learning a high-dimensional function.  The classic approach to discovering
the AS requires gradient information, a fact that severely limits its
applicability.  Furthermore, and partly because of its reliance to gradients,
it is not able to handle noisy observations. The latter is an essential trait if one wants to
be able to propagate uncertainty through stochastic simulators, e.g., through
molecular dynamics codes.  In this work, we develop a probabilistic version of
AS which is gradient-free and robust to observational noise.  Our approach
relies on a novel Gaussian process regression with built-in dimensionality
reduction.  In particular, the AS is represented as an orthogonal projection
matrix that serves as yet another covariance function hyper-parameter to be
estimated from the data. To train the model, we design a two-step maximum
likelihood optimization procedure that ensures the orthogonality of the
projection matrix by exploiting recent results on the Stiefel manifold, i.e., the manifold of matrices with orthogonal columns.
The additional benefit of our probabilistic formulation, is that it
allows us to select the dimensionality of the AS via the Bayesian information
criterion.  We validate our approach by showing that it can discover the right
AS in synthetic examples without gradient information using both noiseless and
noisy observations.  We demonstrate that our method is able to discover the
same AS as the classical approach in a challenging one-hundred-dimensional problem
involving an elliptic stochastic partial differential equation with random
conductivity.  Finally, we use our approach to study the effect of geometric and material uncertainties in
the propagation of solitary waves in a one dimensional granular system.
\end{abstract}

\section{Introduction} \label{sec:intro}

Despite the indisputable successes of modern computational science and
engineering, the increase in the predictive abilities of physics-based models
has not been on a par with the advances in computer hardware. On one hand, we
can now solve harder problems faster. On the other hand, however, the more
realistic we make our models, the more parameters we have to worry about, in
order to be able to describe boundary and initial conditions, material
properties, geometric imperfections, constitutive laws, etc. Since it is
typically impossible, or impractical, to accurately measure every single
parameter of a complex computer code, we have to treat them as uncertain and
model them using probability theory. Unfortunately, the field of uncertainty
quantification (UQ) \cite{ralph_smith_uq_2014, chen_w_2002, auria_galassi_1995,
oberkampf_2004}, which seeks to rigorously and objectively assess the impact of
these uncertainties on model predictions, is not yet mature enough to deal with
high-dimensional stochastic spaces.

The most straightforward UQ approaches are powered by Monte Carlo (MC)
sampling \cite{liu_2001, robert_cp_2004}. In fact, standard MC, as well as
advanced variations, are routinely applied to the uncertainty propagation (UP)
problem \cite{morokoff1995,barth2011,kuo2012}, model calibration
\cite{tarantola2004,bilionis2015}, stochastic optimization \cite{spall2003,
bilionis2012a, bilionis2013a}, involving complex physical models. Despite the
remarkable fact that MC methods convergence rate is independent of the number
of stochastic dimensions, realistic problems typically require tens or hundreds of
thousands of simulations. As stated by A. O'Hagan, this slow convergence is due
to the fact that ``Monte Carlo is fundamentally unsound" \cite{ohagan1987}, in
the sense that it fails to learn exploitable patterns from the collected data.
Thus, MC is rarely ever useful in UQ tasks involving expensive computer codes.

To deal with expensive computer codes, one typically resorts to surrogates of
the response surface.  Specifically, one evaluates the computer code on a
potentially adaptively selected, design of input points, uses the result to
build a cheap-to-evaluate version of the response surface, i.e., a surrogate. Then, 
 he/she replaces all the occurrences of the true computer code in the UQ problem
formulation with the constructed surrogate.  The surrogate may be based on a
generalized polynomial chaos expansion \cite{xiu_karniadakis_2002, xiu_2005,
xiu_2007,babuska_2010, sankaran_2011}, radial basis functions
\cite{gutmann_2001, regis_2005}, relevance vector machines
\cite{bilionis_zabaras_2012}, adaptive sparse grid
collocation\cite{ma_zabaras_2009}, Gaussian processes (GP)
\cite{currin1988, sacks1989, currin1991, bilionis_zabaras_2012, bilionis_2012,
anitescu_lockwood_2012, bilionis_2013, bilionis_2014}, etc.  For relatively 
low-dimensional stochastic inputs, all these methods outperform MC, in the sense 
that they need considerably fewer evaluations of the expensive computer code in
order to yield satisfactorily convergent results.

In this work, we focus on Bayesian methods and, in particular, on GP regression
\cite{rasmussen_gpml_2005}.  The rationale behind this choice is due to the
special ability of the Bayesian formalism to quantify the epistemic uncertainty
induced by the limited number of simulations. In other words, it makes it
possible to produce error bars for the results of the UQ analysis, see
\cite{ohagan1991, ohagan1999, oakley2002, oakley2004, bilionis_2012,
bilionis_zabaras_2012, bilionis_2013, bilionis_2014, chen2015} and
\cite{bilionis2015c} for a recent review focusing on the uncertainty
propagation problem.  This epistemic uncertainty is the key to developing
adaptive sampling methodologies, since it can be used to rigorously quantify
the expected information content of future simulations.  For example, see
\cite{mackay1992,balaprakash2013} for adaptive sampling targeted to overall
surrogate improvement, \cite{jones2001} and \cite{emmerich2005} for single- and
multi-objective global optimization, respectively, and \cite{bilionis_2012} for
the uncertainty propagation problem.

Unfortunately, standard GP regression, as well as practically any generic UQ
technique, is not able to deal with high stochastic dimensions. This is due to
the fact that it relies on the Euclidean distance to define input-space
correlations.  Since the Euclidean distance becomes uninformative as the
dimensionality of the input space increases \cite{bengio_2005}, the number of
simulations required to learn the response surface grows exponentially. This is
known as the \emph{curse of dimensionality}, a term coined by R. Bellman
\cite{bellman_1956}. In other words, blindly attempting to learn generic
high-dimensional functions is a futile task. Instead, research efforts are
focused on methodologies that can identify and exploit some special structure
of the response surface, which can be discovered from data.

The simplest way to address the curse of dimensionality is to use a variable
reduction method, e.g., sensitivity analysis \cite{saltelli_2008,smith2014} or
automatic relevance determination \cite{mackay1992, neal1996, neal1998}.  Such
methods rank the input features in order of their ability to influence the
quantity of interest, and, then, eliminate the ones that are unimportant. Of
course, variable reduction methods are effective only when the dimensionality
of the input is not very high and when the input variables are,
more or less, uncorrelated.  The common case of functional inputs, e.g., flow
through porous media requires the specification of the permeability and the
porosity as functions of space, cannot be teated directly with variable
reduction methods. In such problems one has to start with a 
dimensionality reduction of the functional input.  For example, 
if the input uncertainty is described via a Gaussian random field,
dimensionality reduction can be achieved via a truncated Karhunen-Lo\`eve
expansion (KLE) \cite{ghanem2003}.  If the stochastic input model is to be
built from data, one may use principal component analysis (PCA)
\cite{pearson_1901}, also known as empirical KLE, or even non-linear
dimensionality reduction maps such as kernel PCA \cite{ma2011}. The end goal of
dimensionality reduction techniques is the construction of a low dimensional
set of uncorrelated features on which variable reduction methods, or
alternative methods, may be applied. Note that even though the new features
are lower dimensional than the original functional inputs, they are still
high-dimensional for the purpose of learning the response surface.

A popular example of an exploitable feature of response surfaces that can be
discovered from data is additivity.  Additive response surfaces can be
expressed as the sum of one-variable terms, two-variable terms, and so on,
interpreted as interactions between combinations of input variables.  Such
representations are inspired from physics, e.g., the Coulomb potential
of multiple charges, the Ising model of statistical mechanics.  Naturally,
this idea has been successfully applied to the problem of learning the energy
of materials as a function of the atomic configuration.  For example, in
\cite{bilionis2013a} the authors use this idea to learn the quantum mechanical
energy of binary alloys on a fixed lattice by expressing it as the sum of
interactions between clusters of atoms, a response surface with thousands of
input variables.  The approach has also been widely used by the computational
chemistry community,  where it is known as high-dimensional model
representation (HDMR) \cite{Rabitz1999, Alis2001, Li2001a, Li2001b}.  The UQ
community has been embracing and extending HDMR \cite{chowdhury2009, ma2010},
sometimes referring to it by the name functional analysis of variance (ANOVA)
\cite{wei2014, zhang2014}. It is possible to model additive response surfaces
with a GP by choosing a suitable covariance function. The first such effort 
can be traced to \cite{plate1999} and has been recently revisited
by \cite{kaufman2010, durrande2011, duvenaud2011, gilboa2013, nguyen2014}.  By
exploiting the additive structure of response surfaces one can potentially deal
with a few hundred to a few thousand input dimensions.  This is valid, of
course, only under the assumption that the response surface does have an
additive structure with a sufficiently low number of important terms.

Another example of an exploitable response surface feature is active subspaces
(AS) \cite{russi_2010}.  An AS is a low-dimensional linear manifold of the input space
characterized by maximal response variation. It aims at discovering orthogonal
directions in the input space over which the response varies maximally, ranking
them in terms of importance, and keeping only the most significant ones.
Mathematically, an AS is described by an orthogonal matrix that projects the
original inputs to this low-dimensional manifold. 
The classic framework for
discovering the AS was laid down by Constantine \cite{constantine_2014,
constantine_2014_1, constantine_2013, constantine_gleich_2014}.  One builds a positive-definite 
matrix that depends upon the gradients of the response surface. 
The most important eigenvectors of this matrix form the aforementioned projection matrix. The
dimensionality of the AS is identified by looking for sharp changes in the eigenvalue
spectrum, and retaining only the eigenvectors corresponding to the highest
eigenvalues.  Once the AS is established, one proceeds by: 1) Projecting all the
inputs to the AS; 2) Learning the map between the projections and the quantity
of interest. The latter is known as the \emph{link function}.
The framework has been successfully applied to a variety of
engineering problems \cite{wang_constantine_2011, constantine_2011,
constantine_zaha_2014, constantine_trent_2014, constantine_2015}. 

One of the major drawbacks of classic AS methodology is that it relies on
gradient information. Even though, in principle, it is possible to compute the
gradients either by deriving the adjoint equations \cite{plessix2006} or by
using automatic differentiation \cite{andreas_2008}, in many cases of interest
this is not practical, since implementing any of these two approaches requires a significant amount of time 
for software development, validation and verification.  This is an undesirable scenario when one
deals with existing complex computer codes with decades of development
history. The natural alternative of employing numerical differentiation is also
not practical for high-dimensional input, especially when the underlying
computer code is expensive to evaluate and/or when one has to perform the
analysis using a restricted computational budget. The second major drawback of the 
classic AS methodology is its difficulty in dealing
with relatively large observational noise, since that would require a 
unifying probabilistic framework. 
This drawback significantly limits the applicability of AS to
important problems that include noise. For example, it cannot be used
in conjunction with high-dimensional experimental data, or response surfaces 
that depend on stochastic models e.g., molecular dynamics.

The ideas of AS methodologies are reminiscent of the partial least squares
(PSL) \cite{geladi1986} regression scheme, albeit it is obvious that the two
have been developed independently stemming from different applications. AS
applications focus on computer experiments \cite{constantine_2014,
constantine_2014_1, constantine_2013, constantine_gleich_2014}, while PSL has
been extensively used to model real experiments with high-dimensional
inputs/outputs in the field of chemometrics \cite{wold1999, brereton2003,
ferreira2015}. PSL not only projects the input to a lower dimensional space
using an orthogonal projection matrix, but, if required, it can do the same
to a high-dimensional output. It connects the reduced input to the reduced
output using a linear link function. All model parameters are identified by
minimizing the sum of square errors. PSL does not require gradient information
and, thus, addresses the first drawback of AS.  Furthermore, it also addresses,
to a certain extent, the second drawback, namely the inability of AS to cope
with observational noise, albeit only if the noise level is known a priori or
fitted to the data using cross validation. As all non-Bayesian techniques, PSL
may suffer from overfitting and from the inability to produce robust predictive
error bars.  Another disadvantage of PSL is the assumption that the link map is
linear, a fact that severely limits its applicability to the study of realistic
computer experiments. The latter has been addressed by the locally
weighted PSL \cite{kim2011}, but at the expense of introducing an excessive
amount of parameters.

In this work, we develop a probabilistic version of AS that addresses both its
major drawbacks. That is, our framework is gradient-free (even though it can
certainly make use of gradient information if this is available), and it can
seamlessly work with noisy observations. It relies on a novel Gaussian process
(GP) regression methodology with built-in dimensionality reduction. In
particular, we treat the orthogonal projection matrix of AS as yet another
hyper-parameter of the GP covariance function. That is, our proposed covariance
function internally projects the high-dimensional inputs to the AS, and then
models the similarity of the projected inputs. We determine all the
hyper-parameters of our model, including the orthogonal projection matrix, by
maximizing the likelihood of the observed data. To achieve this, we devise a
two-step optimization algorithm guaranteed to converge to a local maximum of
the likelihood. The algorithm iterates between the optimization of the
projection matrix (keeping all other hyper-parameter fixed) and the optimization of all
other hyper-parameters (keeping the projection matrix fixed), until a convergence
criterion is met.  To enforce the orthogonality constraint on the projection
matrix, we exploit recent results on the description of 
the Stiefel manifold, i.e., the set of matrices with orthogonal columns. The
optimization of the other hyper-parameters is carried out using BFGS
\cite{byrd1995}. The addendum of our probabilistic approach is that it allows
us to select the dimensionality of the AS using the Bayesian information
criterion (BIC) \cite{murphy2012}.
  
This paper is organized as follows. In \sref{gp_reg}, we briefly introduce GP
regression, followed by a discussion of the classic, gradient-based, AS approach (\sref{classic_approach}) 
and the proposed gradient-free approach in (\sref{grad_free_approach}). 
\sref{exa_synthetic} verifies our approach in a series of synthetic examples with known AS 
as well as the robustness of 
our methodology to observational noise.
In \sref{exa_elliptic}, we use a one-hundred-dimensional stochastic 
elliptic partial differential equation (PDE) to demonstrate that the proposed 
approach discovers the same AS as the classic approach - even without 
gradient information. In \sref{exa_gc}, we use our approach to study the effect of geometric and material uncertainties in the propagation of solitary waves through a one dimensional granular system. We present our conclusions in \sref{conclusion}.


\section{Methodology}
\label{sec:metho}

Let $f:\R^D\rightarrow \R$ be a multivariate response surface with $D \gg
1$. Intuitively, $f(\cdot)$ accepts an \emph{input}, $\bx\in\R^D$, and responds
with an \emph{output} (or \emph{quantity of interest} (QoI)), $f(\bx)$. We can
measure $f(\bx)$ by querying an \emph{information source}, which can be either
a computer code or a physical experiment. Furthermore, we allow for noisy
information sources. That is, we assume that instead of measuring $f(\bx)$
directly, we measure a noisy version of it $y = f(\bx) + \epsilon$, where
$\epsilon$ is a random variable. In physical experiments, measurement noise may
rise from our inability to control all influential factors or from irreducible
(aleatory) uncertainties. In computer simulations, measurement uncertainty may
rise from quasi-random stochasticity, or chaotic behavior.

The ultimate goal of this work, is to efficiently propagate
uncertainty through $f(\cdot)$. That is, given a probability density function
(PDF) on the inputs:
\begin{equation}
\label{eqn:input_pdf}
\bx \sim p(\bx),
\end{equation}
we would like to compute the statistics of the output.
Statistics of interest are the \emph{mean}
\begin{equation}
\label{eqn:mean_out}
\mu_f = \int f(\bx)p(\bx)d\bx,
\end{equation}
the \emph{variance},
\begin{equation}
\label{eqn:var_out}
\sigma_f^2 = \int \left(f(\bx) - \mu_f \right)^2p(\bx)d\bx,
\end{equation}
and the PDF of the output, which can be formally written as

\begin{equation}
\label{eqn:output_pdf}
f \sim p(f) = \int \delta\left(f - f(\bx) \right)p(\bx) d\bx,
\end{equation}
where $\delta(\cdot)$ is Dirac's $\delta$-function.
We refer to this problem as the \emph{uncertainty propagation} (UP) problem.

The UP problem is particularly hard when obtaining information about $f(\cdot)$
is expensive. In such cases, we are necessarily restricted to a limited set of
observations. Specifically, assume that we have queried the information
source at $N$ input points,
\begin{equation}
    \bX = \left\{\bx^{(1)}, \dots, \bx^{(N)} \right\},
    \label{eqn:observed_input}
\end{equation}
and that we have measured
\begin{equation}
    \by = \left\{y^{(1)}, \dots, y^{(N)} \right\}.
    \label{eqn:observed_output}
\end{equation}
We consider the following pragmatic interpretation of the UP problem:
What is the best we can say about the statistics of the QoI, given the limited
data in $\calD$? The core idea behind our approach, and also behind most
popular approaches in the current literature, is to replace the expensive
response surface, $f(\cdot)$, with a cheap to evaluate surrogate learned from
$\bX$ and $\by$.

As discussed in \sref{intro}, the fact that we are working in a
high-dimensional regime, $D \gg 1$, causes insurmountable difficulties
unless $f(\cdot)$ has some special structure that we can discover and
exploit.  In this work, we assume that the response surface has, or can be
well-approximated with the following form:
\begin{equation} 
f(\bx) \approx g\left(\bW^T\bx\right),
\label{eqn:f}
\end{equation} 
where the matrix $\bW\in\R^{D\times d}$ projects the high-dimensional
input space, $\R^D$, to the low-dimensional \emph{active subspace},
$\R^d, d\ll D$, and $g:\R^d\rightarrow\R$ is a $d$-dimensional function known
as the \emph{link} function.
Without loss of generality, we may assume that the columns
of $\bW$ are orthogonal. Mathematically, we write $\bW\in V_d\left(\R^D\right)$, where
$V_d\left(\R^D\right)$ is the set of $D\times d$ matrices with orthogonal columns,
\begin{equation}
    V_d\left(\R^D\right) := \left\{\bA\in\R^{D\times d}: \bA^T\bA = \bI_d \right\},
    \label{eqn:stiefel}
\end{equation}
with $\bI_d$ the $d\times d$ unit matrix.
$V_d\left(\R^D\right)$ is also known as the
\emph{Stiefel manifold}.
Note that the representation of \qref{f} is arbitrary up to rotations and
relabeling of the active subspace coordinate system.
Intuitively, we expect that there is a $d$-dimensional subspace of $\R^D$
over which $f(\cdot)$ exhibits most of its variation. If $d$ is indeed much
smaller than $D$, then the learning problem is significantly simplified.

The goal of this paper is to construct a framework for the determination of the
dimensionality of the active subspace $d$, the orthogonal projection matrix
$\bW$, and of the low dimensional map $g(\cdot)$ using only the observations
$\{\bX, \by\}$. Once these elements are identified,
then one may use the constructed surrogate in any uncertainty quantification
task, and, in particular, in the UP problem.
We achieve our goal by following a probabilistic approach, in which
$f(\cdot)$ is represented as a GP with $\bW$ built into its covariance
function and determined by maximizing the likelihood of the model.

\subsection{Gaussian process regression}
\label{sec:gp_reg}

In this section we provide a brief, but complete, description of GP regression.
Since, in later subsections, we use the concept in two different settings, here
we attempt to be as generic as possible so that what we say is applicable to
both settings.  Towards this end, we consider the problem of learning an arbitrary
response surface $h(\cdot)$ which takes inputs $\bq\in\R^l$, assuming that we
have made the, potentially noisy, observations:
\begin{equation}
    \bt = \left\{t^{(1)}, \dots, t^{(N)} \right\},
    \label{eqn:t_output}
\end{equation}
at the input points:
\begin{equation}
    \bQ = \left\{\bq^{(1)}, \dots, \bq^{(N)} \right\}.
    \label{eqn:q_input}
\end{equation}

The philosophy behind GP regression is as follows.  A GP defines a probability
measure on a function space, i.e., a random field. This probability measure
corresponds to our prior beliefs about the response surface.  GP regression
uses Bayes rule to combine these prior beliefs with observations. The result of
this process is a \emph{posterior} GP which is simultaneously compatible with
our beliefs and the data. We call this posterior GP a \emph{Bayesian
surrogate}. If a point-wise surrogate is required, one may use the median of
the posterior GP. Predictive error bars, corresponding to the epistemic
uncertainty induced by limited data, can be derived using the variance of the
posterior GP. To materialize the GP regression program we need three
ingredients: 1) A description of our prior state of knowledge about the
response surface (\sref{gp_prior}); 2) A model of the measurement process 
(\sref{gp_likelihood}); and 3) A characterization of our posterior state of
knowledge (\sref{gp_posterior}). In \sref{gp_fitting} we discuss how the 
posterior of the model can be approximated via maximum likelihood.

\subsubsection{Prior state of knowledge}
\label{sec:gp_prior}
Prior to seeing any data, we model our state of knowledge about $h(\cdot)$ by
assigning to it a GP prior. We say that $h(\cdot)$ is a GP with mean function
$m(\cdot;\btheta)$ and covariance function $k(\cdot,\cdot;\btheta)$, and write:
\begin{equation} 
    h(\cdot) | \btheta \sim \GP(h(\cdot)|m(\cdot;\btheta), k(\cdot, \cdot;\btheta)).
    \label{eqn:gp_prior}
\end{equation}
The parameters of the mean and the covariance function, $\btheta\in\bTheta$, are
known as the \emph{hyper-parameters} of the model.

Our prior beliefs about the response are encoded in our choice of the mean and
covariance functions, as well as in the prior we pick for their
hyper-parameters:
\begin{equation}
    \btheta \sim p(\btheta).
    \label{eqn:theta_prior}
\end{equation}
The mean function is used to model any generic trends of the response surface,
and it can have any functional form. If one does not have any knowledge about
the trends of the response, then a reasonable choice is a zero mean function.
The covariance function, also known as the covariance kernel, is the most
important part of a GP. Intuitively, it defines a nearness or similarity
measure on the input space. That is, given two input points, their covariance
models how close we expect the corresponding outputs to be.  A valid covariance
function must be positive semi-definite and symmetric. Throughout the present work 
we use the Matern-$32$ covariance kernel:
\begin{equation}
k_{\mbox{mat}}(\bq,\bq';\btheta)=s^2 \left( 1 + \sqrt{3} \sum_{i=1}^l\frac{\left(q_i - q_i'\right)^2}{\ell_i^2} \right) \exp\left(-\sqrt{3} \sum_{i=1}^l\frac{\left(q_i - q_i'\right)^2}{\ell_i^2}\right)
\label{eqn:mat_cov}
\end{equation}
where $\btheta = \{s, \ell_1,\dots,\ell_l\}$, with $s > 0$ being the signal
strength and $\ell_i > 0$ the length scale of the $i$-th input.  The Matern-32
covariance function corresponds to the a priori belief that the response
surface is both continuous and differentiable.  For more on covariance functions see Ch.~4 of
Rasmussen \cite{rasmussen_gpml_2005}. 

Given an arbitrary set of inputs $\bQ$, see \qref{q_input}, \qref{gp_prior}
induces by definition a Gaussian prior on the corresponding response outputs:
\begin{equation}
    \bh = \left\{h\left(\bq^{(1)}\right), \dots, h\left(\bq^{(N)}\right)\right\}.
\end{equation}
Specifically, $\bh$ is a priori distributed according to:
\begin{equation}
    \bh | \bQ, \btheta \sim \calN\left(\bh \middle| \bmm, \bK\right),
    \label{eqn:h_prior}
\end{equation}
where $\calN\left(\cdot|\bm{\mu}, \bm{\Sigma}\right)$ is the PDF of a multivariate
Gaussian random variable with mean vector $\bm{\mu}$ and covariance matrix
$\bm{\Sigma}$, $\bmm := \bmm(\bQ;\btheta)\in\R^N$ is the mean function evaluated at all points
in $\bQ$, 
\begin{equation}
    \bmm = \bmm(\bQ;\btheta) = \left(
        \begin{array}{c}
        m\left(\bq^{(1)}; \btheta\right) \\ \vdots \\ m\left(\bq^{(N)};\btheta\right)
        \end{array}
    \right),
\end{equation}
and $\bK:=\bK(\bQ,\bQ;\btheta)\in\R^{N\times N}$ is the \emph{covariance matrix}, a special case of
the more general \emph{cross-covariance matrix} $\bK(\bQ, \hat\bQ;\btheta)\in\R^{N\times\hat N}$,
\begin{equation}
    \bK(\bQ,\hat\bQ;\btheta) = \left(
        \begin{array}{ccc}
            k\left(\bq^{(1)}, \hat{\bq}^{(1)};\btheta \right) & \dots & k\left(\bq^{(1)}, \hat{\bq}^{(\hat N)};\btheta \right)\\
        \vdots & \ddots & \vdots\\
            k\left(\bq^{(N)}, \hat{\bq}^{(1)};\btheta \right) & \dots & k\left(\bq^{(N)}, \hat{\bq}^{(\hat N)};\btheta \right)
        \end{array}
    \right),
\end{equation}
defined between $\bQ$, \qref{q_input}, and an arbitrary set of $\hat{N}$ inputs
$\hat{\bQ} = \left\{\hat{\bq}^{(1)},\dots,\hat{\bq}^{(\hat{N})}\right\}$.

\subsubsection{Measurement process}
\label{sec:gp_likelihood}

The Bayesian formalism requires that we explicitly model the measurement
process that gives rise to the observations $\bt$ of \qref{t_output}. The
simplest such model is to assume that measurements are independent of each
other, and that they are distributed normally about $h(\cdot)$ variance $s_n^2$.
That is,
\begin{equation}
    t^{(i)} \vert h\left(\bq^{(i)}\right), s_n
    \sim \calN\left( t^{(i)} \middle| h\left(\bq^{(i)}\right), s_n^2\right).
    \label{eqn:gp_like}
\end{equation}
Note that $s_n>0$ is one more hyper-parameter to be determined from the data,
and that we must also assign a prior to it:
\begin{equation}
    s_n \sim p(s_n).
    \label{eqn:s_n_prior}
\end{equation}
The assumptions in \qref{gp_like} can be relaxed to allow for heteroscedastic
(input dependent) noise \cite{goldberg1998,plagemann2008}, but this is beyond
the scope of this work. Using the independence assumption, we get:
\begin{equation}
    \bt | \bh, s_n \sim \calN\left(\bt \middle| \bh, s_n^2\bI_N\right).
    \label{eqn:gp_full_like}
\end{equation}
Using the sum rule of probability theory and standard properties of Gaussian
integrals, we can derive the \emph{likelihood} of the observations given the
inputs:
\begin{equation}
    \bt | \bQ, \btheta, s_n \sim \calN\left(\bt \middle| \bmm, \bK + s_n^2\bI_N\right).
    \label{eqn:gp_final_like}
\end{equation}

\subsubsection{Posterior state of knowledge}
\label{sec:gp_posterior}

Using Bayes rule to combine the prior GP, \qref{gp_prior}, with the likelihood,
\qref{gp_final_like}, yields the \emph{posterior} GP:
\begin{equation}
    h(\cdot) | \bQ, \bt, \btheta, s_n \sim \GP\left(h(\cdot) \middle| \tilde{m}(\cdot), \tilde{k}(\cdot,\cdot)\right),
    \label{eqn:gp_posterior}
\end{equation}
where the \emph{posterior} mean and covariance functions are
\begin{equation}
    \tilde{m}(\bq) := \tilde{m}(\bq;\btheta) = m(\bq;\btheta) + \bK(\bq, \bQ;\btheta)\left(\bK + s_n^2\bI_N\right)^{-1}\left(\bt - \bmm\right),
    \label{eqn:gp_posterior_mean}
\end{equation}
and 
\begin{equation}
    \tilde{k}(\bq,\bq') := \tilde{k}(\bq, \bq';\btheta,s_n) = k(\bq, \bq';\btheta) - \bK(\bq,\bQ;\btheta)\left(\bK + s_n^2\bI_N\right)^{-1}\bK(\bQ,\bq;\btheta),
    \label{eqn:gp_posterior_covariance}
\end{equation}
respectively. The posterior of the hyper-parameters is obtained by combining
Eqn.'s (\ref{eqn:theta_prior}) and (\ref{eqn:s_n_prior}) with \qref{gp_full_like} using Bayes rule, i.e.,
\begin{equation}
    \btheta, s_n | \bQ, \bt \sim p(\bt | \bQ, \btheta, s_n) p(\btheta)p(s_n).
    \label{eqn:hyper_posterior}
\end{equation}

Eqn.'s (\ref{eqn:gp_posterior}) and (\ref{eqn:hyper_posterior}) fully quantify
our state of knowledge about the response surface after seeing the data.
However, in practice it is more convenient to work with the \emph{predictive
probability density} at a single input $\bq$ conditional on the
hyper-parameters $\btheta$ and $s_n$, namely: 
\begin{equation}
    h(\bq) | \bQ, \bt, \btheta, s_n \sim \calN\left(h(\bq)\middle | \tilde{m}(\bq), \tilde{\sigma}(\bq)\right),
\end{equation}
where $\tilde{m}(\bq) = \tilde{m}(\bq;\btheta)$ is the predictive mean given 
in \qref{gp_posterior_mean}, and
\begin{equation}
    \tilde{\sigma}^2(\bq) := \tilde{k}(\bq,\bq;\btheta,s_n),
\end{equation}
is the \emph{predictive variance}. Note that the predictive mean can be used as
a point-wise surrogate of the response surface, while the predictive variance
can be used to derive point-wise predictive error bars.

\subsubsection{Fitting the hyper-parameters}
\label{sec:gp_fitting}

Ideally, one would like to characterize the posterior of the hyper-parameters,
see \qref{hyper_posterior} using sampling techniques, e.g., a Markov chain
Monte Carlo (MCMC) algorithm \cite{metropolis1953, hastings1970, haario2006}.
Here, we opt for a much simpler approach by approximating
\qref{hyper_posterior} with a $\delta$-Dirac function centered at the
hyper-parameters that maximize the likelihood \qref{gp_final_like}. For issues
of numerical stability, we prefer to work with the logarithm of the likelihood:
\begin{equation}
    \calL(\btheta, s_n; \bQ, \bt) := \log p(\bt | \bQ, \btheta, s_n).
    \label{eqn:gp_log_like}
\end{equation}
and determine the hyper-parameters by solving the following
optimization problem:
\begin{equation}
    \btheta^*, s_n^* = \arg\max_{\btheta,s_n}\calL(\btheta, s_n; \bQ, \bt),
    \label{eqn:like_opt}
\end{equation}
subject to any constraints imposed on the hyper-parameters(see Ch. 5 of
\cite{rasmussen_gpml_2005}). 
According to Eq. (\ref{eqn:gp_final_like}), the log-likelihood is
\begin{equation}
\log p(\bt|\bQ,\btheta,s_n) = -\frac{1}{2}(\bt-\bmm)^T \left(\bK + s_n^2\bI_N\right)^{-1}(\bt-\bmm) - \frac{1}{2} \log |\bK + s_n^2\bI_N| - \frac{N}{2} \log 2\pi.
\label{eqn:log_like}
\end{equation}
The derivative of the log-likelihood with respect to any arbitrary parameter $\phi$, where $\psi = s_n$ or $\theta_i$, is:
\begin{equation}
    \frac{\partial}{\partial \psi} \calL(\btheta, s_n; \bQ, \bt)= \frac{1}{2} \tr \left[  \left( \left(\bK + s_n^2\bI_N\right)^{-1}(\bt-\bmm) \left(\left(\bK + s_n^2\bI_N\right)^{-1}(\bt-\bmm)\right)^T - \left(\bK + s_n^2\bI_N\right)^{-1}\right) \frac{\partial \left(\bK + s_n^2\bI_N\right)}{\partial\psi} \right] .
\label{eqn:log_like_der}
\end{equation}

This point estimate of the hyper-parameters is known as the \emph{maximum likelihood estimate} (MLE). The approach is justified if the prior is relatively flat and the likelihood is sharply picked. Unless otherwise stated, in this work we solve the optimization problem of \qref{like_opt} via the BFGS optimization algorithm \cite{byrd1995} increasing the chances of finding the global maximum by restarting the algorithm multiple times from random initial points. 

\subsection{Gradient-based approach to active subspace regression}
\label{sec:classic_approach}
In this section, we discuss the classic approach to discovering the active
subspace using gradient information\cite{constantine_2013, constantine_2014, constantine_2014_1, constantine_zaha_2014, constantine_gleich_2014, constantine_trent_2014, constantine_2015, lakaczyk_2014, dow_wang_2013}.
Recall that we are dealing with a high-dimensional response surface, and that
we would like to approximate it as in \qref{f}. The classic approach does this
in two steps. First, it identifies the projection matrix $\bW\in V_d\left(\R^D\right)$
using gradient information (\sref{classic_find_as}).
Second, it projects all inputs to the AS, and then uses GP regression to learn
the map between the projected inputs and the output (\sref{classic_fit_gp}).

Note that the classic approach is not able to deal with noisy measurments.
Therefore, in this subsection, we assume that our measurements of $f(\bx)$ are
exact. That is, we work under the assumption that each $y^{(i)}$ in 
\qref{observed_output} is
\begin{equation}
    y^{(i)} = f\left(\bx^{(i)}\right),
    \label{eqn:classic_observed_output}
\end{equation}
for $i=1,\dots,N$. Also, since it requires gradient information, we assume that
we have observations of the gradient of $f(\cdot)$ at each one of the input
points, i.e., in addition to $\bx$ and $\by$ of \qref{observed_input} and \qref{observed_output}
respectively,
we have access to:
\begin{equation}
    \bG = \left\{\bg^{(1)}, \dots, \bg^{(N)} \right\},
    \label{eqn:observed_gradients}
\end{equation}
where
\begin{equation}
    \bg^{(i)} = \nabla f\left(\bx^{(i)}\right)\in\R^D,
    \label{eqn:observed_single_gradient}
\end{equation}
and $\nabla f(\cdot)$ is the gradient of $f(\cdot)$,
\begin{equation}
    \nabla f(\cdot) =
    \left(\frac{\partial f(\cdot)}{\partial x_1}, \dots, \frac{\partial f(\cdot)}{\partial x_D}\right).
    \label{eqn:response_gradient}
\end{equation}

\subsubsection{Finding the active subspace using gradient information}
\label{sec:classic_find_as}
Let $\rho(\bx)$ be a PDF on the input space, which can be different from the
PDF of the UP problem given in \qref{input_pdf}, and define the matrix
\begin{equation}
    \bC := \int (\nabla f(\bx)) (\nabla f(\bx))^T \rho(\bx)d\bx.
    \label{eqn:C}
\end{equation}
Since $\bC$ is symmetric positive definite, it admits the form
\begin{equation}
    \bC = \bV \bLambda \bV^T,
    \label{eqn:diagonalized_C}
\end{equation}
where  $\bLambda = \diag(\lambda_1, \cdots, \lambda_D)$ is a diagonal matrix
containing the eigenvalues of $\bC$ in decreasing order, $\lambda_1 \ge \dots \ge \lambda_D \ge 0$,
and $\bV\in\R^{D\times D}$ is an orthonormal matrix whose columns correspond to
the eigenvectors of $\bC$. The classic approach suggests separating the $d$
largest eigenvalues from the rest,
\begin{equation*}
\bLambda = 
\begin{bmatrix}
    \bLambda_1 && \bm{0}  \\
    \bm{0} && \bLambda_2
\end{bmatrix},
\quad
\bV = 
\begin{bmatrix}
\bV_1 && \bV_2
\end{bmatrix},
\end{equation*}
(here $\bLambda_1 = \diag(\lambda_1,\dots,\lambda_d), \bV_1 = [\bv_{11}\dots \bv_{1d}]$,
and $\bLambda_2, \bV_2$ are defined analogously),
and setting the projection matrix to
\begin{equation}
    \bW = \bV_1.
    \label{eqn:classical_projection_matrix}
\end{equation}
Intuitively, $\bV$ rotates the input space so that the directions associated
with the largest eigenvalues correspond to directions of maximal function
variability. See \cite{constantine_2014} for the theoretical
justification.

It is impossible to evaluate \qref{C} exactly. Instead, the usual practice is
to approximate the integral via Monte Carlo. That is, assuming that the
observed inputs are drawn from $\rho(\bx)$, one approximates $\bC$ using
the observed gradients, see \qref{observed_gradients}, by:
\begin{equation}
    \bC_N = \frac{1}{N}\sum_{i=1}^N \bg^{(i)}\left(\bg^{(i)}\right)^T.
\end{equation}
In practice, the eigenvalues and eigenvectors of $\bC_N$ are found using the
singular value decomposition (SVD) \cite{golub1996} of $\bC_N$. The
dimensionality $d$ is determined by looking for sharp drops in the spectrum of
$\bC_N$.

\subsubsection{Finding the map between the active subspace and the response}
\label{sec:classic_fit_gp}

Using the classically found projection matrix, see
\qref{classical_projection_matrix}, we obtain the projected observed
inputs $\bZ\in\R^{N\times d}$:
\begin{equation}
    \bZ = \left\{\bz^{(1)},\dots,\bz^{(N)} \right\},
    \label{eqn:reduced_observed_inputs}
\end{equation}
where
\begin{equation}
    \bz^{(i)} = \bW^T\bx^{(i)}.
    \label{eqn:reduced_observed_input}
\end{equation}
The link function $g(\cdot)$ that connects the AS to the output, see \qref{f},
is identified using GP regression, see \sref{gp_reg}, with response
$h(\cdot)\equiv g(\cdot)$, input points $\bq\equiv\bz$,
observed inputs $\bQ \equiv \bZ$, and observed outputs $\bt \equiv \by$.

\subsection{Gaussian processes regression with built-in dimensionality reduction}
\label{sec:grad_free_approach}

\begin{algorithm}[htb]
\caption{Two-step optimization algorithm for the log-likelihood.}
\begin{algorithmic}[1]
	\REQUIRE Observed inputs $\bX$, observed outputs $\by$, maximum number of iterations $M_l$,
        convergence tolerance $\epsilon_l>0$, 
        initial parameter estimates $\bW_0, \bphi_0$ and $s_{n,0}$.
        \STATE $\calL_0 \leftarrow \calL(\bW_0, \bphi_0, s_{n,0}; \bX, \by)$.
        \FOR{$i=1,\dots,M_l$}
        \STATE Perform $1$ iteration towards the solution of the following optimization problem: \\ $\bW_{i} \leftarrow \underset{\bW\in V_d\left(\R^D\right)}{\arg\max} \calL(\bW, \bphi_{i-1}, s_{n,i-1};\bX,\by)$ \COMMENT{using Alg. \ref{alg:stiefel_opt}}
	\STATE Perform $1$ iteration towards the solution of the following optimization problem: \\ $\bphi_{i}, s_{n,i} \leftarrow \underset{\bphi,s_n}{\arg\min} \calL(\bW_{i}, \bphi, s_{n};\bX,\by)$\COMMENT{using BFGS \cite{byrd1995}}
	\STATE $\calL_{i} \leftarrow \calL(\bW_{i}, \bphi_{i}, s_{n, i};\bX,\by)$
         \IF {$ \frac{\calL_{i} - \calL_{i-1}}{\calL_{i-1}} < \epsilon_l$} 
      			\STATE break
         \ENDIF
	\ENDFOR
	\STATE $\calL_0 \leftarrow \calL_i$
	\FOR{$i=1,\dots,M_l$}
        \STATE Solve the optimization problem stated in step 3 until convergence.
	\STATE Solve the optimization problem stated in step 4 until convergence. 
	\STATE $\calL_{i} \leftarrow \calL(\bW_{i}, \bphi_{i}, s_{n, i};\bX,\by)$
         \IF {$ \frac{\calL_{i} - \calL_{i-1}}{\calL_{i-1}} < \epsilon_l$} 
      			\STATE break
         \ENDIF
	\ENDFOR

	\RETURN $\bW_{i}, \bphi_{i}, s_{n, i}$
\end{algorithmic}
\label{alg:alt_alg}
\end{algorithm}

As mentioned in \sref{intro} the classic approach to AS-based GP regression,
see \sref{classic_approach}, suffers from two major drawbacks: 1) It relies on
gradient information; and 2) It cannot deal seamlessly with measurement noise.
In this section, we propose a probabilistic, unifying view of AS that is able
to overcome these difficulties.

Our approach is based on novel covariance function on the high-dimensional
input space:
\begin{equation}
    k_{\mbox{AS}}:\R^D\times\R^D\times V_d\left(\R^D\right)\times \bPhi \rightarrow \R,
    \label{eqn:as_cov_func_def}
\end{equation}
with form:
\begin{equation}
    k_{\mbox{AS}}(\bx, \bx'; \bW, \bphi) = k_d(\bW^T\bx, \bW^T\bx'; \bphi),
    \label{eqn:as_cov_def}
\end{equation}
where $k_d:\R^d\times\R^d\times\bphi\rightarrow\R$ is a standard covariance
function on the low-dimensional space parameterized by $\bphi\in\bPhi$.
In words, the high-dimensional covariance function, \qref{as_cov_def}, first
projects the inputs to the AS and, then, assesses the similarity of the
projected inputs using the low-dimensional covariance function $k_d(\cdot,
\cdot;\bphi)$. Note that he high-dimensional covariance
function is parameterized by both the orthonormal projection matrix $\bW$ and
the hyper-parameters $\bphi$ of the low-dimensional covariance function.

To appreciate the unifying character of our approach note that the way to
proceed is verbatim the generic GP regression approach of \sref{gp_reg} with
response $f(\cdot)\equiv h(\cdot)$,
input points $\bq\equiv \bx$, observed inputs $\bQ\equiv\bX$,
observed ouputs $\bt\equiv\by$,
covariance hyper-parameters $\btheta = \{\bW, \bphi\}$
taking values in $\bTheta \equiv V_d\left(\R^D\right)\times \bPhi$, and covariance
function $k(\cdot,\cdot;\btheta) \equiv k_{\mbox{AS}}(\cdot,\cdot;\bW, \bphi)$.
The only difficulty that we face, albeit non-trivial, is that the likelihood
maximization of \qref{like_opt} must take into account the constraint that the
projection matrix is orthonormal, $\bW\in V_d\left(\R^D\right)$.
The rest of this methodology is concerned with this optimization problem.
In particular, \sref{iter_like_opt} discusses the overall optimization
algorithm, \sref{stiefel_opt} the optimization over the Stiefel manifold,
and \sref{BIC} the selection of the AS dimensionality.

\subsubsection{Iterative two-step likelihood maximization}
\label{sec:iter_like_opt}
As mentioned earlier, the optimization problem that we have to solve for the
determination of the covariance hyper-parameters $\btheta = \{\bW, \bphi\}$ 
and the noise variance $s_n^2$, is given by \qref{like_opt} subject to the
constrain that $\bW\in V_d\left(\R^D\right)$.
To solve this problem we devise an iterative two-step
optimization algorithm guaranteed to converge to a local optimum.
The fist step keeps $\bphi$ and $s_n$ fixed, and performs 1 iteration
towards the optimization of the log-likelihood
over $\bW\in V_d\left(\R^D\right)$ (see \sref{stiefel_opt} for the details).
The second step, keeps $\bW$ fixed, and performs 1 iteration towards 
the optimization of the log-likelihood over $\bphi$ and $s_n$ using the BFGS algorithm \cite{byrd1995}.
We iterate between these two steps until the relative change in log-likelihood
falls below a threshold $\epsilon_l>0$.
Finally, we repeat the two step iteration process again, this time without 
constraining the number of iterations of each optimization process to 1.
We observe that this additional step forces the objective function 
to find a better local minimum. 
The procedure is outlined in Algorithm~\ref{alg:alt_alg}.
In order avoid getting trapped in a local optimum, we restart the algorithm
from multiple random initial points, 
$\btheta_0 = \{\bW_0, \bphi_0, s_{n,0}\}$ in Algorithm~\ref{alg:alt_alg},
and select the overall optimum.
To initialize $\bW_0$ we sample uniformly the Stiefel manifold $V_d\left(\R^D\right)$ using
Algorithm~\ref{alg:uniform_stiefel}.

\begin{algorithm}[htb]
    \caption{Uniform sampling of $V_d\left(\R^D\right)$ (for justification, see Bartlett decomposition theorem \cite{opac-b1079079}).}
    \label{alg:uniform_stiefel}
    \begin{algorithmic}[1]
        \REQUIRE Number of rows $D$ and number of columns $d$.
        \STATE Sample a random matrix $\mathbf{A}\in\R^{D\times d}$ with independent
        normally distributed entries:
        $$
        a_{ij} \sim \calN(a_{ij} | 0, 1),\;\text{for}\;i=1,\dots,D,j=1,\dots,d.
        $$
        \STATE Compute the QR-factorization of $\mathbf{A}$:
        $$\mathbf{A} = \mathbf{Q}\mathbf{R}.$$
        \RETURN $\bQ$
    \end{algorithmic}
\end{algorithm}

\subsubsection{Maximizing the likelihood with respect to the projection matrix}
\label{sec:stiefel_opt}

In this subsection, we consider the problem of maximizing the log-likelihood
with respect to $\bW\in V_d\left(\R^D\right)$ keeping the covariance hyper-parameters
$\bphi$ and the noise variance $s_n^2$ fixed. This is one of the steps required
by Algorithm~\ref{alg:alt_alg}. For notational convenience, define the function:
\begin{equation}
    \label{eqn:f_function}
    \mathcal{F}(\bW) := \mathcal{L}(\bW, \bphi, s_n; \bX, \by),
\end{equation}
where $\bphi$ and $s_n$ are supposed to be fixed. The optimization problem
we wish to solve is:
\begin{equation}
    \bW^* = \underset{\bW\in V_d\left(\R^D\right)}{\arg\max}\mathcal{F}(\bW).
    \label{eqn:stiefel_opt_problem}
\end{equation}
What follows requires the gradient of $\mathcal{F}(\cdot)$
with respect to $\bW$. This can be found from \qref{log_like_der} by setting $\psi \equiv w_{ij}$,
where $w_{ij}$ is the $(i,j)$ element of $\bW$,
and noticing that:
\begin{equation}
    \frac{\partial}{\partial w_{ij}}  k_{\mbox{AS}}\left(\bx, \bx'; \bW, \bphi\right) =
    \frac{\partial}{\partial z_j}\left[k_d(\bW^T\bx, \bW^T\bx';\bphi)\right]x_i +
    \frac{\partial}{\partial z_j'}\left[k_d(\bW^T\bx, \bW^T\bx';\bphi)\right]x_i',
    \label{eqn:grad_F}
\end{equation}
for the covariance function $k_{\mbox{AS}}(\cdot,\cdot;\bW,\bphi)$ introduced in \qref{as_cov_def},
where $\frac{\partial}{\partial z_j}$ denotes the partial derivative with
respect to the $j$-coordinate of the low-dimensional covariance function $k_d(\cdot,\cdot;\bphi)$.

\qref{stiefel_opt_problem} is a hard problem because of non-convexity as well as the
difficulty of preserving the orthogonality constraints.
We approach it using the procedure described in \cite{wen_2013},
a gradient ascent scheme on the Stiefel manifold in which
orthogonality is ensured via a Crank-Nicholson-like update involving the
Cayley transform.
To introduce the scheme, let $\bW\in V_d\left(\R^D\right)$ and define the curve 
\begin{eqnarray}
    \bm{\gamma}(\tau;\bW) = \left(\bI_D - \frac{\tau}{2}\bA(\bW)\right)^{-1}\left(\bI_D + \frac{\tau}{2}\bA(\bW) \right)\bW,
\end{eqnarray}
where
\begin{equation}
    \bA(\bW) := \nabla_{\bW}\mathcal{F}(\bW) \bW - \bW \left( \nabla_{\bW}\mathcal{F}(\bW) \right)^T.
\end{equation}
As shown in \cite{wen_2013}, the curve lives in the Stiefel manifold,
i.e., $\bm{\gamma}(\tau) \in V_d\left(\R^D\right)$, and it defines an ascent direction, i.e.,
\begin{equation}
    \frac{\partial }{\partial \tau}\mathcal{F}\left(\bm{\gamma}(\tau=0;\bW)\right) \ge 0.
\end{equation}
Fortunately, $\gamma(\tau;\bW)$ does not require the inversion of a $D\times D$
matrix, but can be computed in $O(d^3)$ flops (see \cite{wen_2013} for the
details). These results, suggest an optimization algorithm that iteratively
maximizes $\mathcal{F}(\cdot)$ over the curve $\bm{\gamma}(\cdot;\bW)$
until the relative change in $\mathcal{F}(\cdot)$ becomes smaller than a threshold
$\epsilon_s>0$. To solver the inner curve search problem, we use the efficient
global optimization (EGO) scheme~\cite{jones1998} which, typically, takes 2-5 evaluations of $\mathcal{F}$ to converge.
Other curve search algorithms could have been used (see \cite{wen_2013}).
See Algorithm~\ref{alg:stiefel_opt}.

\begin{algorithm}[htb]
    \caption{Stiefel manifold optimization}
\begin{algorithmic}[1]
    \REQUIRE Initial parameter $\bW_0$, maximum step size $\tau_{\max}>0$, maximum number of iterations $M_s$, tolerance $\epsilon_s>0$,
    all the fixed parameters required to evaluate $\mathcal{F}(\cdot)$ of \qref{f_function}.
            \STATE $\mathcal{F}_0 \leftarrow \mathcal{F}(\bW_0)$.
        \FOR{$i=1,\dots,M_s$}
        \STATE $\tau_{i} \leftarrow \underset{\tau\in[0,\tau_{\max}]}{\arg\max}\mathcal{F}\left(\bm{\gamma}(\tau;\bW_{i-1})\right)$ \COMMENT{Using EGO~\cite{jones1998}}
            \STATE $\bW_i \leftarrow \bm{\gamma}(\tau_{i};\bW_{i-1})$
                \STATE $\mathcal{F}_{i} = \mathcal{F}(\bW_{i})$
                \IF {$\frac{\mathcal{F}_{i}-\mathcal{F}_{i-1}}{\mathcal{F}_{i-1}} <  \epsilon_s$} 
      			\STATE break 
     		 \ENDIF
        \ENDFOR
   \RETURN $\bW_i$				
\end{algorithmic}
\label{alg:stiefel_opt}
\end{algorithm}

\subsubsection{Identification of active subspace dimension}
\label{sec:BIC}

Bayesian model selection involves assigning a prior on models and deriving
the posterior probability of each model conditioned on observable data~\cite{jaynes2003}.
This process requires the computation of the normalization
constant of the posterior of the hyper-parameters of each model being considered (see \qref{hyper_posterior}).
The logarithm of this normalization constant is known as the \emph{model evidence},
the equivalent of the partition function of statistical mechanics, is notoriously
difficult to calculate~\cite{neal1993,neal2001}.
The Bayesian information criterion ($\BIC)$~\cite[Ch. 4.4.1]{bishop2006} is
a crude, but cheap, approximation to the model evidence (up to an additive constant).
To define it, let $\btheta^*_d = \{\bW^*_d,\bphi^*_d,s^*_{n,d}\}$ be the MAP estimate of the hyper-parameters found
by Algorithm~\ref{alg:alt_alg}. The BIC score of the $d$-dimensional
AS model is:
\begin{equation}
    \BIC_d = \calL(\btheta^*_d;\bX,\by) - \frac{1}{2}\#(\btheta^*_d)\log N,
\label{eqn:BIC_score}
\end{equation}
where $N$ is the number of observations, and $\#(\btheta^*_d)$ is the number of estimated parameters $\btheta^*_d$:
\begin{equation}
    \#(\btheta^*_d) = \#(\bW^*_d) + \#(\bphi_d^*) + \#(s_{n,d}^*) = dD + \#(\bphi_d^*) + 1.
\end{equation}
That is, the BIC is equal to the maximum log-likelihood minus a term that penalizes model complexity.
Typically, $\BIC_d$ increases as a function of $d$.
The sharper the increase of the $\BIC$ from $d$ to $d+1$,
the stronger the evidence that the most complex model is closest to the truth.
Motivated by this, we propose an algorithm that sequentially
increases $d$ until the relative change in $\BIC$ becomes smaller than a
threshold $\epsilon_b>0$. This is summarized in Algorithm~\ref{alg:BIC}.

\begin{algorithm}[htb]
\caption{Identification of active subspace dimension}
\begin{algorithmic}[1]
    \REQUIRE Maximum allowed AS dimensionality $d_{\max}$, tolerance $\epsilon_b$, all the data and parameters required to run Algorithm~\ref{alg:alt_alg}.
        \STATE $\BIC_{0} \leftarrow -\infty$
            \FOR{$d=1,\dots,d_{\max}$}
        \STATE{Find $\btheta_d^*$ by running Algorithm~\ref{alg:alt_alg} for a $d$-dimensional AS}
        \IF{$\frac{\BIC_d - \BIC_{d-1}}{\BIC_{d-1}}\le \epsilon_b$}
                \STATE break
        \ENDIF
       \ENDFOR
\end{algorithmic}
\label{alg:BIC}
\end{algorithm}

\section{Examples}
\label{sec:examples}

We have implemented both the classic approach, \sref{classic_approach}, and
the proposed gradient-free approach, \sref{grad_free_approach}, in Python. Our code
extends the GPy module \cite{gpy} and is publicly available at~\cite{pyaspgp}.
All the numerical
results we present here can be replicated by following the instructions on the
aforementioned website. In all cases, we used the same parameters for our
optimization algorithms.
Specifically, we used $1,000$ restarts of log-likelihood optimization,
Algorithm~\ref{alg:alt_alg}, with parameters $M_l=10000$, $\epsilon_l=10^{-16}$, $m=1$ and $n=1$.
For the Stiefel manifold optimization, Algorithm~\ref{alg:stiefel_opt},
we used $\tau_{\max}=0.1$, $M_s=10000$, and $\epsilon_s=10^{-16}$.
Finally, we used $\epsilon_d=10^{-3}$ in Algorithm~\ref{alg:BIC}.

\sref{exa_synthetic} uses a series of synthetic examples (known projection
matrix and known non-linear link function) to verify that the proposed
approach, \sref{grad_free_approach}, finds the same AS as the classic approach,
\sref{classic_approach}. Our goal is to address the first identified drawback
of classic AS, namely the reliance on gradient information. Furthermore, this
section validates our claim that the proposed methodology is robust to
measurement noise. In \sref{exa_elliptic}, we apply our technique to a standard
UQ benchmark with one hundred input dimensions, a stochastic elliptic partial
differential equation with random conductivity. The results are again compared 
to the classic AS, thereby verifying the agreement between the two in a more
challenging, truly high-dimensional setting. We conclude this section with an
exhaustive uncertainty analysis of a one-dimensional granular crystal with geometric and
material imperfections, see \sref{exa_gc}. The latter is not amenable to the
classic AS approach due to lack of gradient information. Note that, to the best
of our knowledge, this is the first time an uncertainty analysis of this scale
has been performed to a granular crystal.

\subsection{Synthetic response surface with known structure}
\label{sec:exa_synthetic}

Let $f:\R^D\rightarrow\R$ be a response surface of the form:
\begin{equation}
    f(\bx) = g\left(\bW^T\bx\right),
    \label{eqn:synthetic_f}
\end{equation}
with $\bW\in V_d\left(\R^D\right)$, and quadratic link function $g:\R^d\rightarrow\R$,
\begin{equation}
    g(\bz) = \alpha + \bbeta^T\bz + \bz^T\bGamma\bz,
    \label{eqn:synthetic_g}
\end{equation}
with $\alpha\in\R, \bbeta\in\R^d$ and $\bGamma\in\R^{d\times d}$. The gradient
of \qref{synthetic_f} with respect to $\bx$ is:
\begin{equation}
    \nabla f(\bx) = \left(\bbeta + 2\bx^T \bW\bGamma\right)\bW^T.
    \label{eqn:synthetic_f_grad}
\end{equation}
In all the cases considered in this subsection, the number of input dimensions
is ten, $D=10$.  The parameters $\bW, \balpha, \bbeta$ and $\bGamma$ were
randomly generated. Reproducibility is ensured by fixing the random
seed. Due to lack of space, we only give the values of these parameters when
the dimension of the active subspace, $d$, is lower than or equal to two. For
all other cases, we refer the reader to the accompanying website of this paper~\cite{pyaspgp}.
Given a frozen set of $\alpha,\bbeta$ and $\bGamma$, we query the response $f(\cdot)$ at $N$
normally distributed input points and contaminate the measurements with
synthetic zero mean Gaussian noise with standard deviation $s_n>0$. This
results in a collection of inputs, $\bX$ as in \qref{observed_input}, and
outputs, $\by$ as in \qref{observed_output}. When needed, we also collect
gradient data, $\bG$ as in \qref{observed_gradients}, but we do not contaminate
them with noise.

In \sref{exa_synthetic_1d} and \sref{exa_synthetic_2d}, we verify that the
gradient-free approach discovers the underlying 1D and 2D AS
structure, respectively. \sref{exa_synthetic_bic} demonstrates the efficacy of
the BIC as an automatic method dimensionality detection method.
Finally, in \sref{exa_synthetic_noise} we study the robustness of the
gradient-free approach to measurment noise.

\subsubsection{Synthetic response with 1D active subspace}
\label{sec:exa_synthetic_1d}

\begin{figure*}
    \centering
    \subfigure[] {
        \includegraphics[width=80mm]{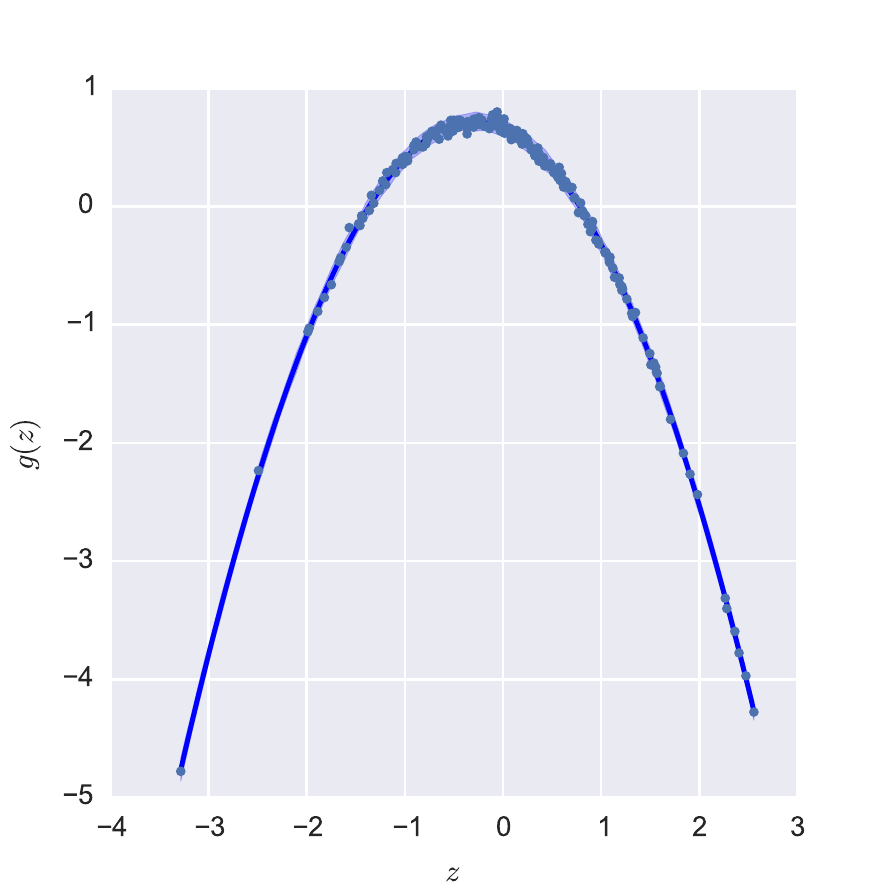}
    }
    \subfigure[] {
        \includegraphics[width=80mm]{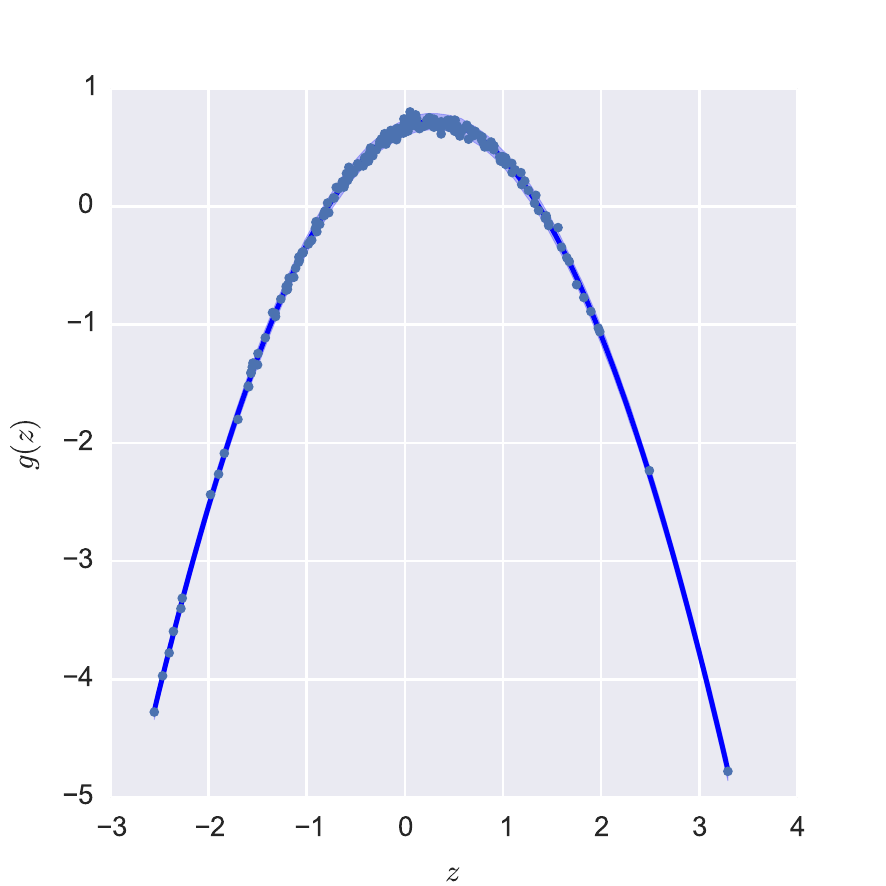}
    }
    \subfigure[] {
        \includegraphics[width=80mm]{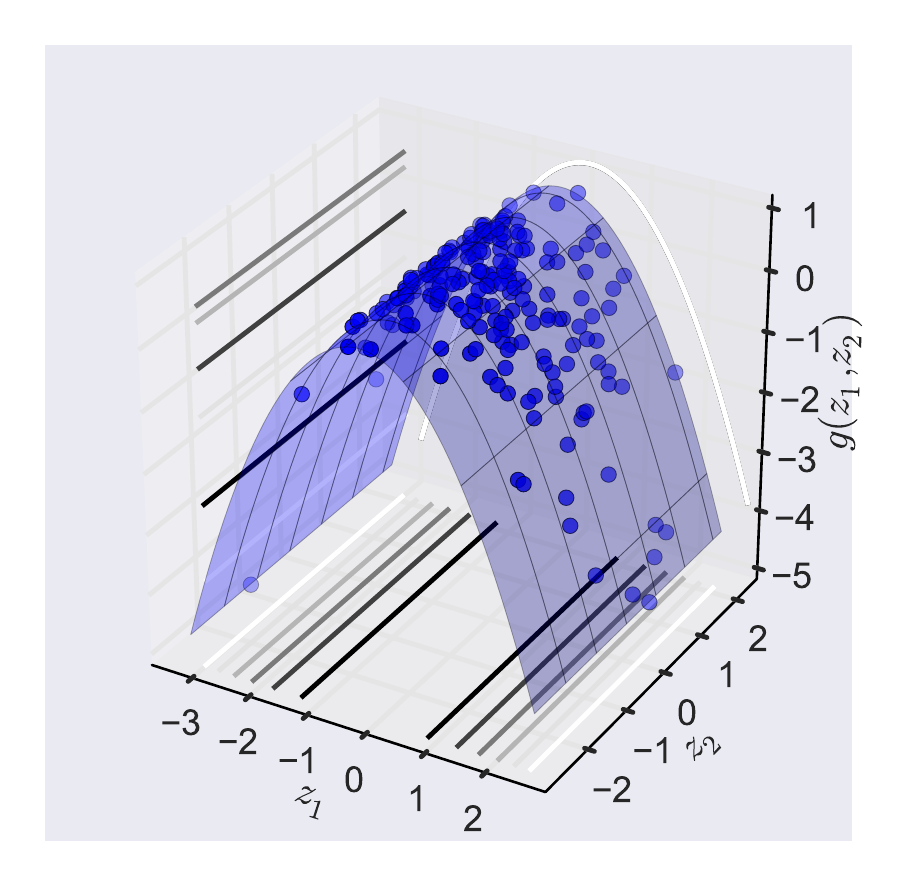}
    }
    \subfigure[] {
        \includegraphics[width=80mm]{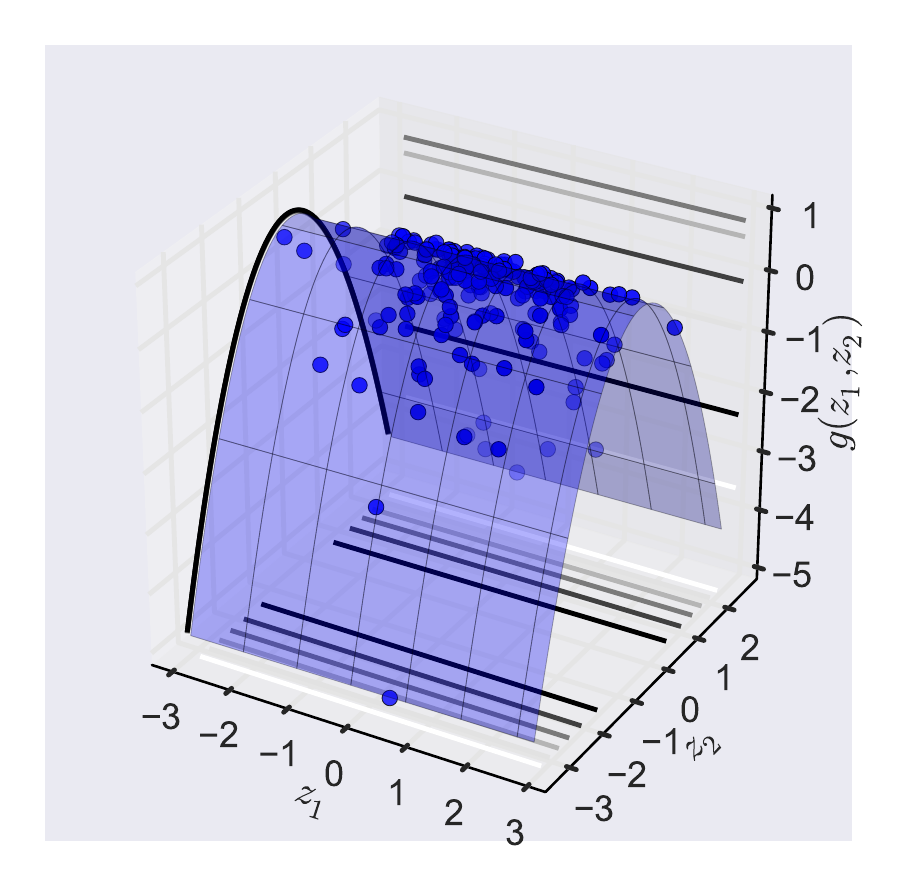}
    }
    \subfigure[] {
        \includegraphics[width=80mm]{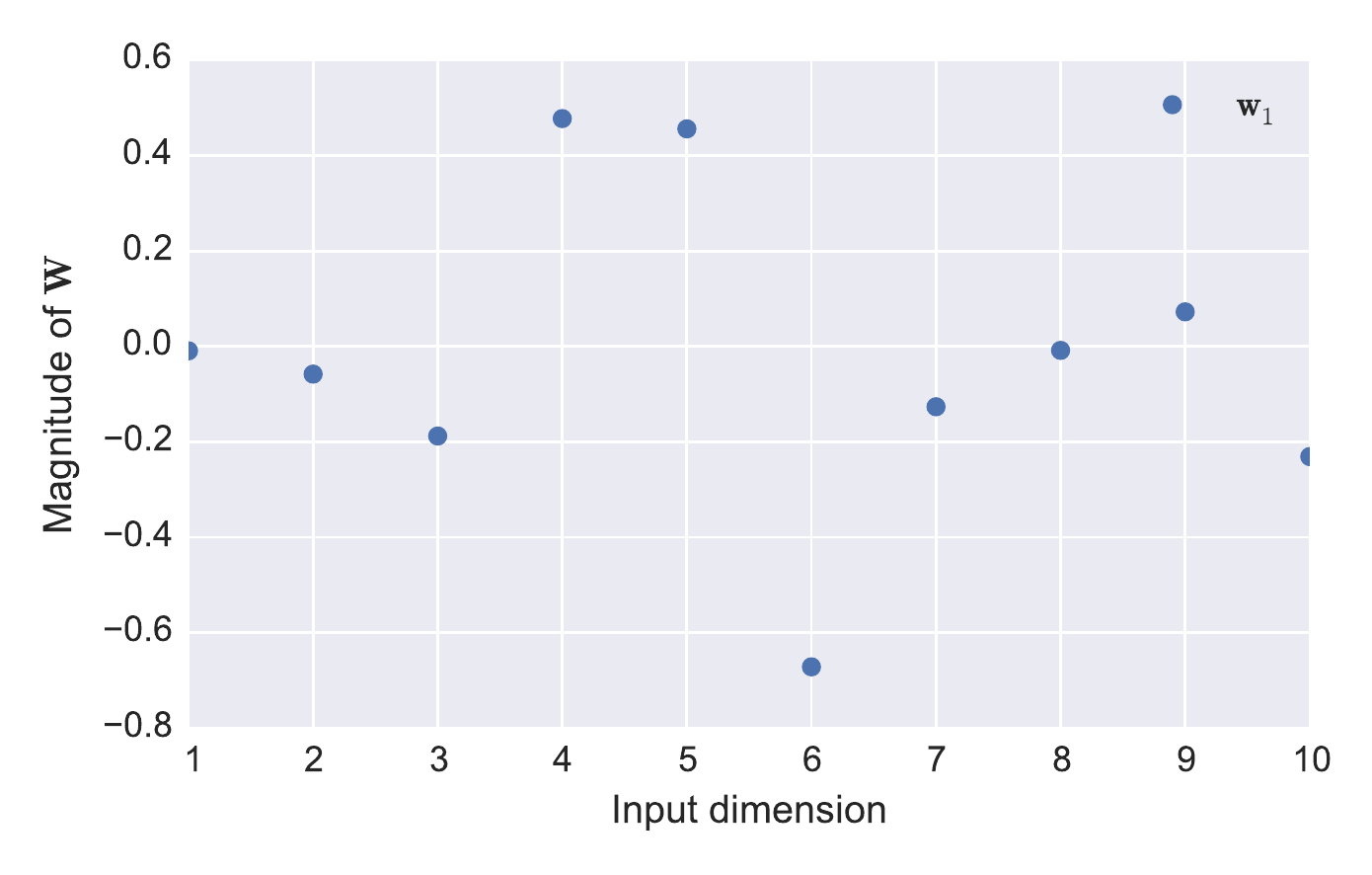}
    }
    \subfigure[] {
        \includegraphics[width=80mm]{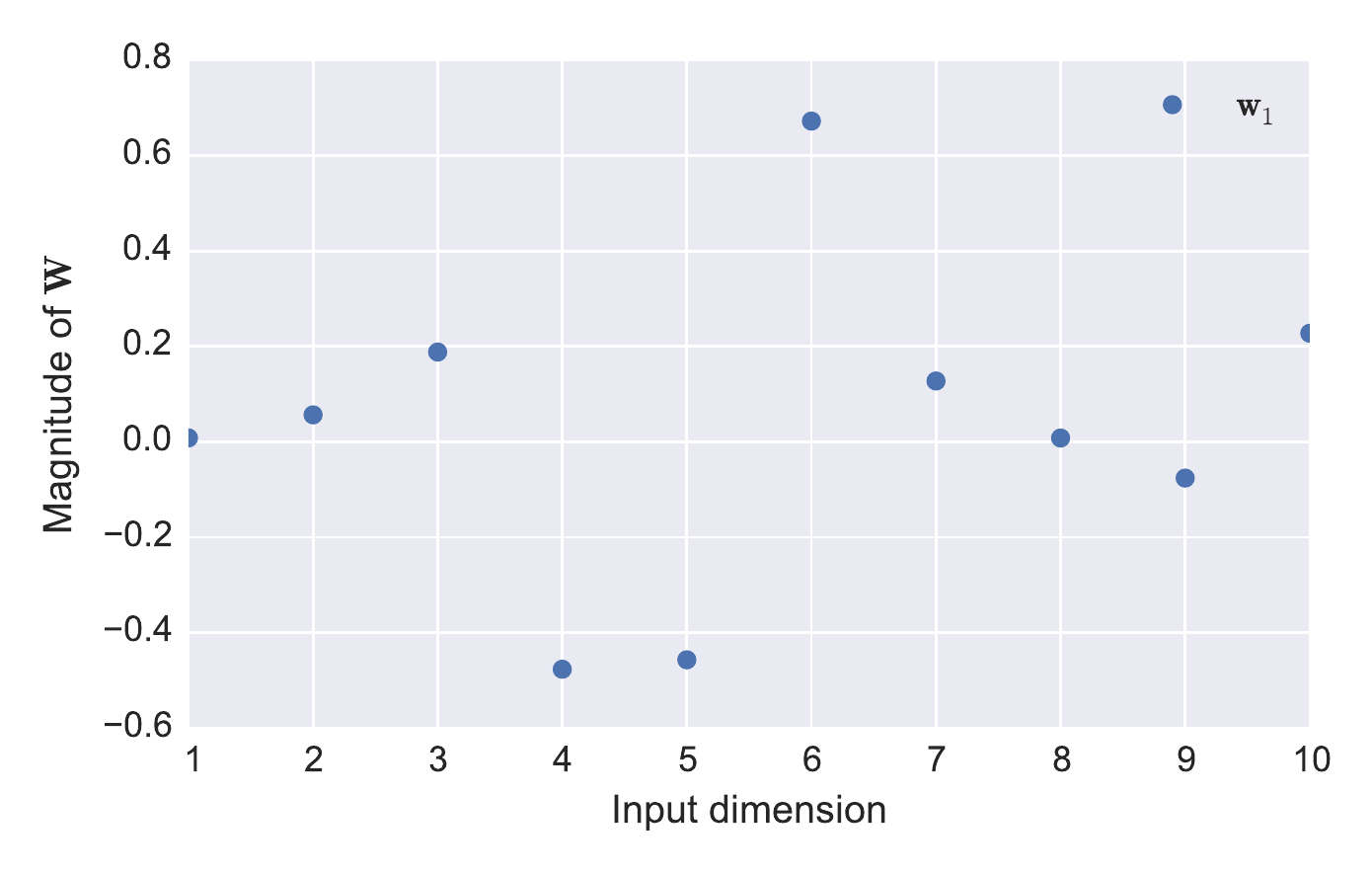}
    }
\caption{
Synthetic example $d=1$. The left and the right columns correspond to results obtained
with the classic and the gradient-free approach respectively. The first and
second rows depict the predictions of each method for the link function
assuming a 1D and 2D underlying AS, respectively, along with a scatter plot of
the projections of 60 validation inputs vs the validation outputs. The third
row visualizes the projection matrix that each method discovers.
}
\label{fig:ex1_1d}
\end{figure*}    

In this example the underlying AS is 1D, $d=1$. The projection matrix is:
\begin{equation}
    \bW = \left(\begin{array}{cccccccccc}
            -0.0091, &
            -0.0579, &
            -0.1877, &
             0.4774, &
             0.4559, &
            -0.6714, &
            -0.1264, &
            -0.0082, &
             0.0724, &
            -0.2308
    \end{array}\right)^T,
    \label{eqn:exa_synthetic_1d_W}
\end{equation}
and the parameters of the link function of \qref{synthetic_g} are:
\begin{equation}
\alpha = -0.16113, \bbeta = (\begin{array}{c}-0.97483\end{array}),\;\text{and}\;
\bGamma = (\begin{array}{c}-1.66526\end{array}).
    \label{eqn:exa_synthetic_1d_g}
\end{equation}
We make $N=140$ observations with noise variance $s_n^2 = 0.1$. 
In this first example, we do not make use of the automatic method for the
detection of the dimensionality of the AS. Rather, we use the plain vanilla
version of both the classic and the gradient-free approaches assuming a
1D or a 2D AS. 
For validation 60 random input/output pairs, not used
in the training process, were generated.
\fref{ex1_1d} compares the results obtained with both methodologies.
The left column corresponds to the classic approach and the right one to the
gradient-free approach.
The first row, Fig.\ref{fig:ex1_1d}(a) and~(b), shows the link function 
learned by each approach assuming $d=1$, along with a $95\%$ prediction interval,
and a scatter plot of the validation input/output pairs.
The quantitative agreement between the two approaches becomes obvious once one
recalls that the representation of \qref{f} is arbitrary up to permutations and
reflections of the reduced dimensions.
This is confirmed by looking at the projection matrices discovered by each
method, shown in \fref{ex1_1d}(e) and (f), respectively.
It is clearly seen that one is the negative image of the other.
In \fref{ex1_1d}(c) and (d), we depict the link function learned by assuming
that $d=2$.
Note that, both methods correctly discovered one completely flat direction.

\subsubsection{Synthetic response with 2D active subspace}
\label{sec:exa_synthetic_2d}

\begin{figure*}
    \centering
    \subfigure[] {
        \includegraphics[width=80mm]{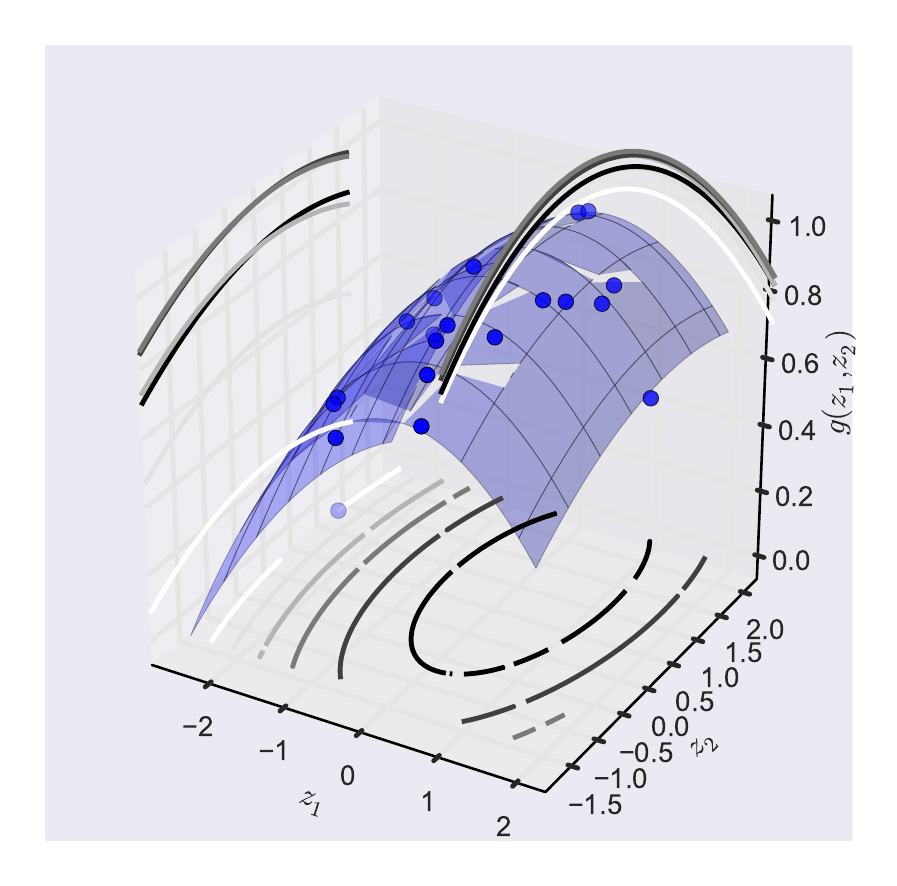}
    }
    \subfigure[] {
        \includegraphics[width=80mm]{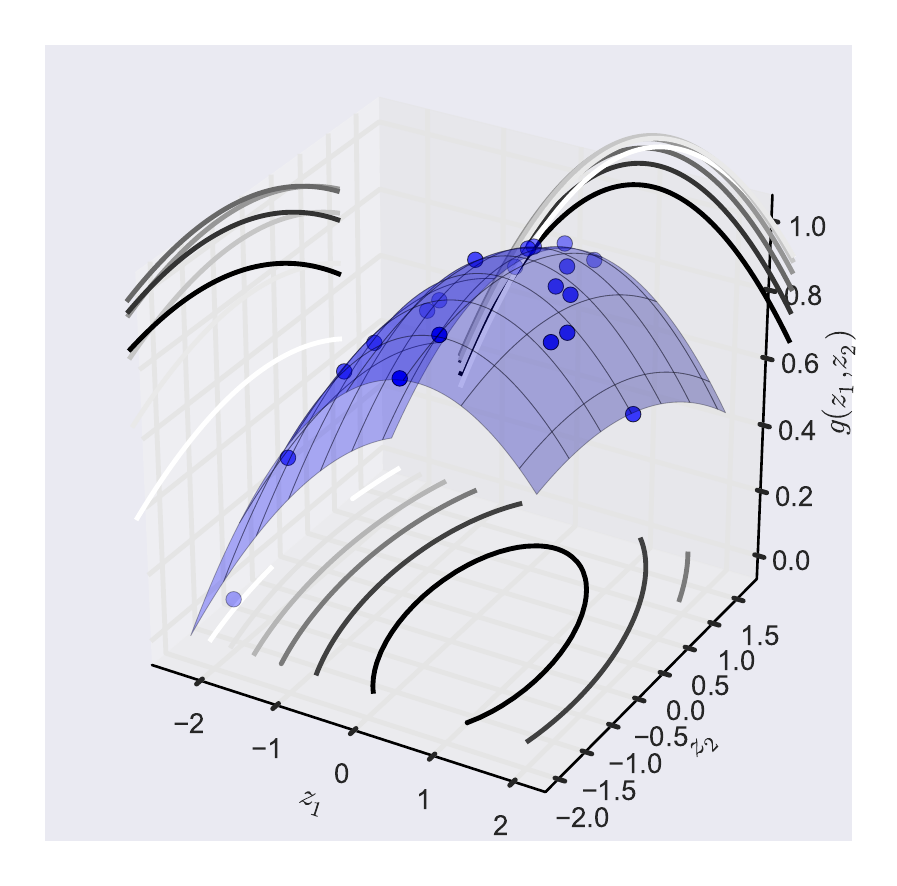}
    }
    \subfigure[] {
        \includegraphics[width=80mm]{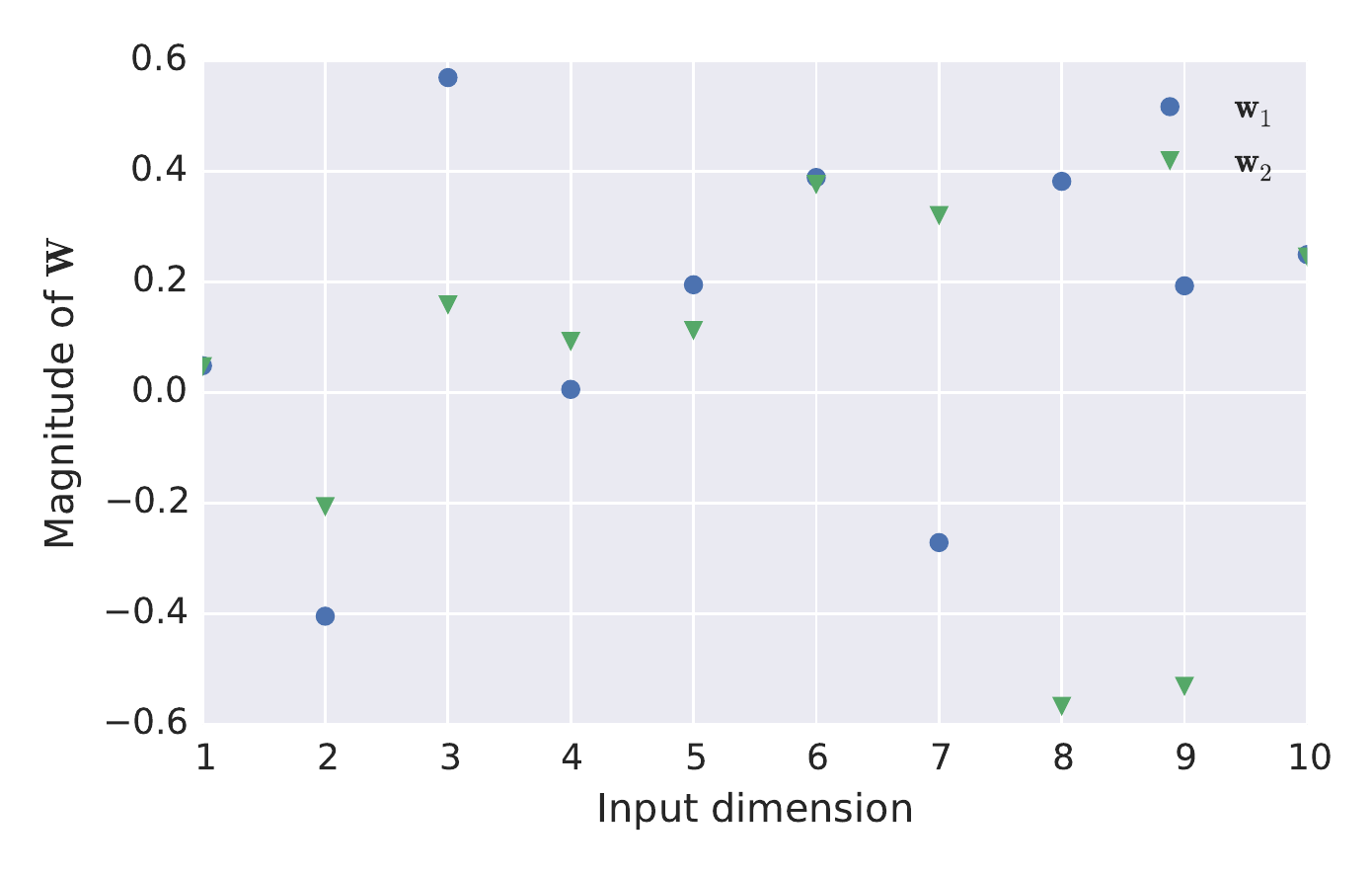}
    }
    \subfigure[] {
        \includegraphics[width=80mm]{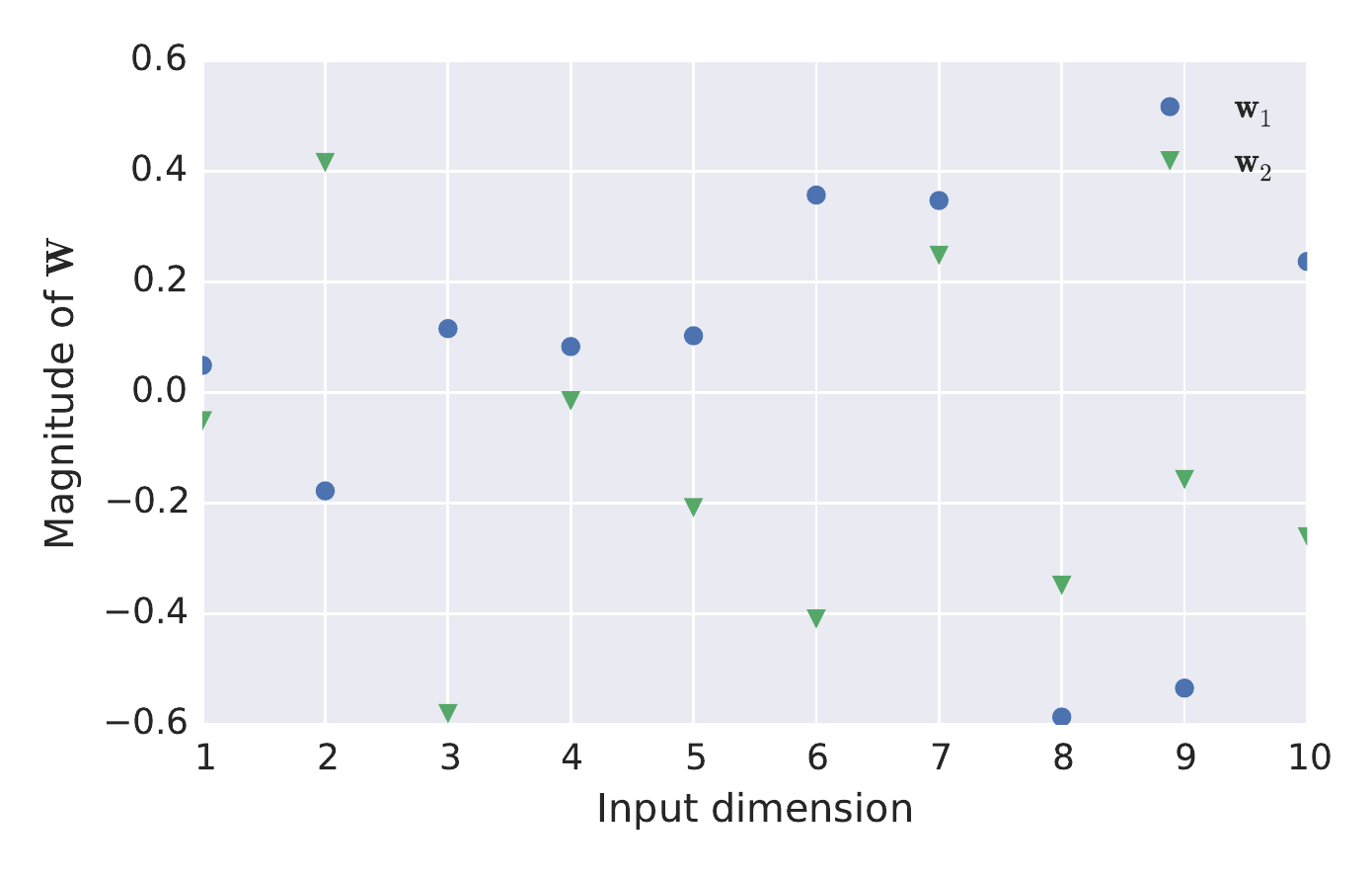}
    }
\caption{
Synthetic example $d=2$. The left and the right columns correspond to results obtained
with the classic and the gradient-free approach respectively. The first
row depicts the predictions of each method for the link function
assuming a 2D underlying AS, along with a scatter plot of
the projections of the 60 validation inputs vs the validation outputs. The second
row visualizes the projection matrix that each method discovers.
}
\label{fig:ex1_2d}
\end{figure*}

In this example the underlying AS is 2D, $d=2$. The projection matrix is:
\begin{equation}
    \bW = \left(\begin{array}{cccccccccc}
            0.00840 & 
            -0.18426& 
            0.34300 & 
            -0.05347& 
            0.08108 & 
            0.06556 & 
            -0.41219& 
            0.65424 & 
            0.48483 & 
            0.03966 \\
            0.0672 & 
            -0.4148 &
            0.4821 &
            0.0755 &
            0.2101 &
            0.5375 &
            0.0781 &
            -0.2002 &
            -0.2912 &
            0.3480
    \end{array}\right)^T,
    \label{eqn:exa_synthetic_2d_W}
\end{equation}
and the parameters of the link function of \qref{synthetic_g} are:
\begin{equation}
\alpha = -0.06976, \bbeta = \left(\begin{array}{c}0.43759 \\ 0.98696041 \end{array}\right),\;\text{and}\;
\bGamma = \left(\begin{array}{cc}
        -0.92567723 & -0.38399783 \\
        -0.41740642 & -0.67655046
\end{array}\right).
    \label{eqn:exa_synthetic_2d_g}
\end{equation}
As in \sref{exa_synthetic_1d}, we make $N=140$ observations with noise
variance $s_n^2 = 0.1$,
We do not make use of Algorithm~\ref{alg:BIC} for the automatic detection of
the dimensionality, but rather set $d=2$.
\fref{ex1_2d} depicts the results.
As before, the left column corresponds to the classic approach and the right
column to the gradient-free approach.
The first row,\fref{ex1_2d}(a) and (b), shows the learned link function
along with a scatter plot of 20 randomly generated input/output pairs.
The second row,\fref{ex1_2d}(c) and (d), shows the learned projection matrices.
The quantitative agreement between the two approaches up
to permutations and reflections of the AS is also obvious.

\subsubsection{Validation of BIC for the identification of the active subspace dimension}
\label{sec:exa_synthetic_bic}

\begin{figure*}
    \centering
    \subfigure[] {
        \includegraphics[width=80mm]{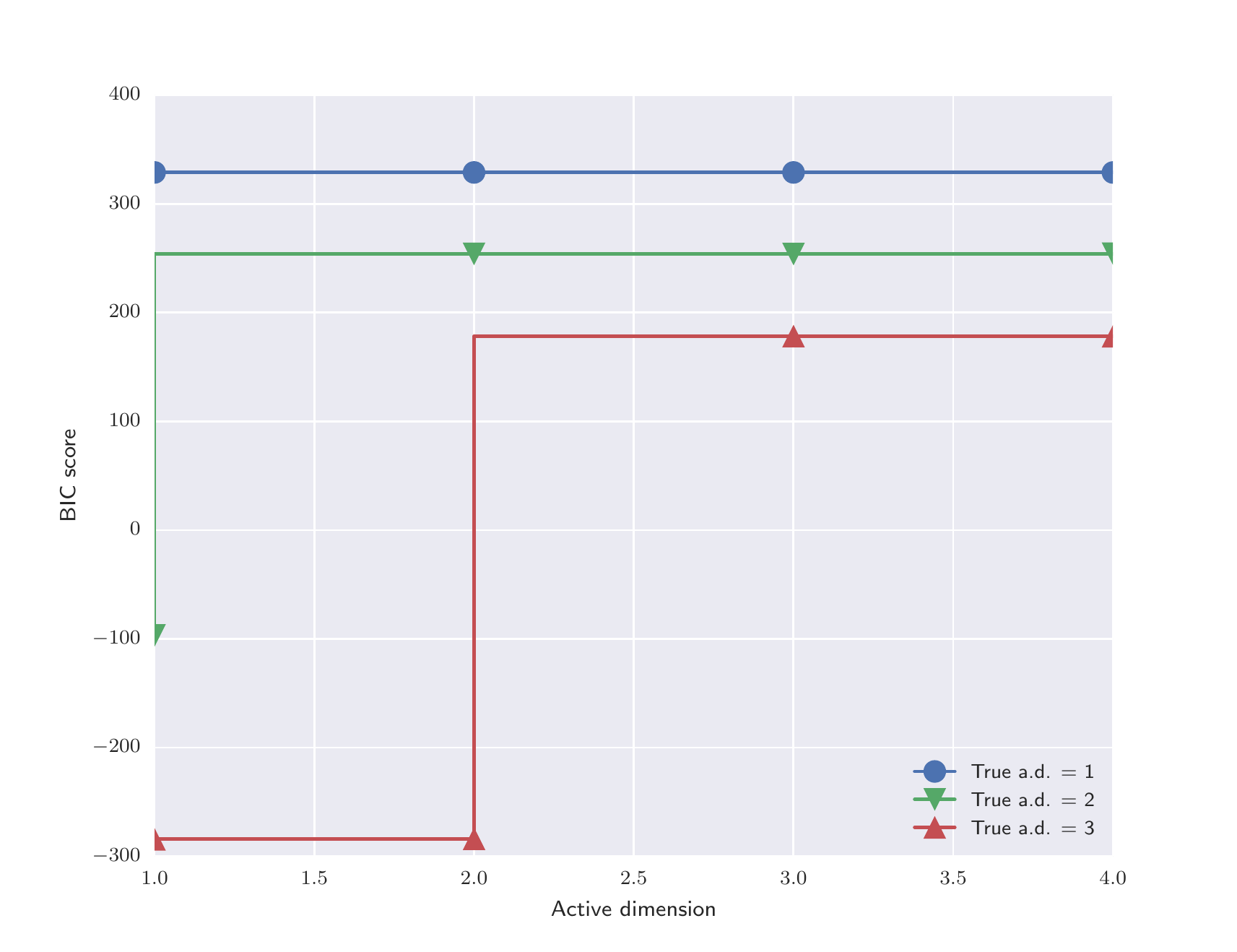}
    }
    \subfigure[] {
        \includegraphics[width=80mm]{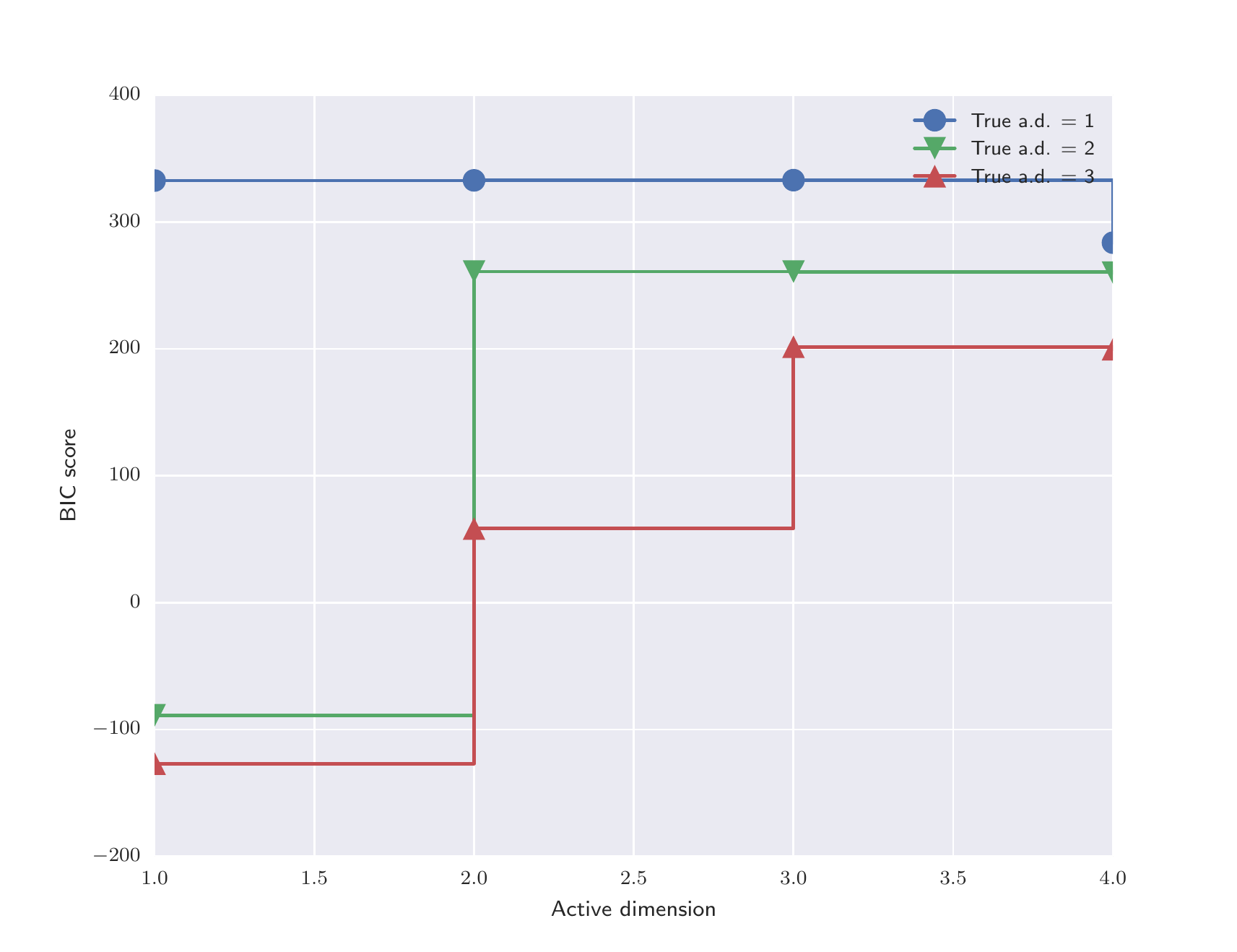}
    }
\caption{
Synthetic example. BIC score as a function of the hypothesized active dimension
for classic model~(a) and the gradient-free model~(b). The different lines
correspond to cases with a 1D (blue, true response as in
\sref{exa_synthetic_1d}),
2D (green, true response as in \sref{exa_synthetic_2d}, and 3D
(red, true response as in details on the accompanying website) true
AS.
}
\label{fig:exa_synthetic_bic}
\end{figure*}  

Here, we verify the effectiveness of the BIC, \sref{BIC}, to automatically
determine the dimensionality of the AS for both the classic and the
gradient-free approach. The hypothesis is that the $\BIC_d$ becomes
flat as a function of $d$ after $d$ exceeds the true AS dimensionality.
This is confirmed numerically in \fref{exa_synthetic_bic} for the cases of a 1D, 2D,
and 3D true AS.
Note that for the 1D and 2D examples, the observations we used
to train the models were the same as in \sref{exa_synthetic_1d} and
\sref{exa_synthetic_2d}.
For the 3D true AS case also has an underlying
response surfaces with randomly generated $\alpha, \bbeta, \bGamma$, and $\bW$.
The values used can be found in the accompanying website.
As before, we used $N=140$ observations.

\subsubsection{Validation of robustness to measurement noise}
\label{sec:exa_synthetic_noise}

\begin{figure*}
    \centering
    \subfigure[]{
        \includegraphics[width=80mm]{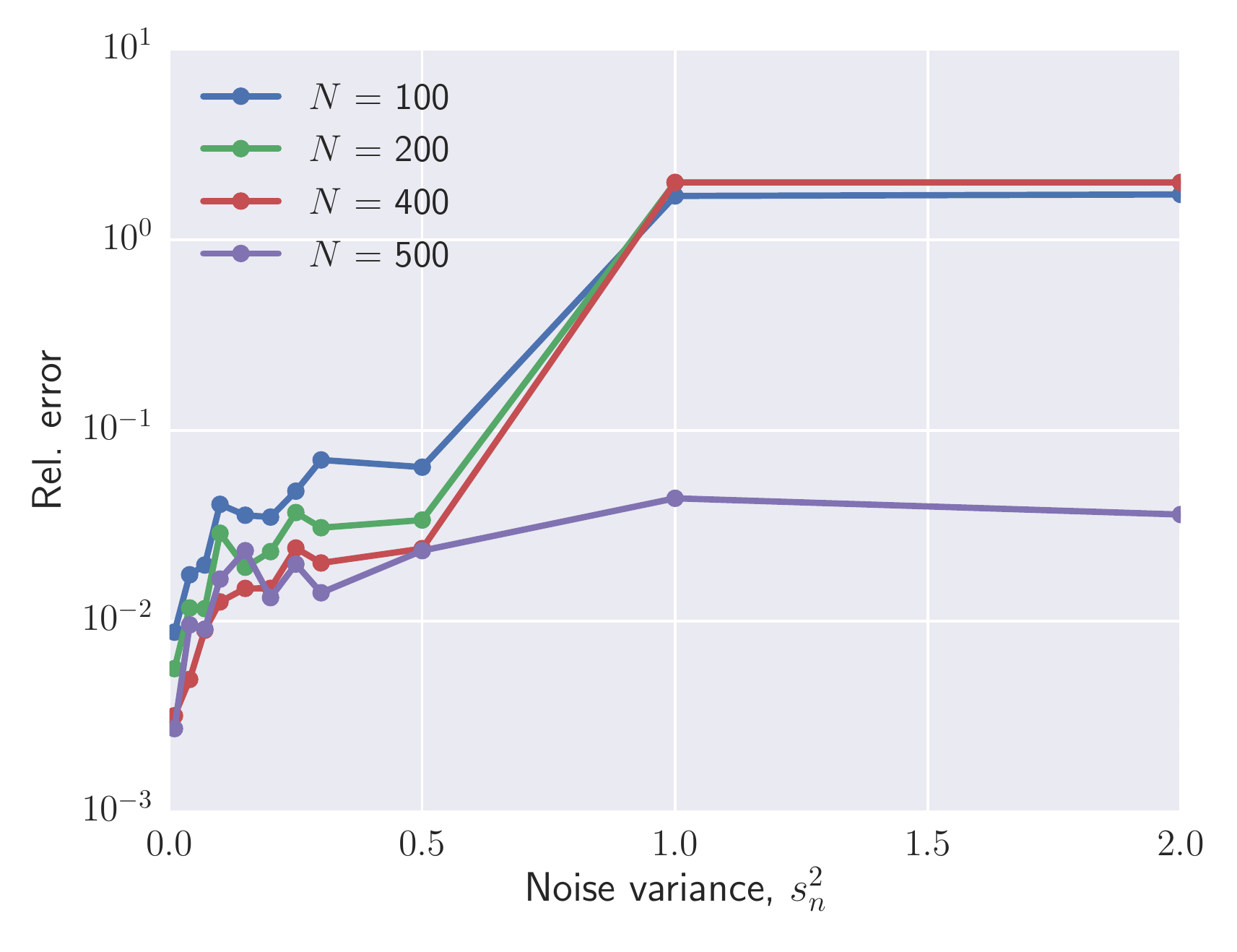}
    }
    \subfigure[]{
        \includegraphics[width=80mm]{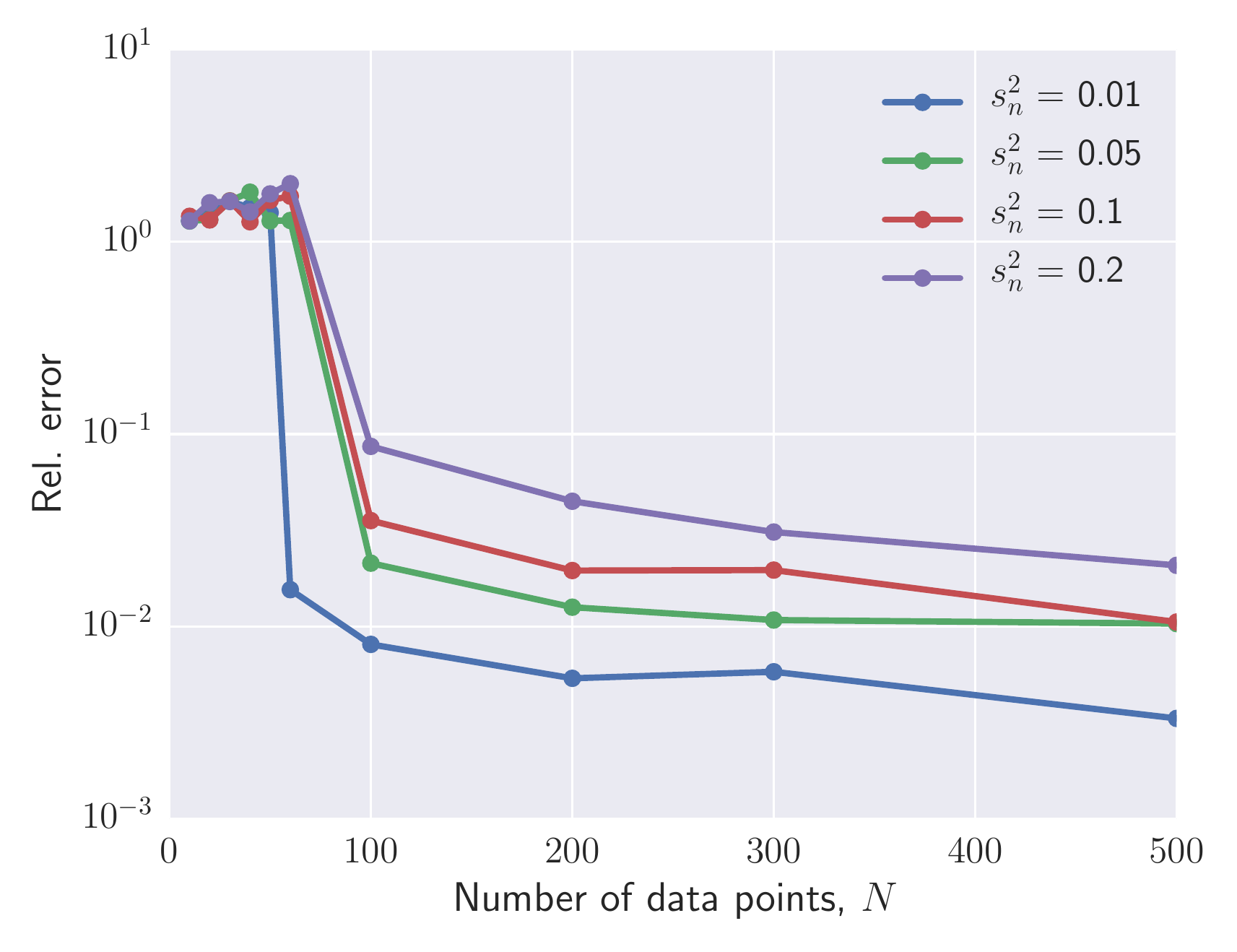}
    }
    \caption{
Synthetic example. Robustness of the proposed approach to measurement noise.
The figure shows the evolution of the relative error in the
determination of the true active subspace as a function of the measurement
noise variance (keeping the number of observations constant)~(a) and as a
function of the number of observations (keeping the measurement noise variance
constant~(b)).
}
\label{fig:exa_synthetic_noise}
\end{figure*}

We conclude this subsection with a study of the robustness of the proposed
scheme to measurement noise. To avoid the non-uniqueness issues mentioned
earlier, we work with the 10D-input-1D-AS
response surface
of \sref{exa_synthetic_1d}. In this case, the arbitrariness can be removed by
making sure that the signs of the estimated and the true projection matrix
match. We want to quantify the ability of the model
to discover the true AS and how this is affected by changes in the measurement
noise, $s_n$, as well as in the number of available observations $N$.
A good measure of this ability is the relative error in the estimation of
the projection matrix:
\begin{equation}
    \epsilon_{\mbox{rel}}(s_n, N) =
    \frac{\|\bW(s_n, N)-\bW\|_F}{\|\bW\|_F},
\label{eqn:exa_synthetic_relerr}
\end{equation}
where $\|\cdot\|_F$ is the Frobenius norm, $\bW(s_n, N)$ is the estimated
projection matrix when $N$ measurements contaminated with zero mean Gaussian
noise of variance $s_n^2$ are used, and $\bW$ is the true projection matrix
given in \qref{exa_synthetic_1d_W}. The results of our analysis are presented
in \fref{exa_synthetic_noise}. \fref{exa_synthetic_noise}(a) plots the relative
error, $\epsilon_{\mbox{rel}}$, as a function of $s_n^2$ for $N=30,100,200$, and
$500$. As expected, we observe that $\epsilon_{\mbox{rel}}$ increases as a
function of $s_n^2$ and that a larger $N$ is required to maintain a given
accuracy. \fref{exa_synthetic_noise}(b) plots the relative error,
$\epsilon_{\mbox{rel}}$, as a function of $N$ for $s_n=0.01, 0.05, 0.1$ and
$0.2$. We note that the method converges to the right answer as $N$ increases,
albeit the rate of convergence decreases for higher noise.

\begin{figure*}
    \centering
    \subfigure[] {
        \includegraphics[width=80mm]{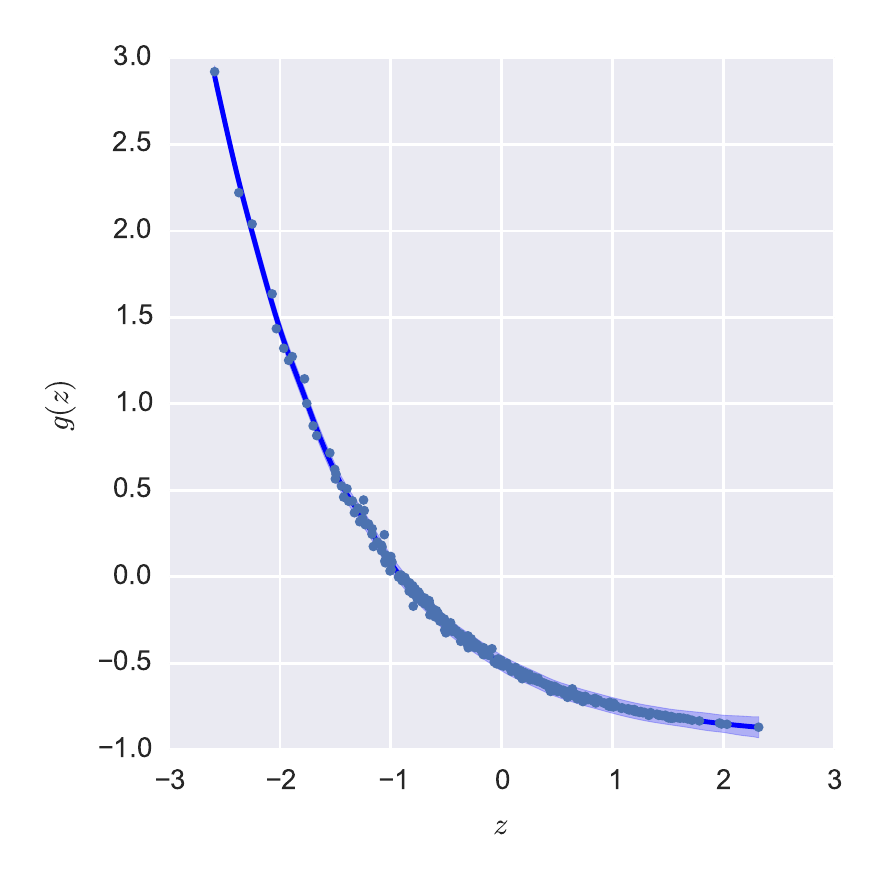}
    }
    \subfigure[] {
        \includegraphics[width=80mm]{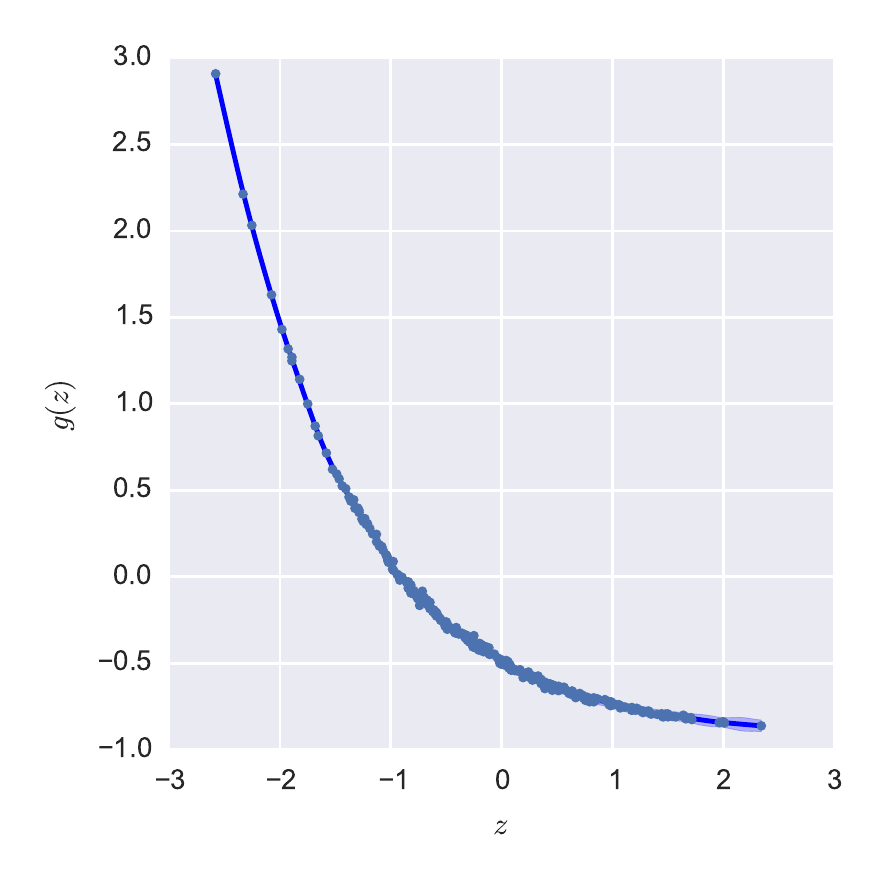}
    }
    \subfigure[] {
        \includegraphics[width=80mm]{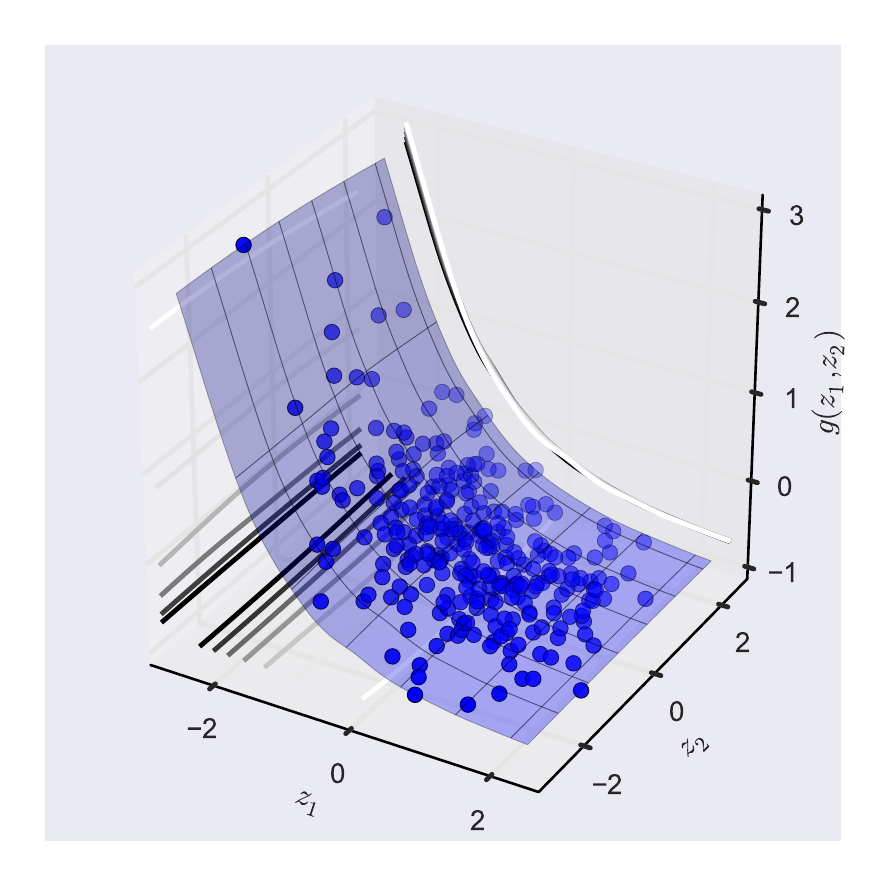}
    }
    \subfigure[] {
        \includegraphics[width=80mm]{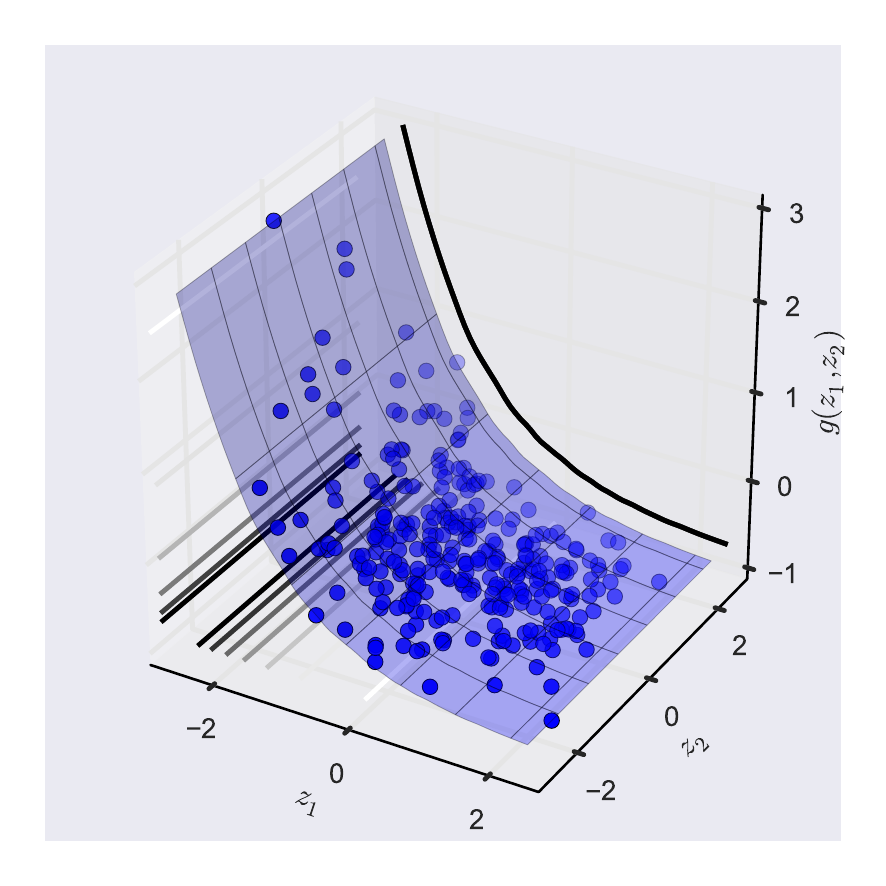}
    }
    \subfigure[] {
        \includegraphics[width=80mm]{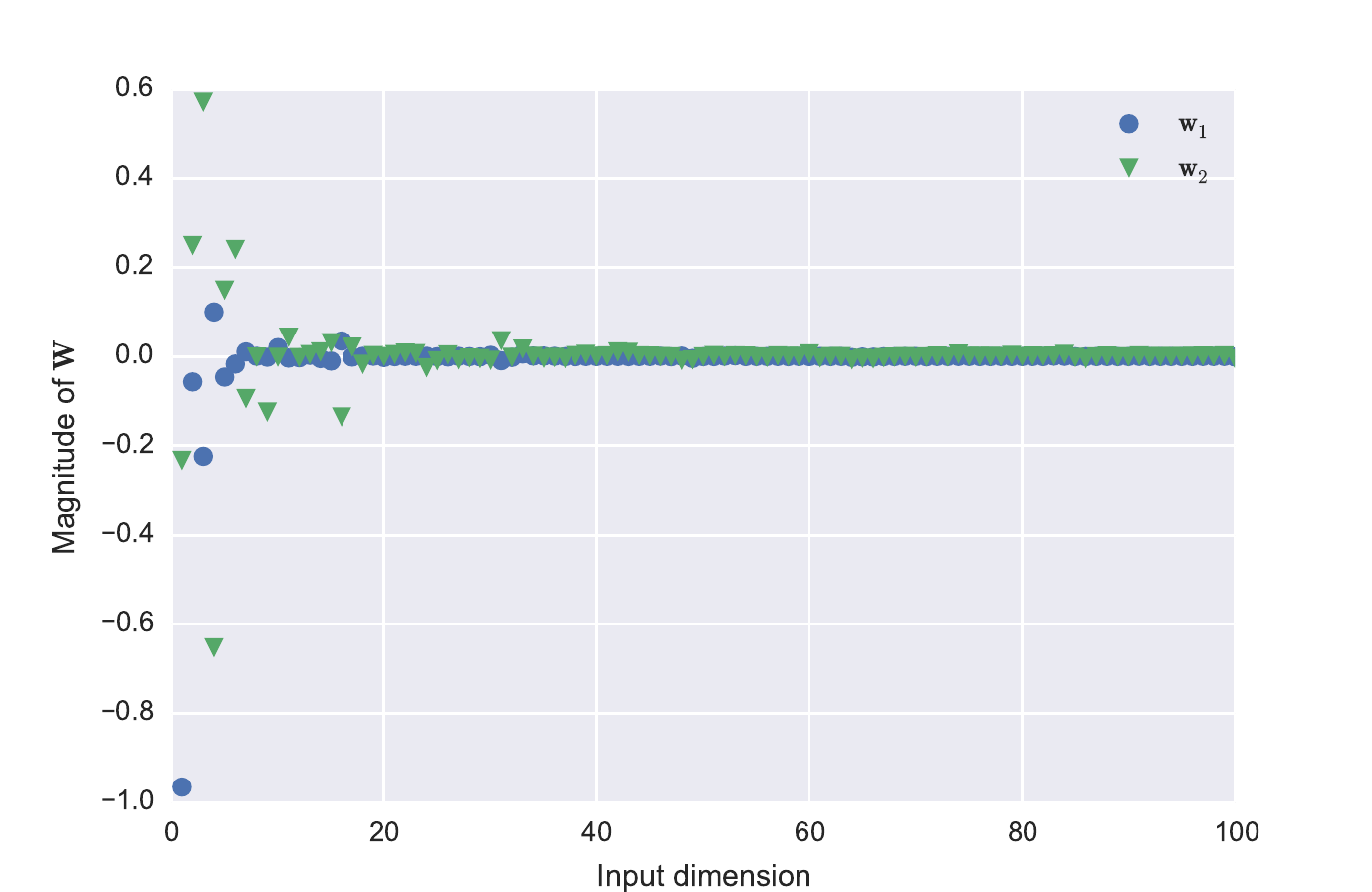}
    }
    \subfigure[] {
        \includegraphics[width=80mm]{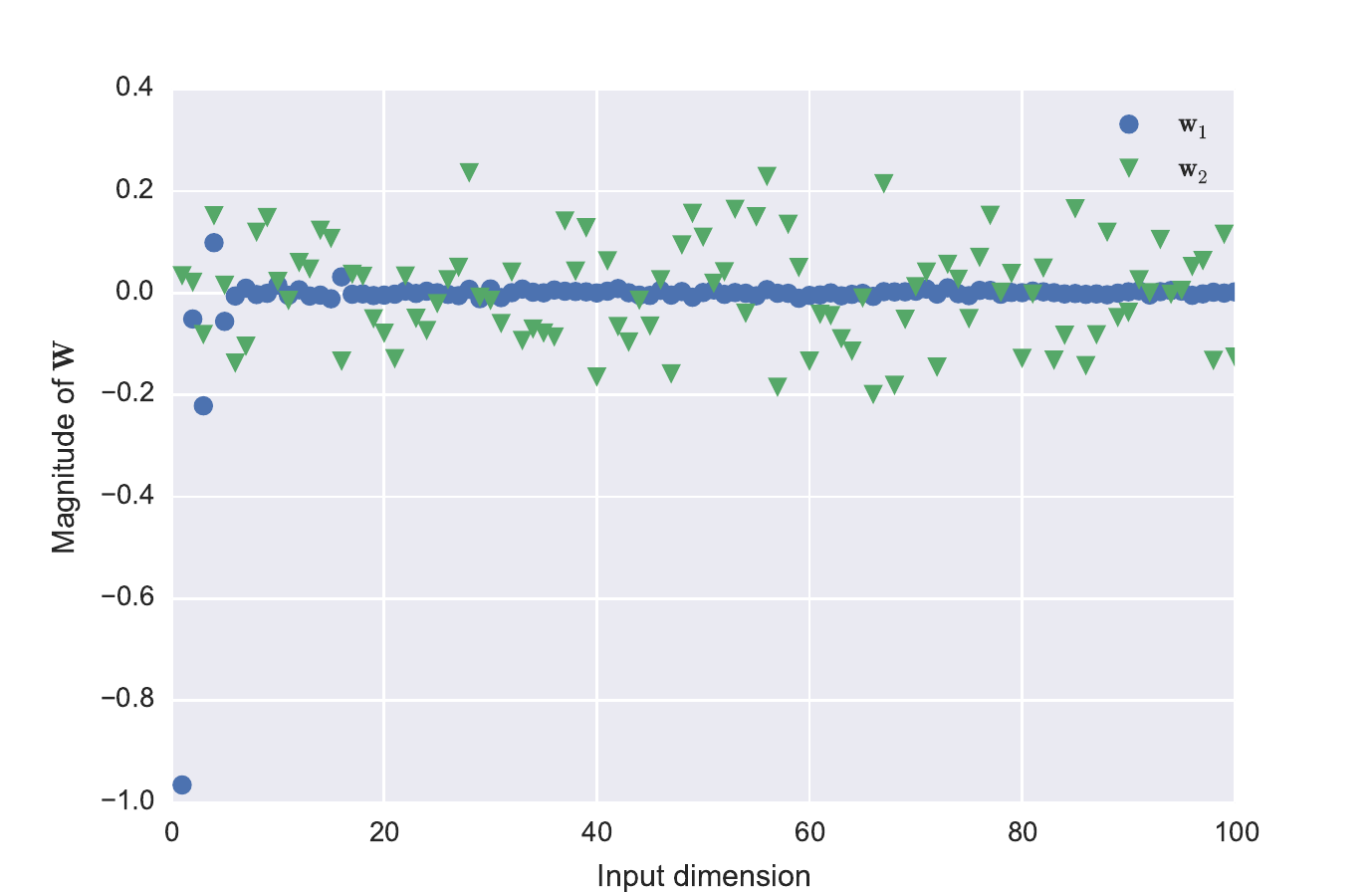}
    }
\caption{
    Elliptic PDE, long correlation length ($\ell=1$).
The left and the right columns correspond to results obtained
with the classic and the gradient-free approach respectively. The first and
second rows depict the predictions of each method for the link function
assuming a 1D and 2D underlying AS, respectively, along with a scatter plot of
the projections of 30 validation inputs vs the validation outputs. The third
row visualizes the projection matrix that each method discovers. }
\label{fig:exa_elliptic_long_corr}
\end{figure*}

\begin{figure*}
    \centering
    \subfigure[] {
        \includegraphics[width=80mm]{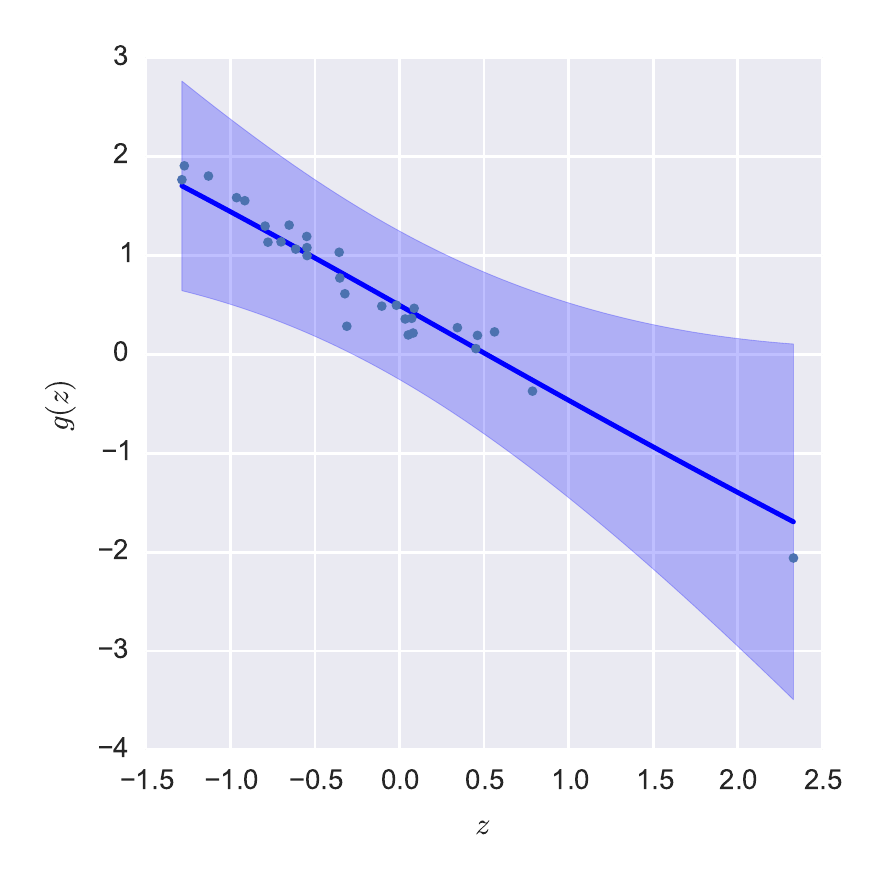}
    }
    \subfigure[] {
        \includegraphics[width=80mm]{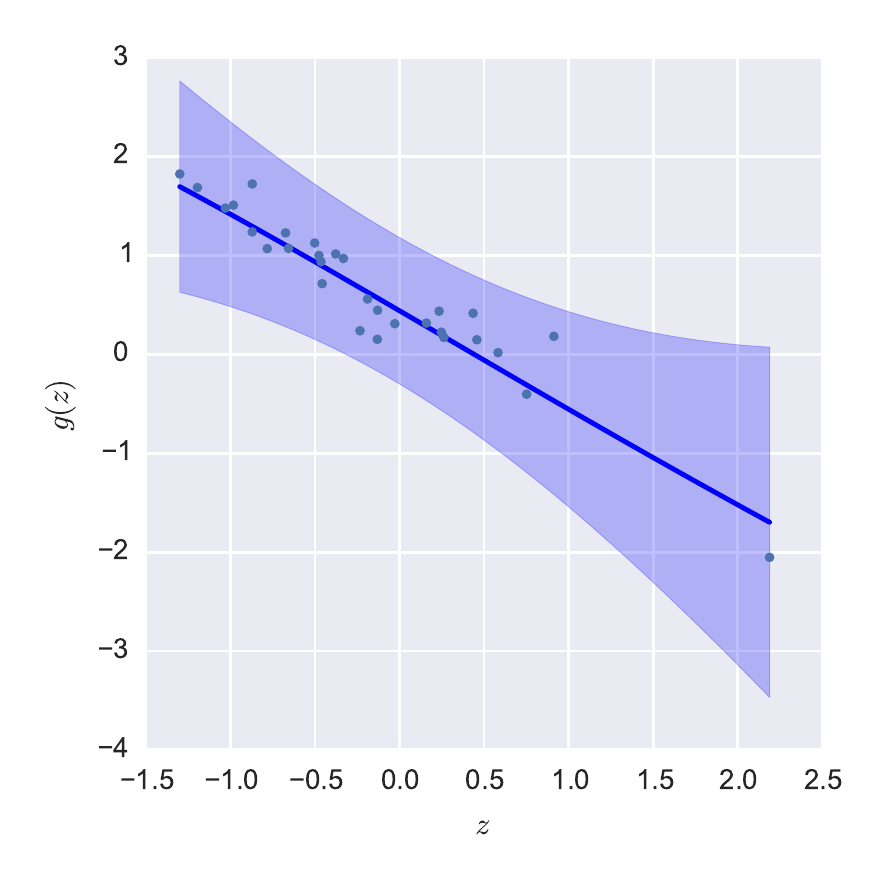}
    }
    \subfigure[] {
        \includegraphics[width=80mm]{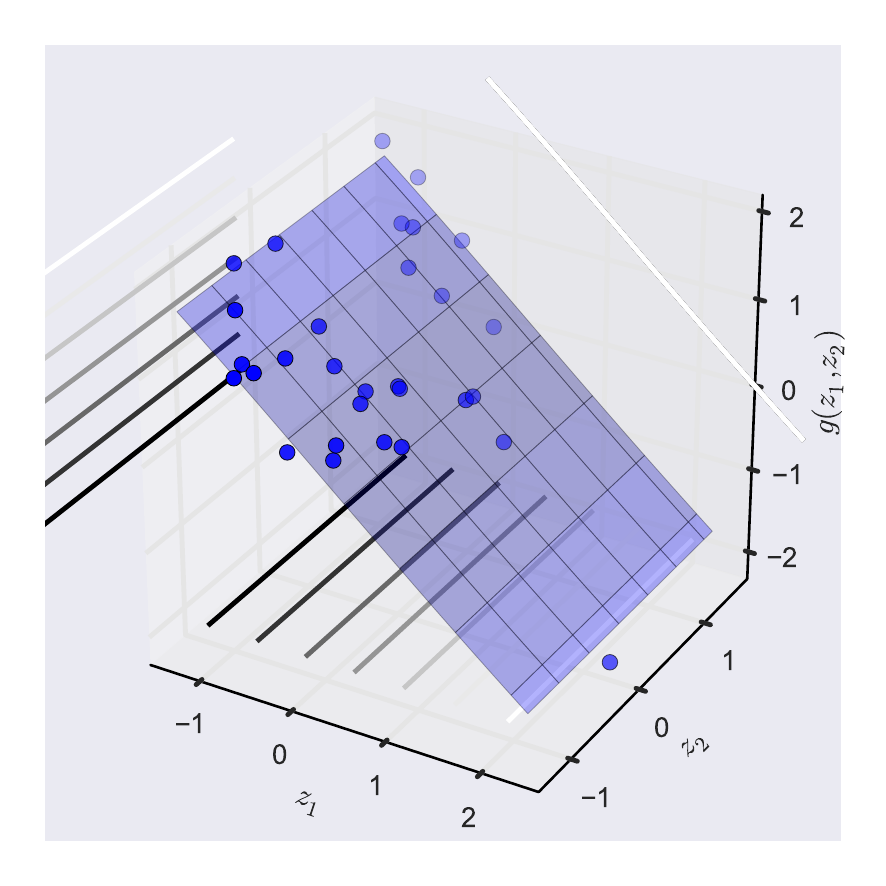}
    }
    \subfigure[] {
        \includegraphics[width=80mm]{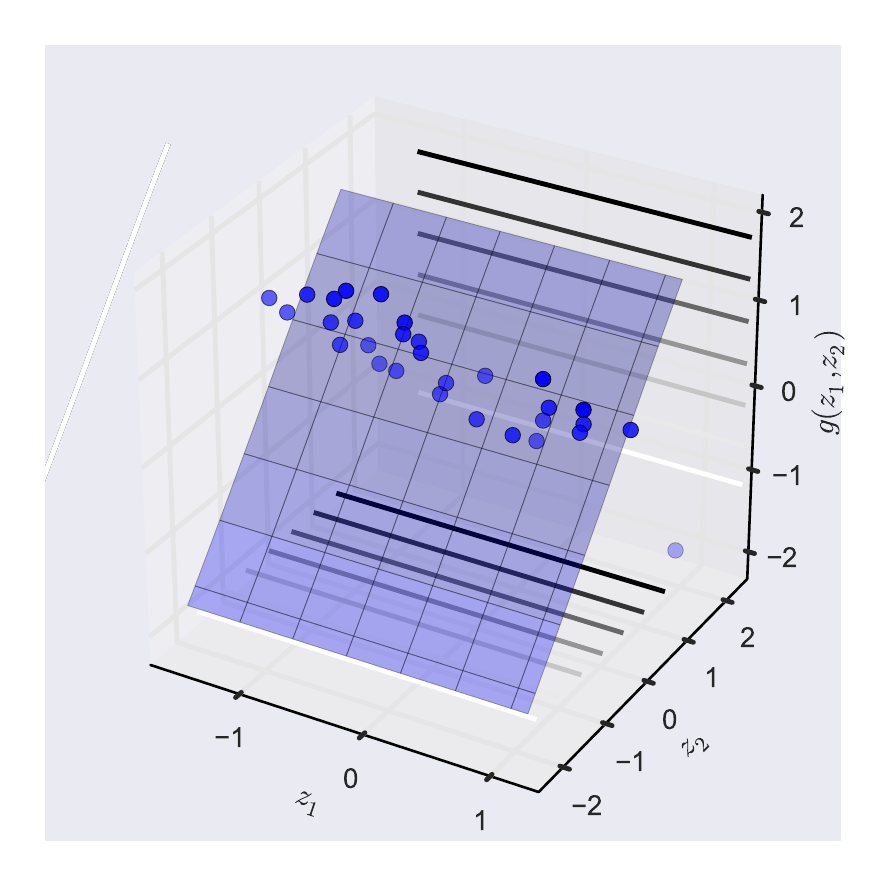}
    }
    \subfigure[] {
        \includegraphics[width=80mm]{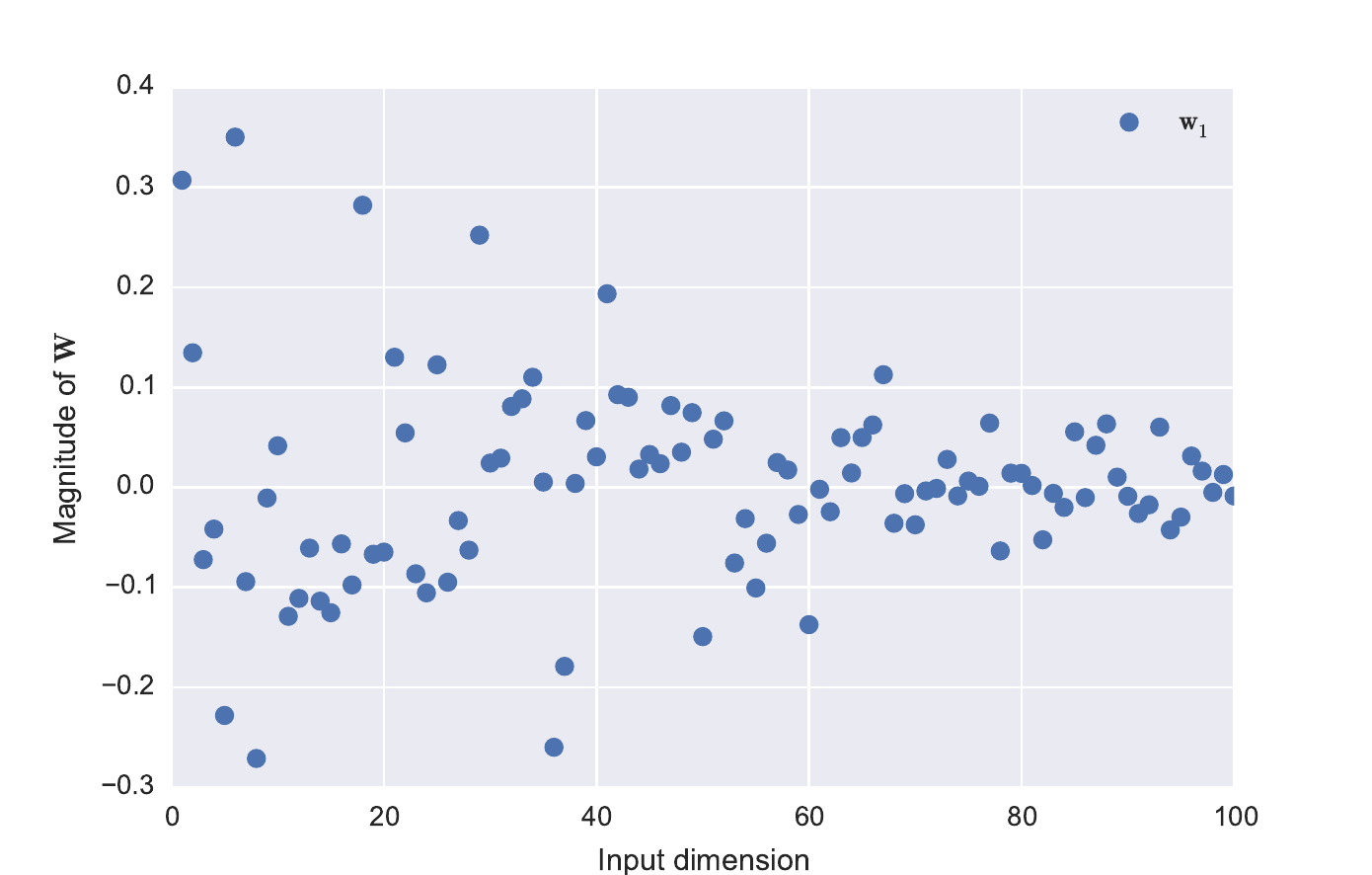}
    }
    \subfigure[] {
        \includegraphics[width=80mm]{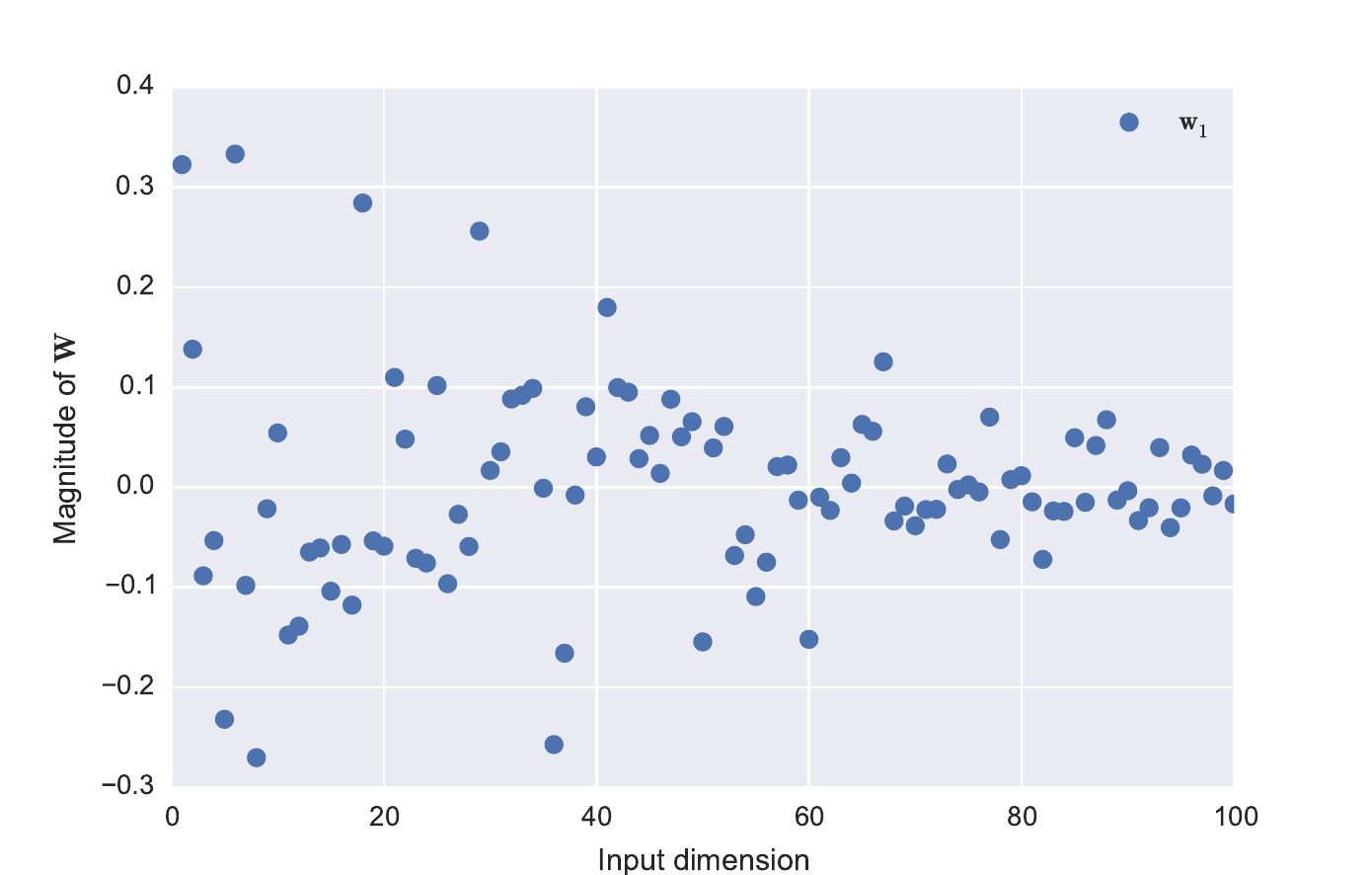}
    }
\caption{
    Elliptic PDE, long correlation length ($\ell=0.01$).
The left and the right columns correspond to results obtained
with the classic and the gradient-free approach respectively. The first and
second rows depict the predictions of each method for the link function
assuming a 1D and 2D underlying AS, respectively, along with a scatter plot of
the projections of 30 validation inputs vs the validation outputs. The third
row visualizes the projection matrix that each method discovers.
  }
\label{fig:exa_elliptic_short_corr}
\end{figure*}

\subsection{Stochastic elliptic partial differential equation}
\label{sec:exa_elliptic}

\begin{figure*}
    \centering
    \subfigure[] {
        \includegraphics[width=80mm]{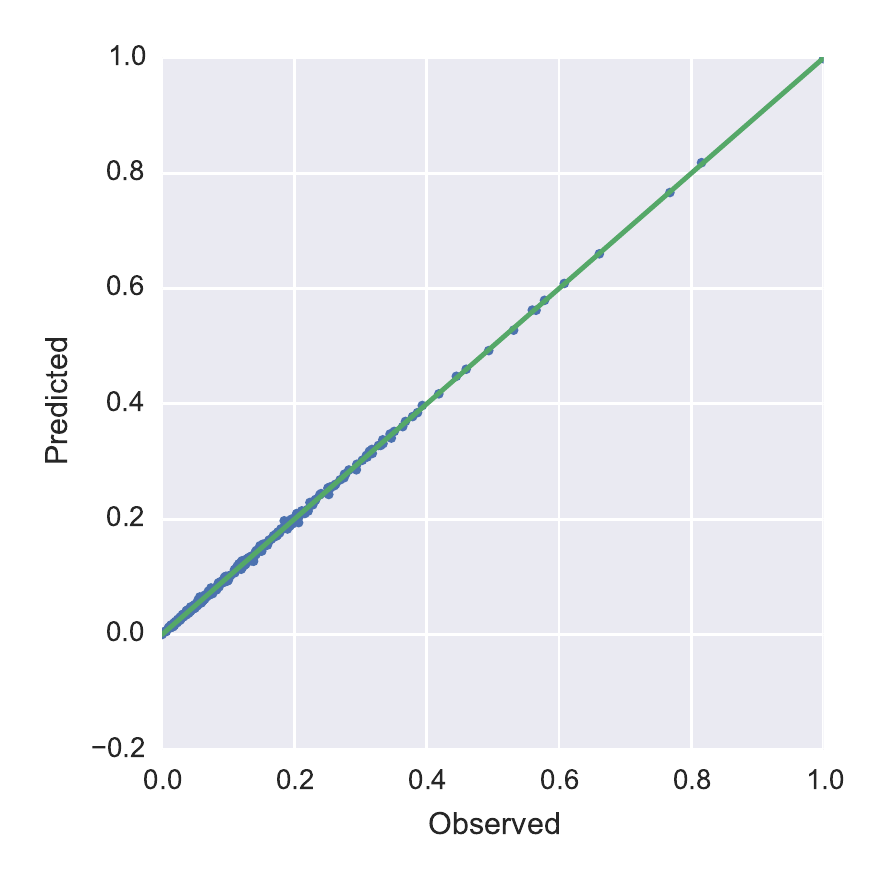}
    }
    \subfigure[] {
        \includegraphics[width=80mm]{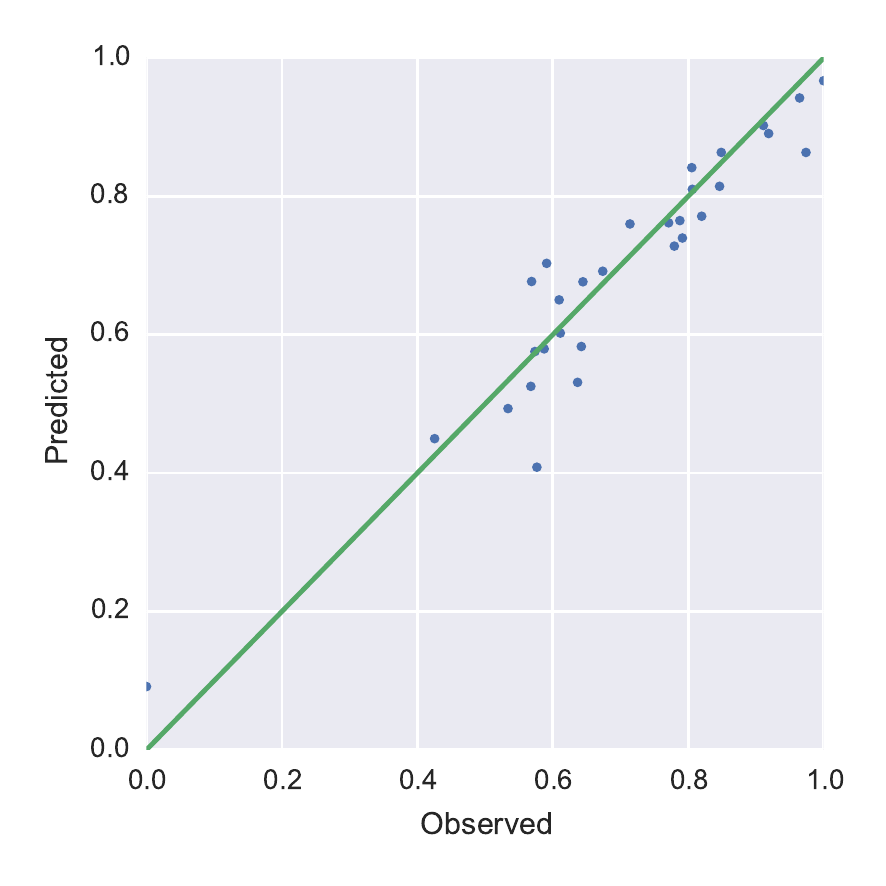}
    }
    \subfigure[] {
    	\includegraphics[width=80mm]{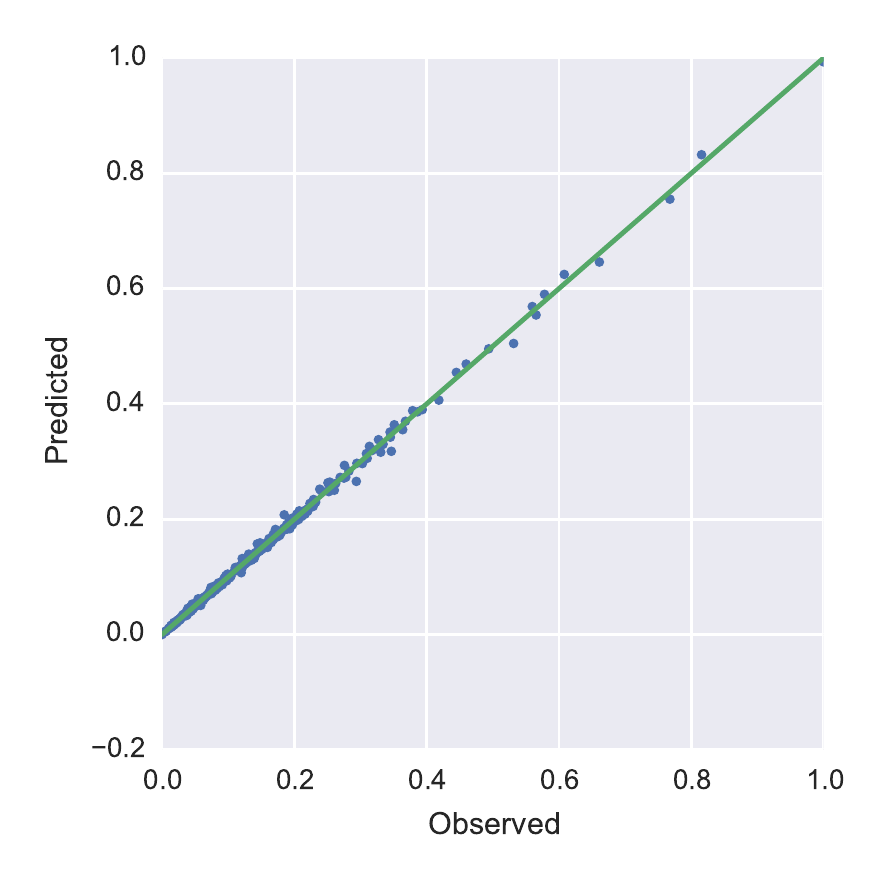}
    }
    \subfigure[] {
    	\includegraphics[width=80mm]{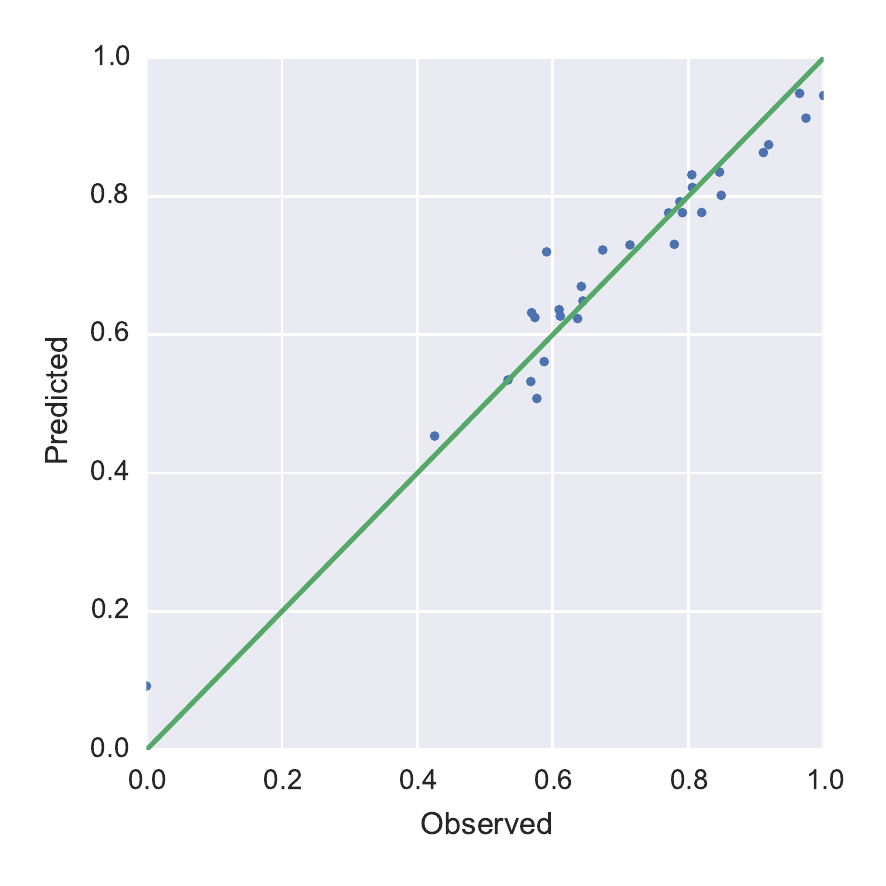}
    }
    \caption{
    Elliptic PDE.
    The dots correspond to true observed responses vs predicted ones for 30
    validation inputs for the long ($\ell=1$, left) and short
    ($\ell=0.01$, right) correlation cases.
    Perfect predictions would fall on
    the green $45^\circ$ line of each subplot. The top row corresponds to the gradient-free approach while the bottom row corresponds to the classic approach.
    }
    \label{fig:exa_elliptic_new_pred_vs_obs}
\end{figure*}

Consider the elliptic partial differential equation \cite{constantine_2014}:
\begin{equation}
    \nabla \cdot \left(c(\bs)\nabla u(\bs)\right) = 1,\;\bs\in\Omega=[0,1]^2,
\label{eqn:elliptic_pdf}
\end{equation}
with boundary conditions
\begin{eqnarray}
    \label{eqn:top_bottom}
    u(\bs) &=& 0,\ \bs \in \Gamma_1,\\
    \label{eqn:left_right}
    \nabla u(\bs)\cdot \mathbf{n} &=& 0,\ \bs \in \Gamma_2,
\end{eqnarray}
where $\Gamma_1$ contains the top, bottom and left boundaries, $\Gamma_2$ denotes the right boundary
of $\Omega$, while $\bm{n}$ is the unit normal vector to the boundary.
We assume that the conductivity field is unknown and model its logarithm as
a Gaussian random field with an exponential correlation function:
\begin{equation}
    C(\bs, \bs';\ell) = \exp\left\{-\frac{|s_1-s_1'| + |s_2-s_2'|}{\ell} \right\},
    \label{eqn:exa_elliptic_corfun}
\end{equation}
with correlation length $\ell>0$. Using a truncated Karhunen-Lo\`eve expansion
(KLE)~\cite{ghanem2003},
the logarithm of the conductivity can be expressed as:
\begin{equation}
    \log \bc(\bs;\bx) := \sum_{i=1}^{100}x_i \sqrt{\lambda_i}\phi_i(\bs),
\end{equation}
where $\lambda_i$ and $\phi_i(\bs)$ are the eigenvalues and eigenfunctions of
the correlation function, \qref{exa_elliptic_corfun}, and $\bx$ is a random
vector modeled as uniformly distributed on $[-1,1]^{100}$, 
i.e., $\bx \sim \calU\left([-1,1]^{100}\right)$. The latter violates the 
theoretical form of the KLE, but guarantees the existence of a solution to the
boundary value problem defined by Eqn.'s
(\ref{eqn:elliptic_pdf})-(\ref{eqn:left_right}) for all $\bx$.
Given any value for $\bx$, the solution of the boundary value problem is
$u(\cdot;\bx)$.

In our analysis, we attempt to learn the following scalar quantity of interest:
\begin{equation} 
    \label{eqn:response} 
    f(\bx) := f[u(\cdot;\bx)] := \frac{1}{|\Gamma_2|}
              \int_{\Gamma_2} u(\bs; \bx) ds,
\end{equation}
using both the classic, \sref{classic_approach},
and the gradient-free approach, \sref{grad_free_approach}.
We examine two cases exhibiting two different correlation lengths.
The first case uses a long correlation length, $\ell=1$, and the second case
a short correlation length $\ell=0.01$. In both cases, we use $N=270$
noiseless observations of input-output pairs for training purposes, while
setting $30$ aside for validation. The data along with the \texttt{MATLAB} code
that generates them, developed by Paul Constantine, can be obtained from
\url{https://bitbucket.org/paulcon/active-subspace-methods-in-theory-and-practice/src}.

\fref{exa_elliptic_long_corr} shows the
results we obtain using the long correlation length.
The first and second rows of this figure depict the discovered link function
under the assumption that $d=1$ and $d=2$, respectively. Note that both
methodologies agree on the most important AS dimension, but slightly disagree
on the second, albeit relatively flat, dimension. A close examination of
the discovered projection matrices, third line of the figure, reveals the
following. The most important column of the classic projection matrix,
$\bw_1$, matches with the corresponding column discovered by the 
gradient free approach. The latter, however, looks like a
``noisy" version of the former. This is reasonable if one takes into account
that the gradient-free approach uses significantly less information than the
classic approach. Finally, we notice that the columns of secondary importance
do not match. This discrepancy is unimportant given that the BIC score
eventually selects a 1D AS. 

\fref{exa_elliptic_short_corr} shows the results for the more challenging
case of the short correlation length. We present the 1D representation of the link function, as discovered by the 
classic and the gradient-free approach, in the first row and show the components of the
projection matrix  estimated by each methodology in the second row. We note that both methodologies 
show similar 1D active subspace representation of the 
surrogate. Indeed, this is the most important dimension as the response should be flat 
along the $2^{nd}$ dimension. We find that the components of projection matrix estimated by
both methods are in qualitative agreement for the 1D surrogate. 
As expected, the BIC score selects the model corresponding to the 1D 
active subspace as the right model. 

\fref{exa_elliptic_new_pred_vs_obs} shows the comparison between the prediction on the test 
inputs and the actual response. The closer the points lie to the green $45^\circ$ line, the more 
accurate the prediction. We make these comparisons for the 1D representation of the link 
function for both the short and long correlation length cases. It appears that the predictive 
capabilities of the classic approach are slightly better than the gradient-free approach. 
A comparison of the RMS error for the predictions by each methodology confirms this although
the difference is essentially negligible given its order of magnitude. We tabulate this data in Table \ref{table:rms}. 

\begin{table}[H]
\begin{center}
\begin{tabular}{|c|c|c|}
\hline
   & $\ell = 1$ & $\ell=0.01$ \\
\hline
Classic approach  & $1.87 \times 10^{-5}$ & $3.2 \times 10^{-6}$ \\
\hline
Gradient-free approach & $2.66 \times 10^{-5}$ & $3.57 \times 10^{-6}$ \\
\hline
\end{tabular}
\end{center}
\caption{Predictive RMS errors for $\ell = 1, 0.01$ corresponding to classic and gradient-free methodologies}
\label{table:rms}
\end{table}

\subsection{Granular Crystals}
\label{sec:exa_gc}

Granular crystals, or tightly packed lattices of solid particles that deform on contact with each other \cite{nesterenko2001}, are strongly nonlinear systems that have attracted significant attention due to their unique dynamics (see, e.g., \cite{porter_2015} and references therein). In particular, a one-dimensional uncompressed chain of elastic spherical particles supports the formation and propagation of solitary waves \cite{sen_2008}, i.e., elastic waves that remain highly localized and coherent while traveling along the chain. This behavior is due to the interplay between nonlinearity and discreteness of the unilateral Herztian contact interaction between the particles in the system. Over the last two decades, extensive experimental, computational and theoretical research has been conducted to advance the understanding of theses systems. For example, experimental techniques have been developed to measured the temporal evolution of the solitary wave \cite{daraio_2005, yang_2014,leonard_2013}, and dissipative \cite{herbold_2007,carretero_2009,gonzalez_2012}, plastic \cite{pal_2013} and nonlocal \cite{gonzalez_2012b} deformation effects between particles have been included in simulations. However, a systematic and thorough uncertainty analysis of theses systems remains elusive, mainly due to the curse of dimensionality. It is worth noting that the high localization of this elastic waves suggests the use of an AS approach, specifically a gradient-free approach due to the lack of gradient information. We present the mathematics of the problem next.

\begin{figure*}[b] 
    \centering
        \includegraphics[width=150mm]{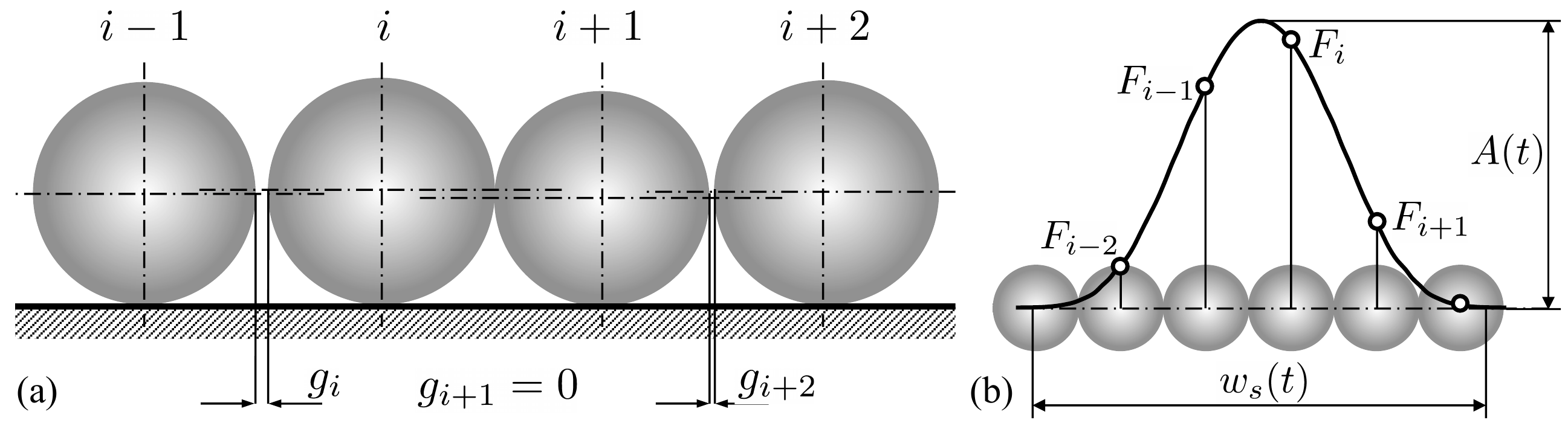}
    \caption{One-dimensional granular crystal.
    (a) Initial equilibrium position of spherical particles in contact with a horizontal flat rigid surface under the action of gravity. Particles are separated from point contact by horizontal positive gaps $g_i$. (b) Traveling solitary wave with amplitud $A(t)$ and width $w_s(t)$ as defined by averaged compressive forces $\tilde{F}_{i}(t)$ acting on particles $i$. }
    \label{fig:GC_schematic}
\end{figure*}

Consider a one-dimensional chain of $n_p\ (=47)$ spherical particles whose displacements from
the initial equilibrium positions are described by the position matrix $\bq =
\left(\bq_1, \bq_2, \cdots, \bq_{n_p}\right)$ with $\bq_i \in \mathbb{R}^3$. The equilibrium position is such that particles are in contact with a horizontal flat rigid surface under the action of gravity. In addition, a positive gap $g_i$ may exist between $i$-th and $i-1$-th particles---$g_1$ corresponds to the gap between the first particle and a rigid back wall. A gap equal to zero corresponds to point contact. Each bead has a radius $R_i$ and Young's modulus $E_i$. The density of the particles is a constant $\rho=7,900\mbox{kg}/\mbox{m}^3$ and the mass of each particle is $m_i=\rho \frac{4}{3}\pi R_{i}^{3}$. A solitary wave forms and propagates through the one-dimensional granular crystal after the $n_p$-th particle strikes the chain with velocity $v_s$.
Therefore, all the parameters of the system are:
\begin{equation}
    \bx = \left(R_1, R_2, \dots, R_{n_p}, E_1, E_2, \dots, E_{n_p}, g_1, g_2, \dots, g_{n_p}, v_s \right)\in\R^{3n_p+1}.
\label{eqn:gran_crys_inputs_gaps}
\end{equation}
In the present effort, we consider the two cases where: a) the particles are in point contact with each other and thus, the system is completely parameterized by the particle radii, the Young's moduli and the striker velocity:
\begin{equation}
\bx = \left(R_1, R_2, \dots, R_{n_p}, E_1, E_2, \dots, E_{n_p}, v_s \right) \in \R^{2n_p + 1},
\label{eqn:gran_crys_inputs}
\end{equation}
and b) where the particles are separated by small gaps and thus, the system is parameterized by $\bx$ as defined in \qref{gran_crys_inputs_gaps}.
The displacement vector satisfies Newton's law of motion:
\begin{equation}
    m_i(\bx)\ddot{\bq}_i = \mathbf{F}^\mathrm{H}_{i-1,i}\left(\bq; \bx \right)+\mathbf{F}^\mathrm{H}_{i+1,i}\left(\bq; \bx \right),
\end{equation}
where $ \mathbf{F}^\mathrm{H}_{k,i}\left(\bq; \bx \right)$  is the unilateral Hertzian contact force between particle $k$ and $i$ \cite{nesterenko2001,yang_2014,gonzalez_2012}. The
initial conditions are:
\begin{eqnarray*}
    \bq_i(0) &=& (0, 0, 0), \\
    \dot{\bq}_i(0) &=& (0, 0, 0),\;\forall\;i = \{1, 2, 3, \cdots, n_p - 1\}, \\
    \dot{\bq}_{n_p}(0) &=& (-v_s, 0, 0).
\end{eqnarray*}
Let $\bq(t;\bx)$ be the solution to this initial value problem.
We are interested in characterizing the properties of the solitary wave propagated
through the granular crystal. To this end, we will be observing an average of the absolute value of the horizontal component of the two unilateral Hertzian contact forces acting on each particle as a function of time for a given set of parameters $\bx$ \cite{daraio_2005},
 \begin{equation}
\tilde{F}_i \left(t; \bx\right) \equiv \dfrac{1}{2} 
	\left[  \mathbf{F}^\mathrm{H}_{i-1,i}\left(\bq\left(t; \bx\right); \bx \right) + \mathbf{F}^\mathrm{H}_{i,i+1}\left(\bq\left(t; \bx\right); \bx \right) \right] 
	\cdot (1, 0, 0)  .
 \end{equation}
That is, for each $\bx$, we obtain, by integrating the equations of motion,
the force at a finite number of timesteps, $0=t_1<\dots< t_{n_t}$, $n_t=6252$.
The output, for each $\bx$, forms a matrix
$\tilde{\mathbf{F}}(\bx):=\{\tilde{F}_i(t_j;\bx): i=1,\dots,n_p, j=1,\dots,n_t\}$.
The dimensionality of the $\tilde{\mathbf{F}}(\bx)$ is $n_p \times n_t$. 
The time step at which the maximum force is observed as the solitary wave passes over particle $i$: 
\begin{equation}
        j_i^*(\bx) = \arg\max_j \tilde{F}_{ij}(\bx).
 \label{eqn:ts_max}
 \end{equation}
 In order to characterize the behavior of the soliton as it propagates over the particle chain, 
 we look at three properties of the soliton as it stands over any given particle - the 
 amplitude $(A_i)$, the time of flight $(t_{\mbox{flight}})$ and full width at half maximum $(f_{\mbox{h}})$. 
 The amplitude of the soliton as it passes over the particle is obtained as follows: 
     \begin{equation}
        A_i(\bx) = \tilde{F}_{ij_i^*(\bx)}(\bx).
        \label{eqn:amp_max}
    \end{equation}
   Then we extract the width, $f_{\mbox{h}}$, as follows:
    \begin{equation}
        f_{\mbox{h}, i}(\bx) = 0.364\  w_s( j_i^*(\bx)) ,
        \label{eqn:fw_hm}
    \end{equation}
   where  $w_s(t)$ is the width of the soliton at any given instance of time $t$.
  Finally, we extract the time of flight of the soliton as it passes over particle $i$:
   \begin{equation}
        t_{\mbox{flight},i}(\bx) = t_{j_i^*(\bx)}.
    \end{equation}

We study these properties of the soliton 
as it travels over the $20^{th}$ and $30^{th}$ particles. Let us denote these quantities as 
$y_1 = t_{\mbox{flight},20}, y_2 = A_{20}, y_3 = f_{\mbox{h}, 20}, 
y_4 = t_{\mbox{flight},30}, y_5 = A_{20}, y_6 = f_{\mbox{h}, 20}$.
We repeat this entire process for $1000$ samples of $\bx$ and 
construct the output vectors $\by_1, \by_2, \by_3, \by_4, \by_5$ and $\by_6$ which 
we define as $\by_i = \{y_{i}^{(1)}, \dots, y_{i}^{(1000)}\}$ such that $\by_i \in \R^{1000}$.
The input in each simulation is the vector of parameters shown in \qref{gran_crys_inputs_gaps} and 
\qref{gran_crys_inputs} for cases with and without inter particle gaps respectively.
Thus with 1000 of these input vectors we contruct the input design matrices $\bX_1 \in \R^{1000 \times 2n_p + 1}$
and $\bX_2 \in \R^{1000 \times 3n_p + 1}$ for the cases with and without point contacts, respectively. 

We now proceed to apply our proposed gradient-free AS approach
to build a cheap-to-evaluate surrogate for propagating 
uncertainty through this system. 
We train the model on 1000 observations with inputs sampled using a Latin Hypercube design 
within the range $(180 \mbox{GPa}, 220 \mbox{GPa})$ for Young's moduli input, 
$(8.57\mbox{mm}, 10.47\mbox{mm})$ for radii input, $(1.125\mbox{m/s}, 1.375\mbox{m/s})$ for impact velocities input 
and gaps $g_i$ such that 90\% of the time $g_i = 0$ (i.e. there is no gap between the $i-1^{th}$ and $i^{th}$ particles)
and the remaining 10\% of the time there $g_i = 0.001 R_i$.
Note that we construct a different 
AS for each one of the cases. We use 100 out-of-sample data to test the predictive 
accuracy of the surrogate.
We consider all possible combinations of 
data-sets $\calD_{j} = \{ \bX_i, \by_{j}\}, \ \ \forall i \in \{1, 2\}, \ \forall j \in \{1, 2, 3, 4\}$, 
and build the corresponding surrogates.

\begin{figure*}
    \centering
    \subfigure[] {
        \includegraphics[width=80mm]{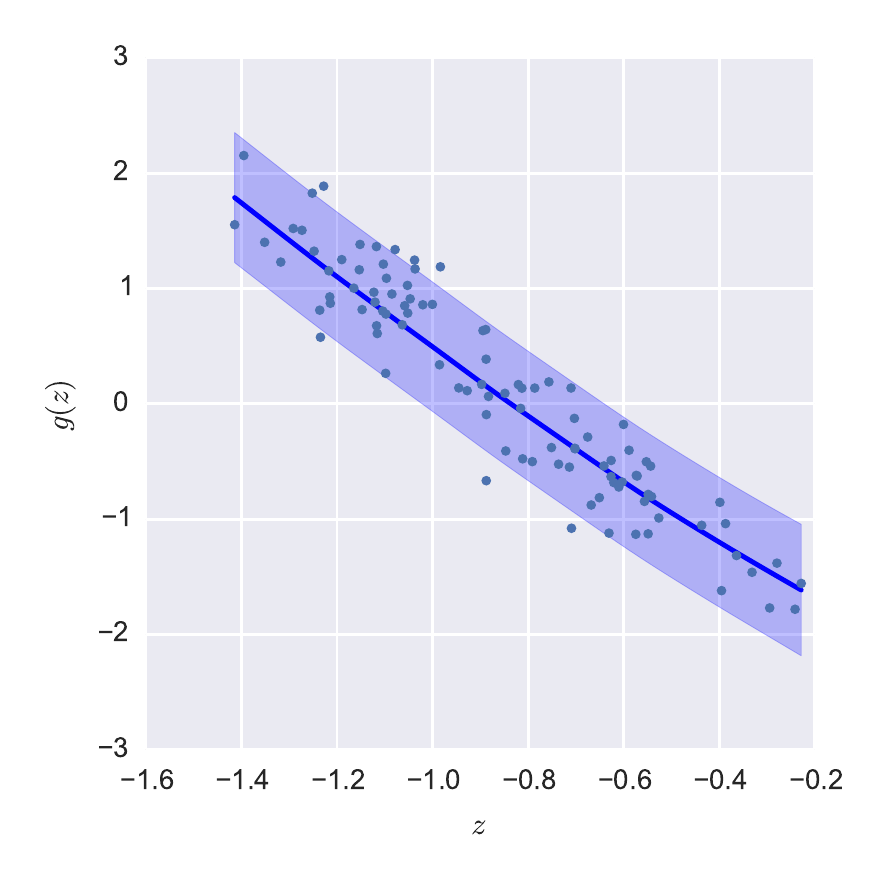}
    }
    \subfigure[] {
        \includegraphics[width=80mm]{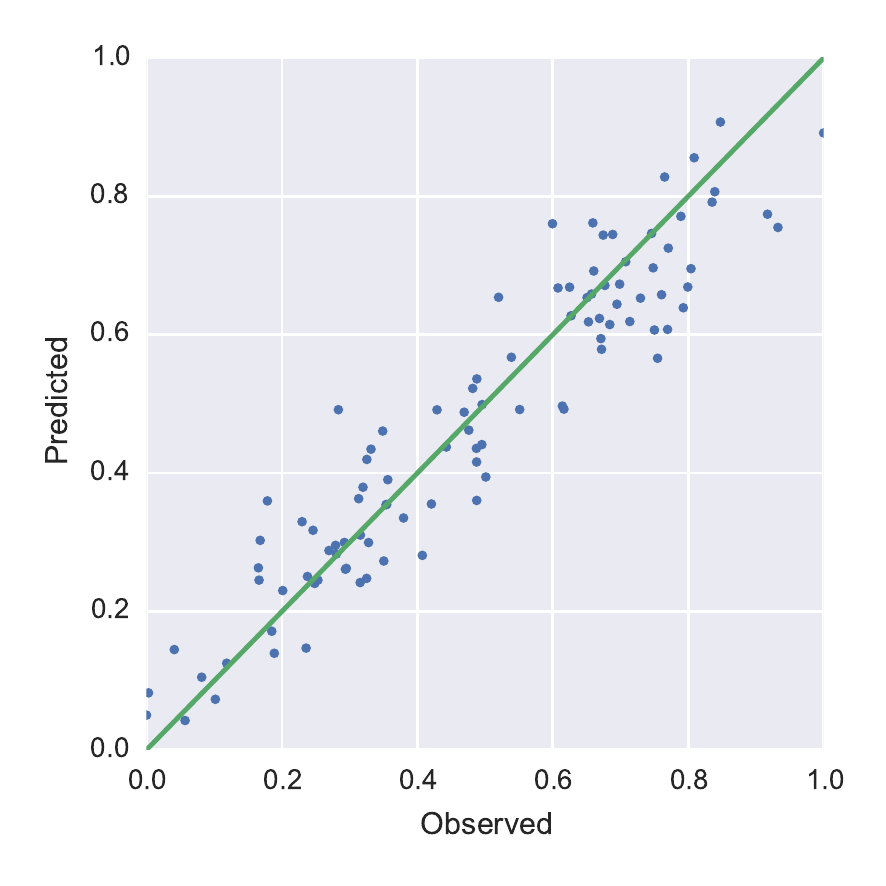}
    }
    \subfigure[] {
    	\includegraphics[width=80mm]{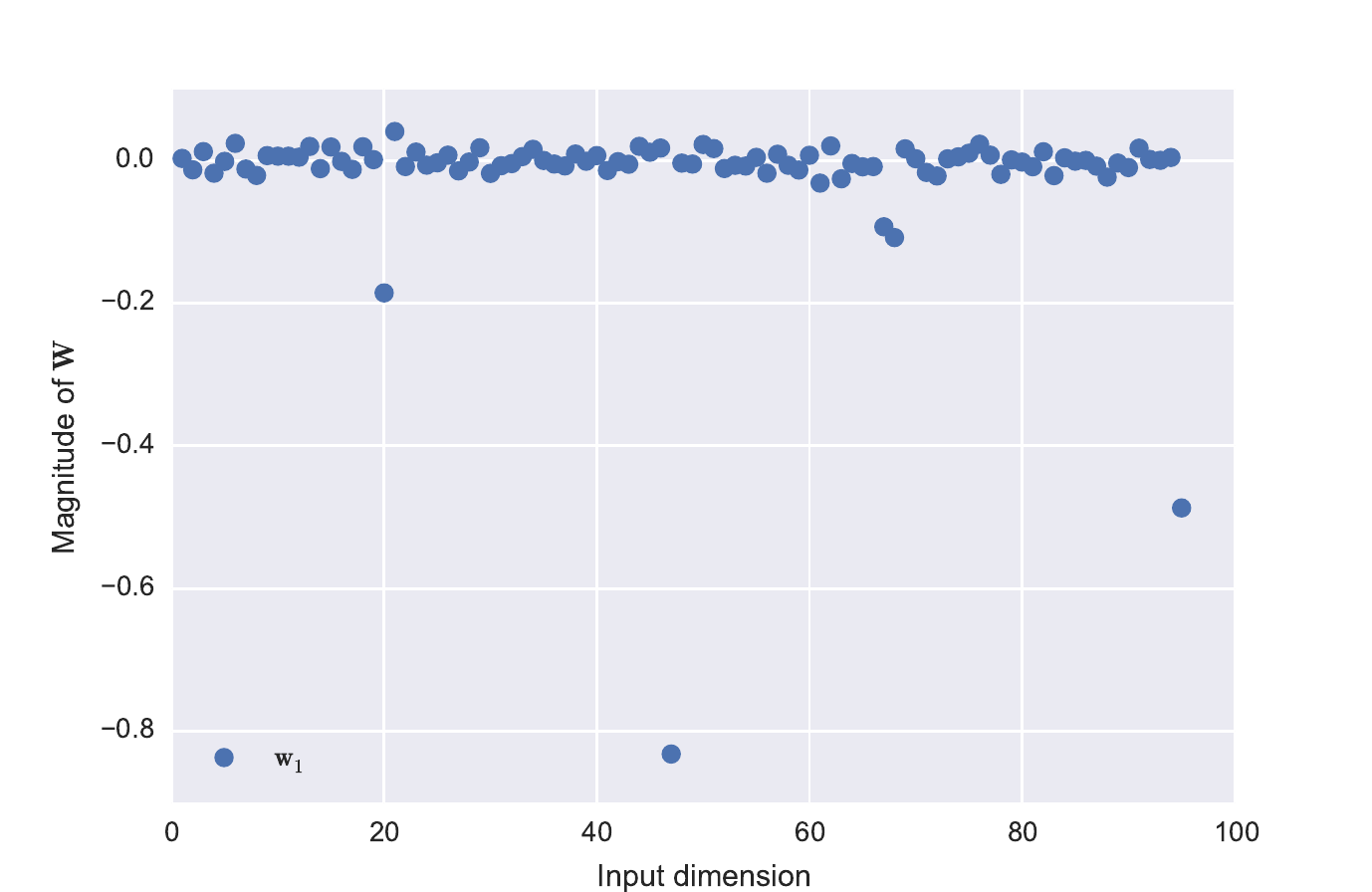}
    }
    \caption{
    One-dimensional granular crystal without gaps - amplitude of the soliton over the $20^{th}$ particle.
    The first plot shows the response surface in the active subspace. the second plot depicts the test observations vs model prediction plot. The final plot depicts the components of the projection matrix.
    }
    \label{fig:p_20_amp}
\end{figure*}

\begin{figure*}
    \centering
    \subfigure[] {
        \includegraphics[width=80mm]{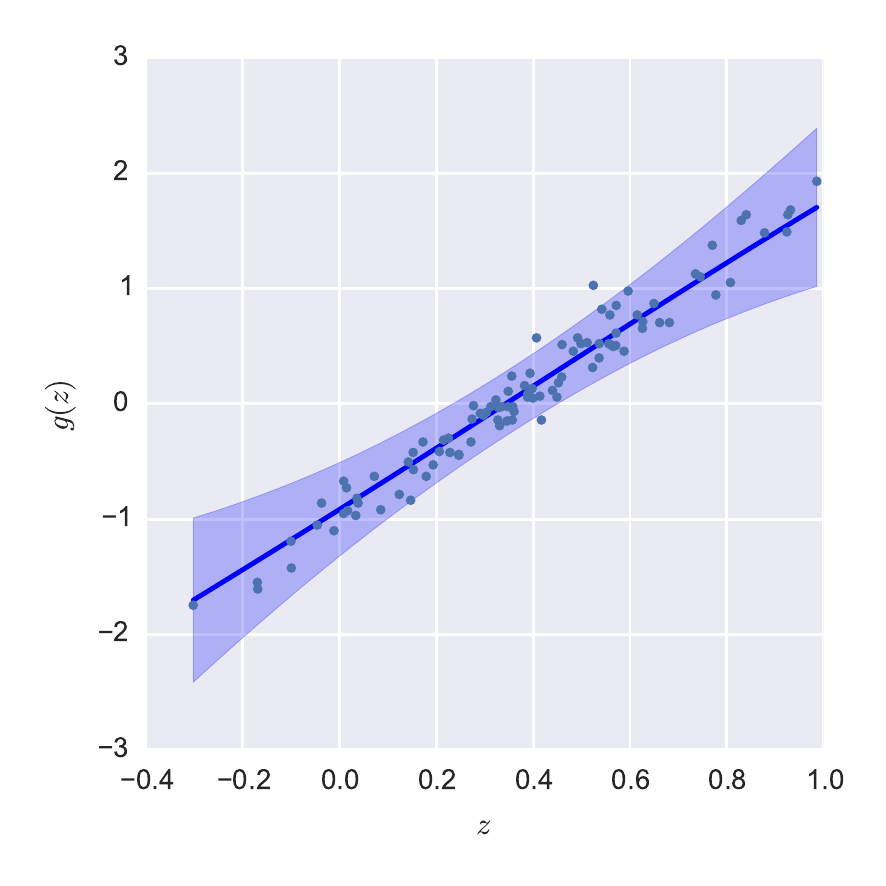}
    }
    \subfigure[] {
        \includegraphics[width=80mm]{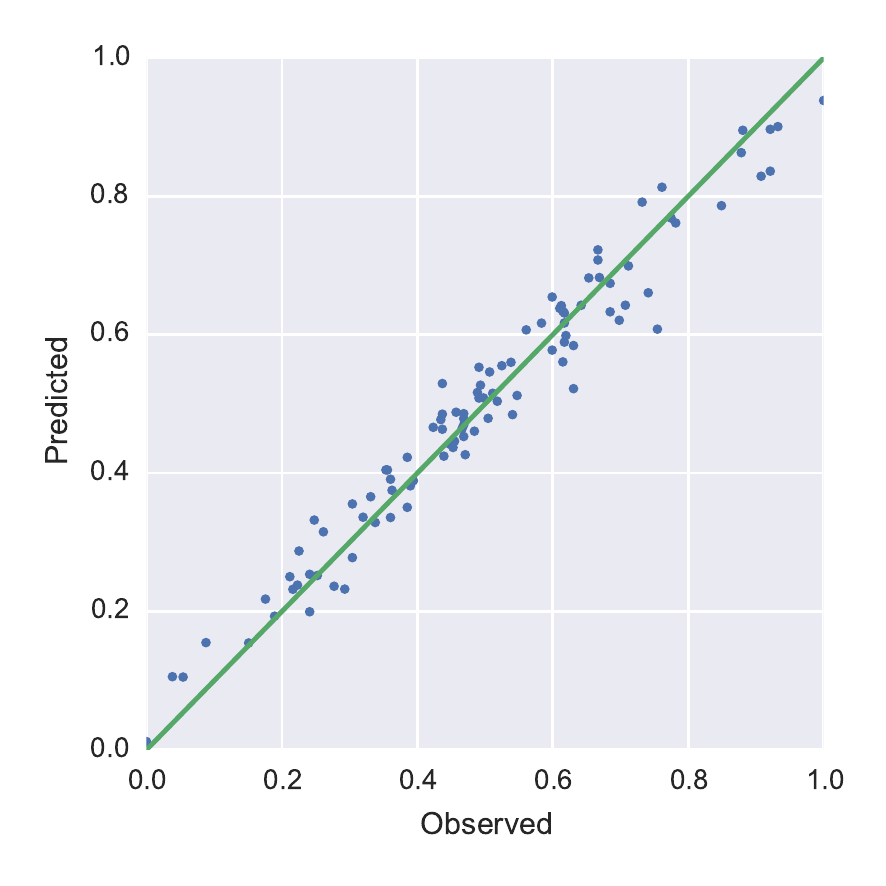}
    }
    \subfigure[] {
    	\includegraphics[width=80mm]{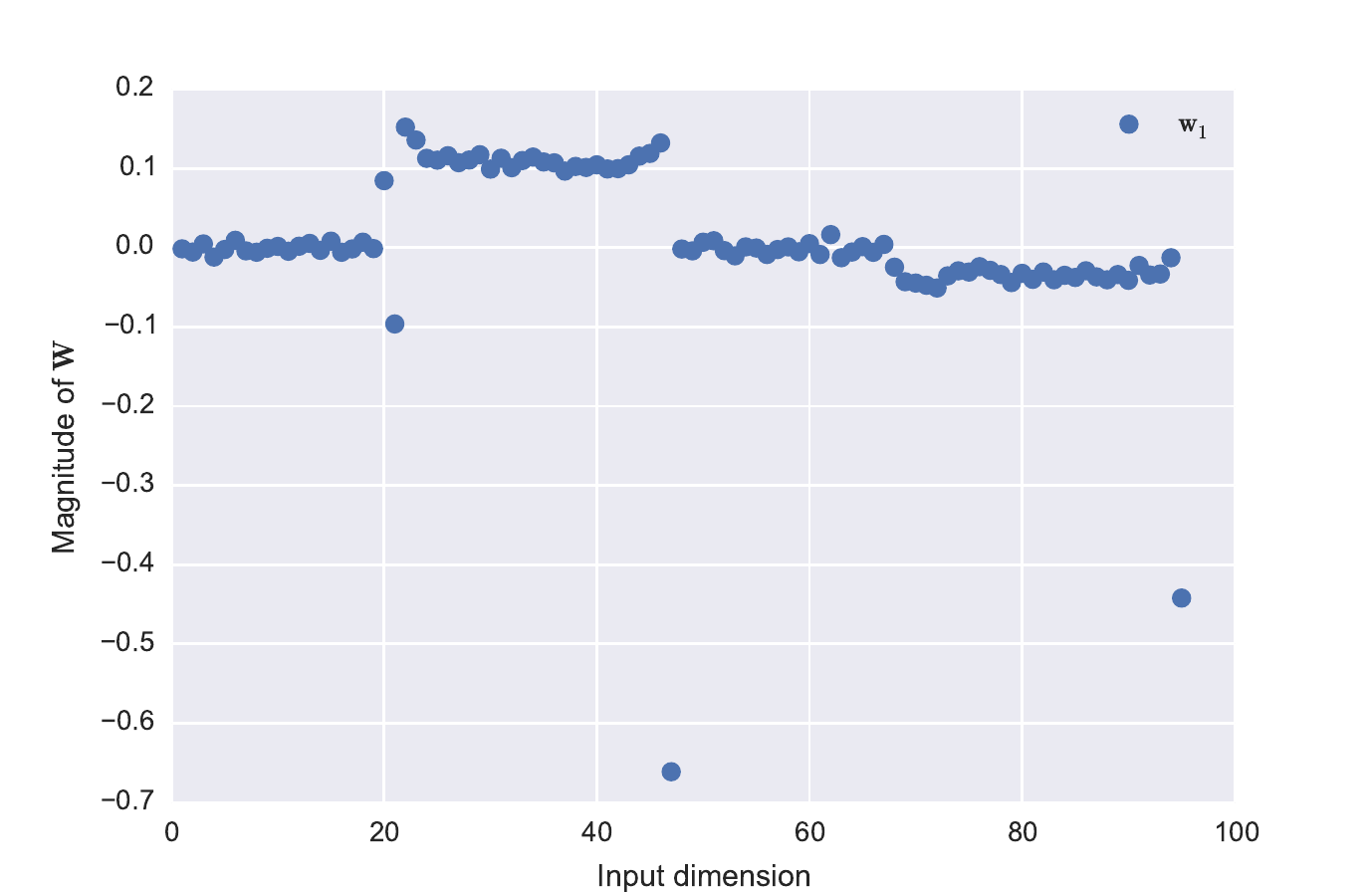}
    }
    \caption{
    One-dimensional granular crystal without gaps - time of flight of the soliton over the $20^{th}$ particle.
    The first plot shows the response surface in the active subspace. the second plot depicts the test observations vs model prediction plot. The final plot depicts the components of the projection matrix.
    }
    \label{fig:p_20_tof}
\end{figure*}

\begin{figure*}
    \centering
    \subfigure[] {
        \includegraphics[width=80mm]{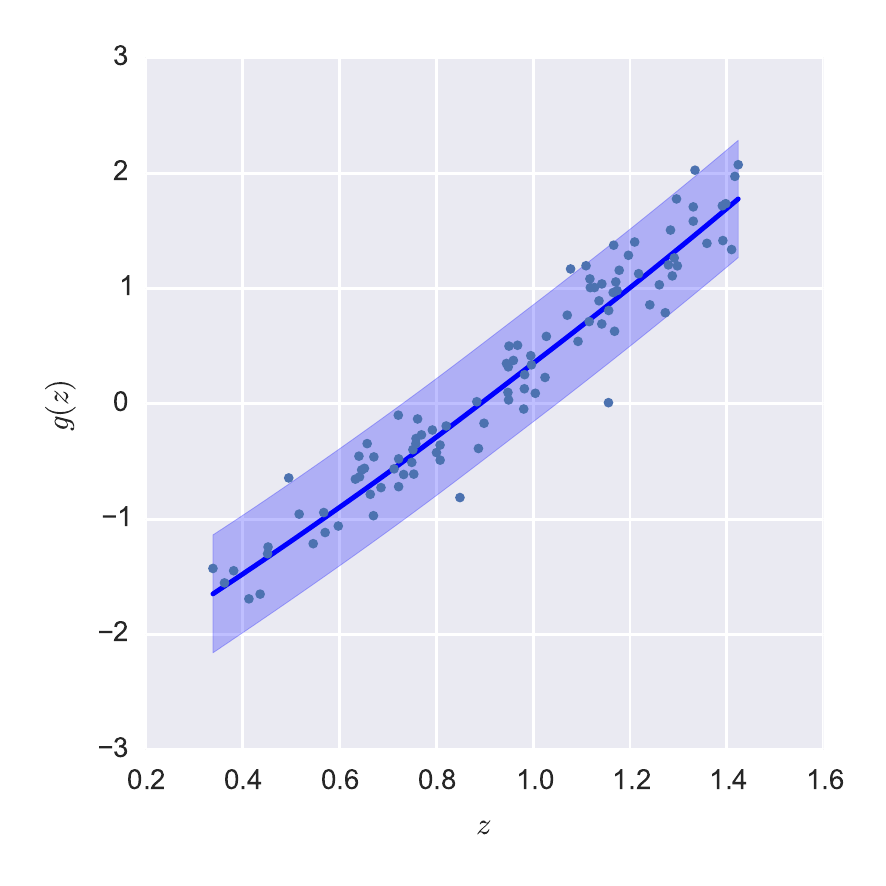}
    }
    \subfigure[] {
        \includegraphics[width=80mm]{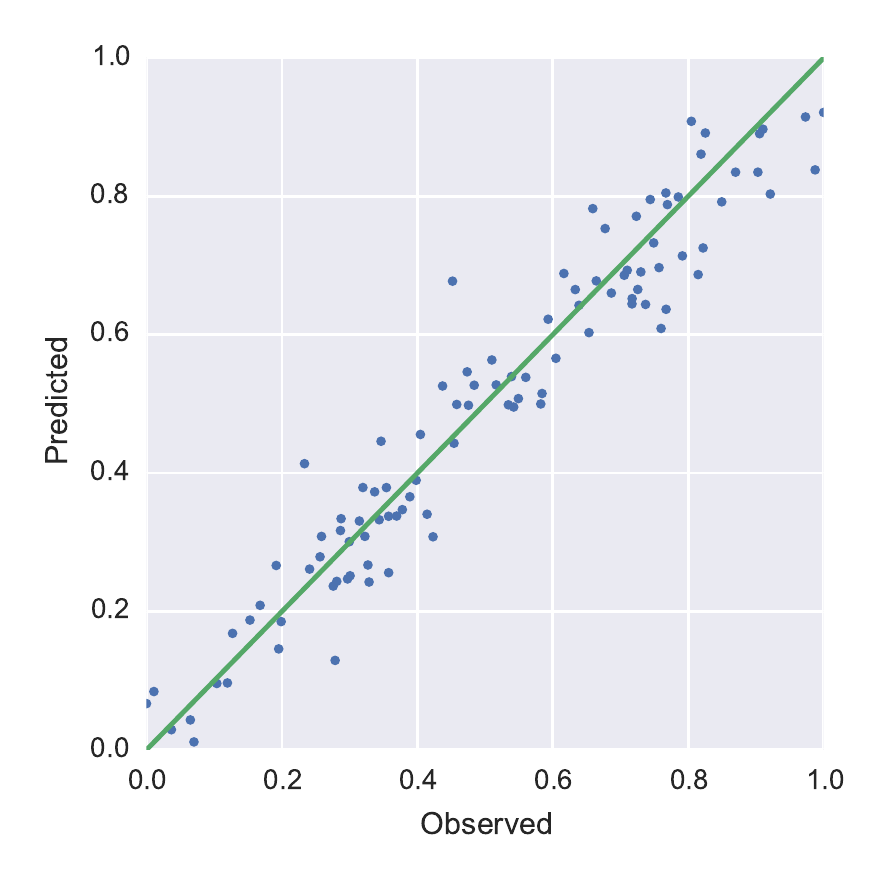}
    }
    \subfigure[] {
    	\includegraphics[width=80mm]{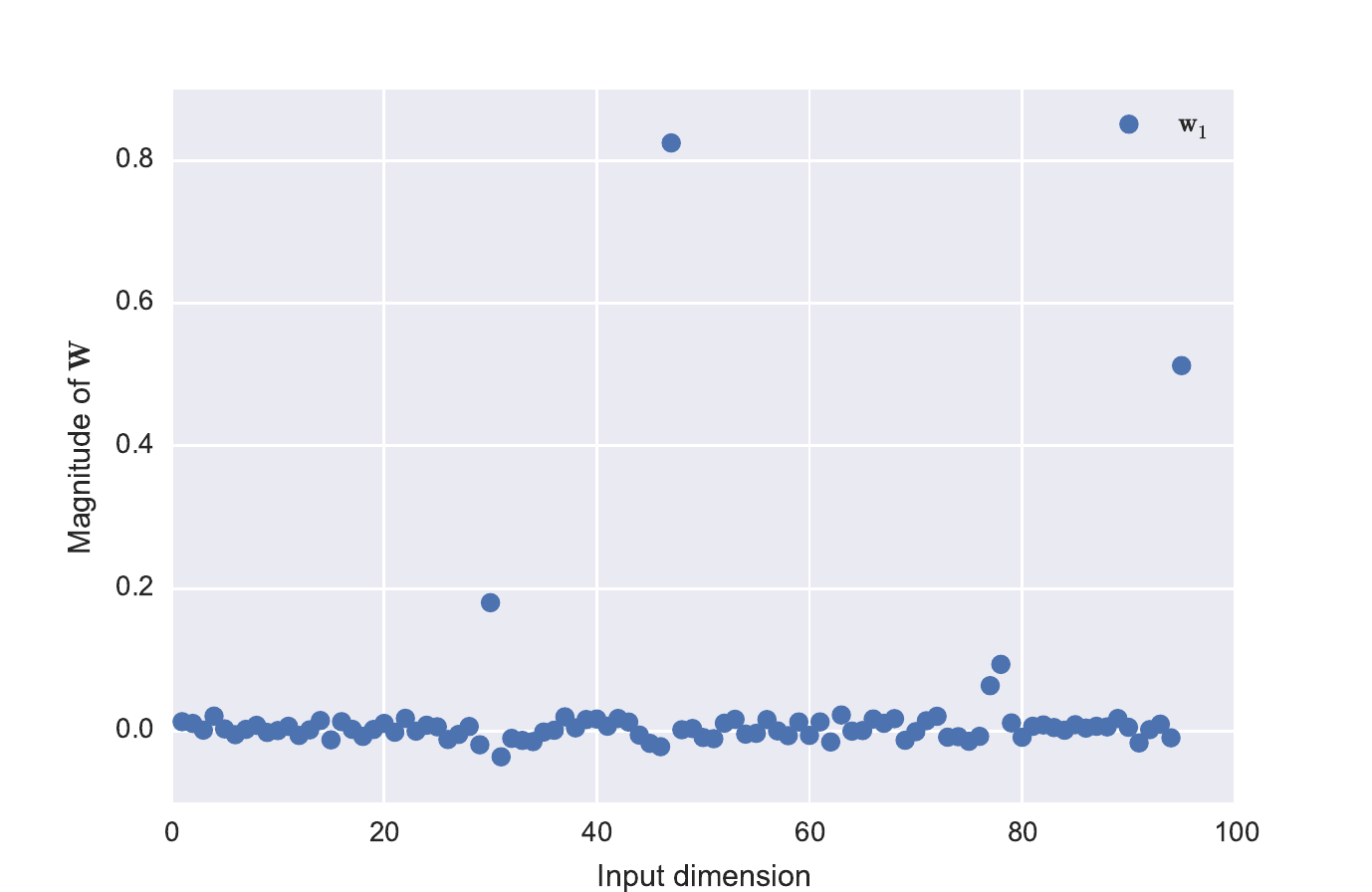}
    }
    \caption{
    One-dimensional granular crystal without gaps - amplitude of the soliton over the $30^{th}$ particle.
    The first plot shows the response surface in the active subspace. the second plot depicts the test observations vs model prediction plot. The final plot depicts the components of the projection matrix.
    }
    \label{fig:p_30_amp}
\end{figure*}

\begin{figure*}
    \centering
    \subfigure[] {
        \includegraphics[width=80mm]{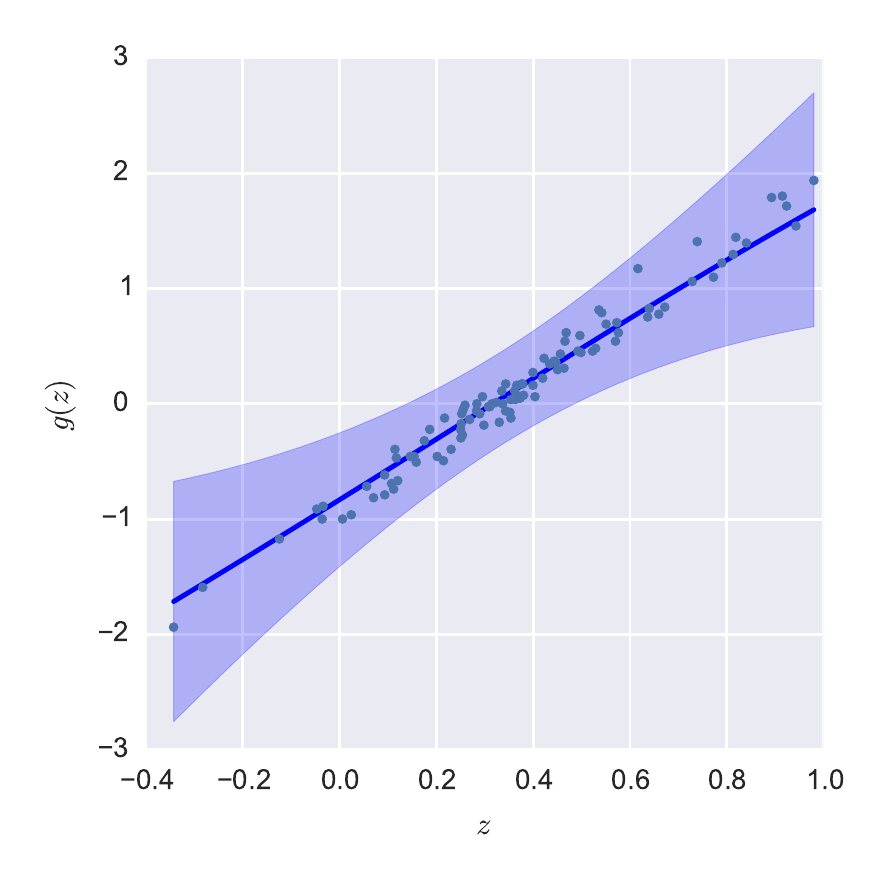}
    }
    \subfigure[] {
        \includegraphics[width=80mm]{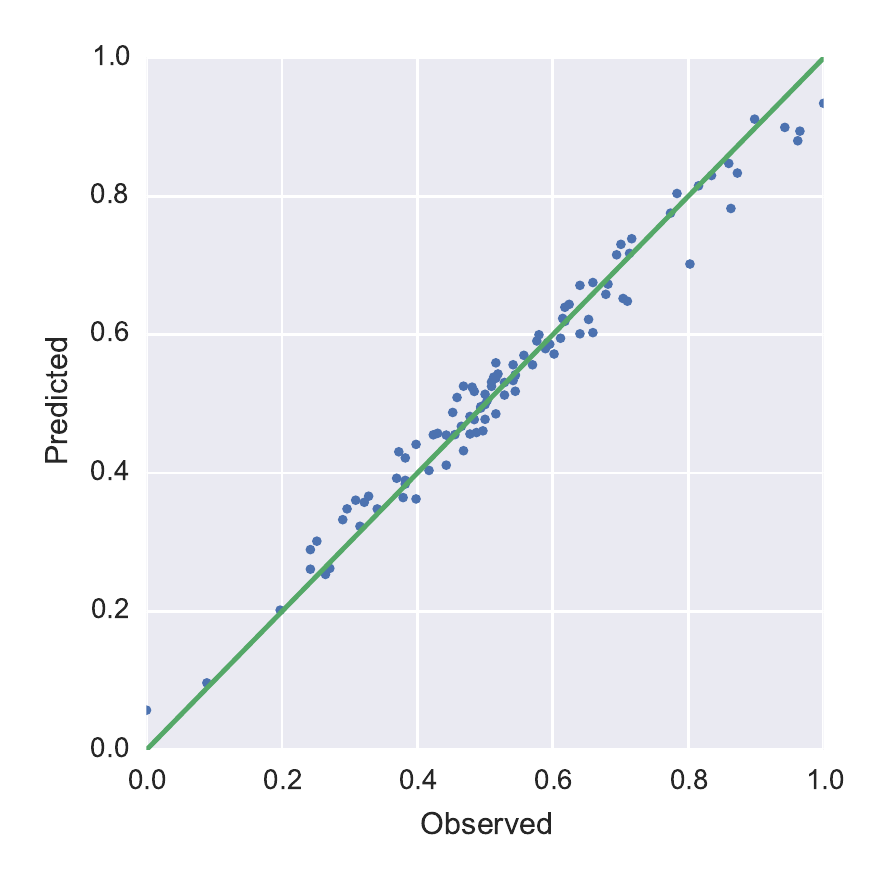}
    }
    \subfigure[] {
    	\includegraphics[width=80mm]{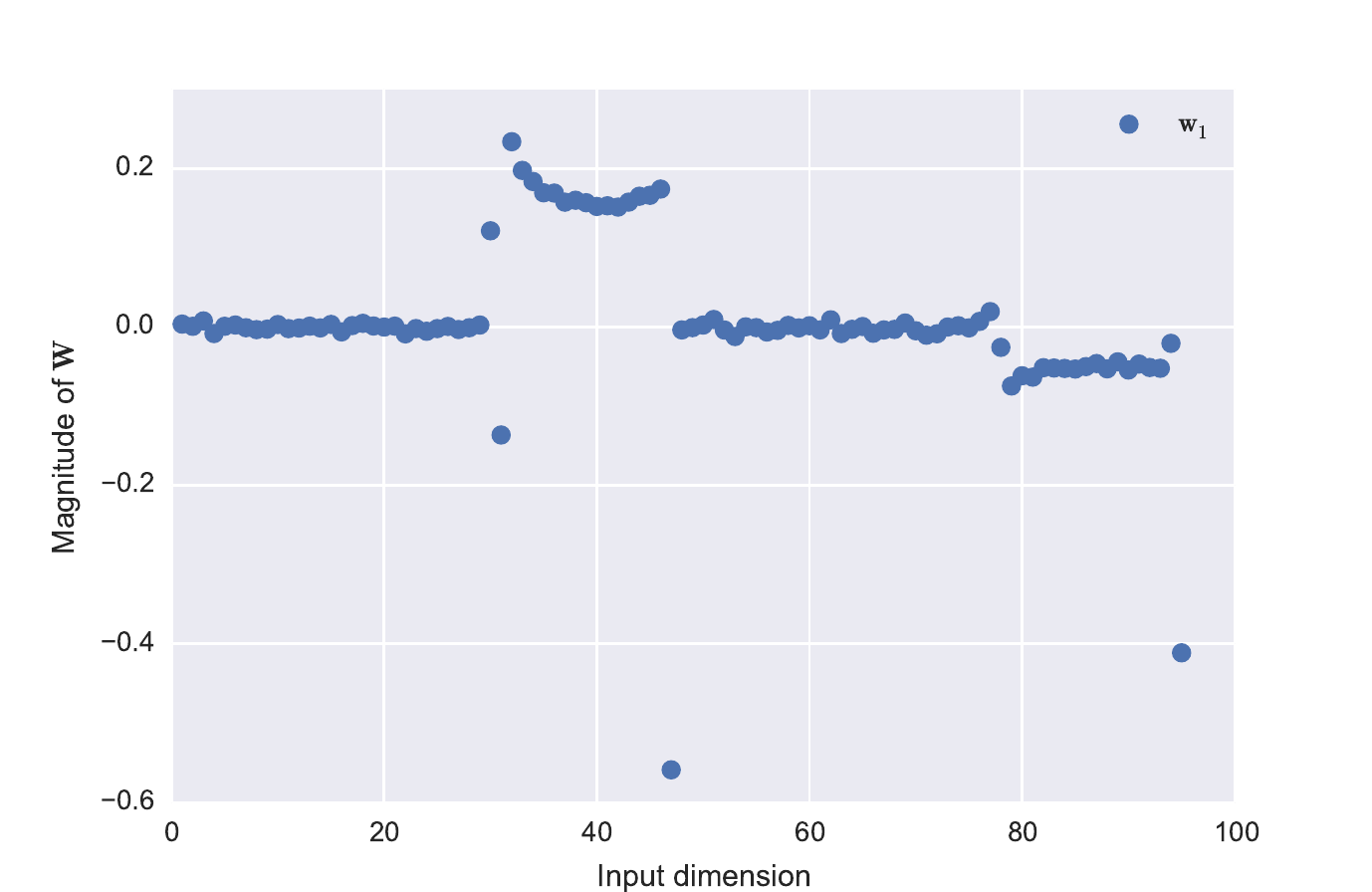}
    }
    \caption{
    One-dimensional granular crystal without gaps - time of flight of the soliton over the $30^{th}$ particle.
    The first plot shows the response surface in the active subspace. the second plot depicts the test observations vs model prediction plot. The final plot depicts the components of the projection matrix.
    }
    \label{fig:p_30_tof}
\end{figure*}

\begin{figure*}
    \centering
    \subfigure[] {
        \includegraphics[width=80mm]{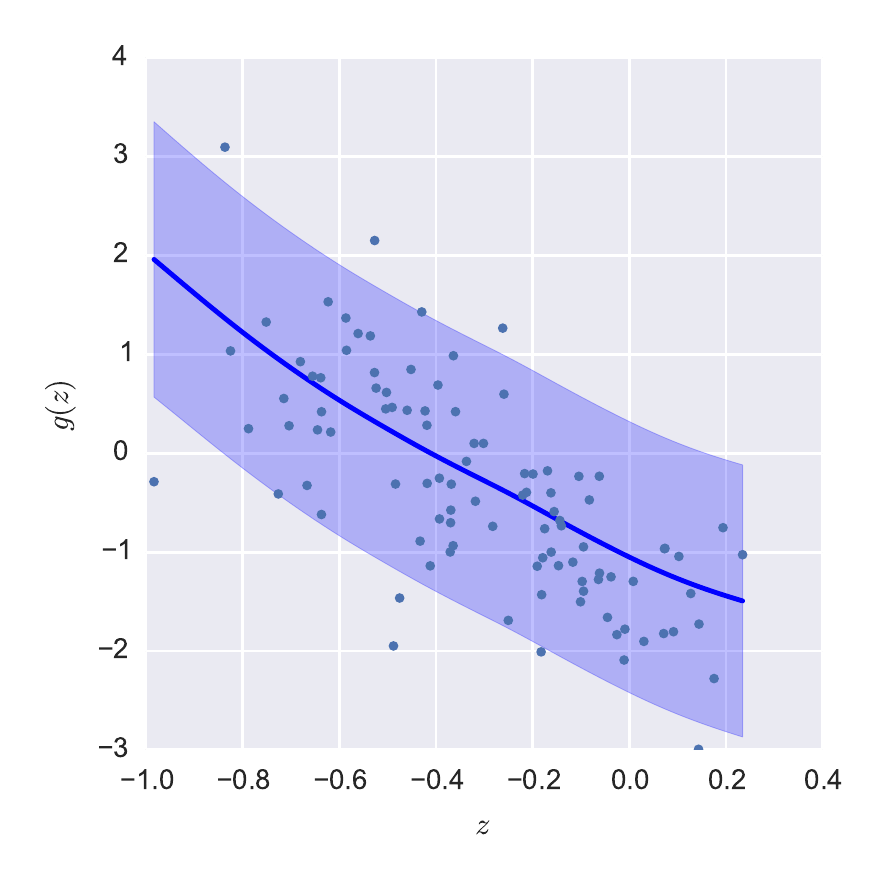}
    }
    \subfigure[] {
        \includegraphics[width=80mm]{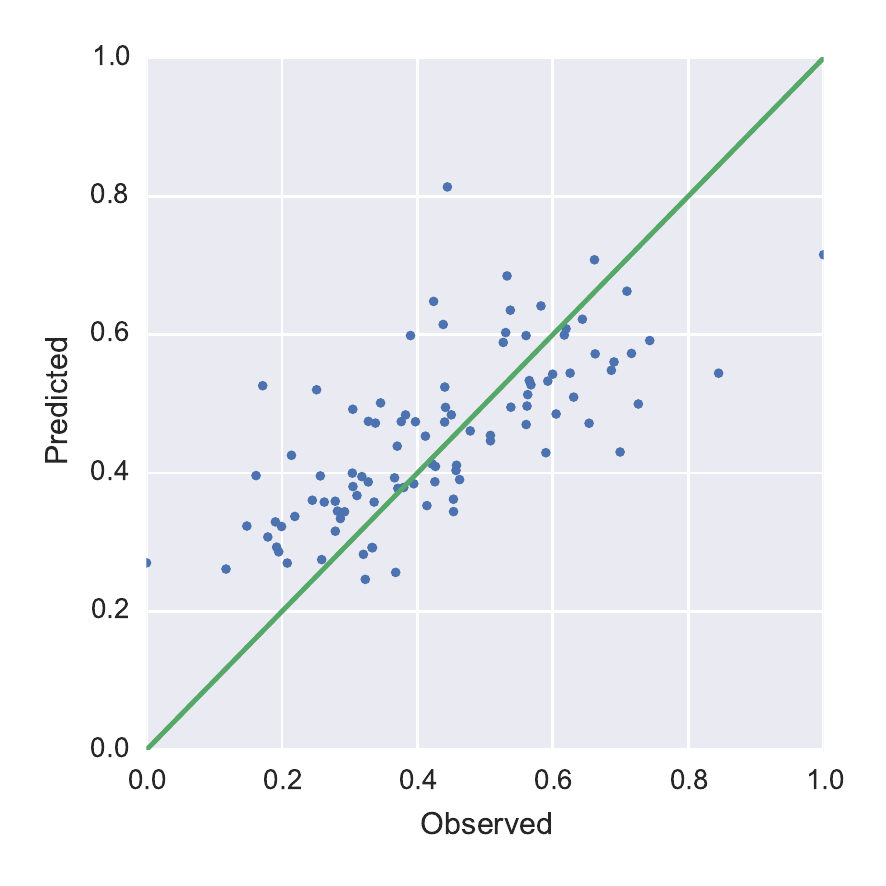}
    }
    \subfigure[] {
    	\includegraphics[width=80mm]{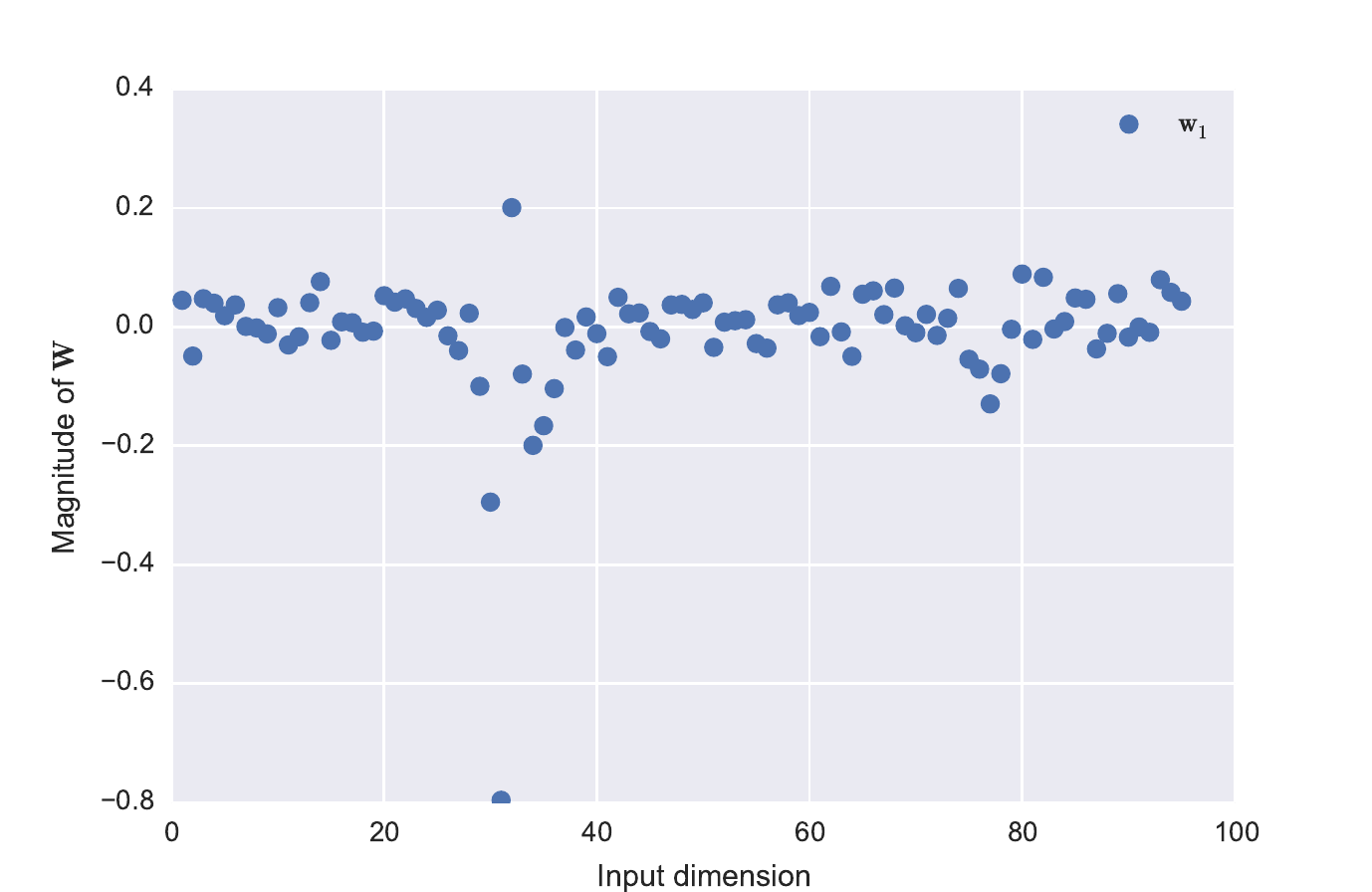}
    }
    \caption{
    One-dimensional granular crystal without gaps - full width at half maximum of the soliton over the $30^{th}$ particle.
    The first plot shows the response surface in the active subspace. the second plot depicts the test observations vs model prediction plot. The final plot depicts the components of the projection matrix.
    }
    \label{fig:p_30_fwhm}
\end{figure*}

\begin{figure*}
    \centering
    \subfigure[] {
        \includegraphics[width=80mm]{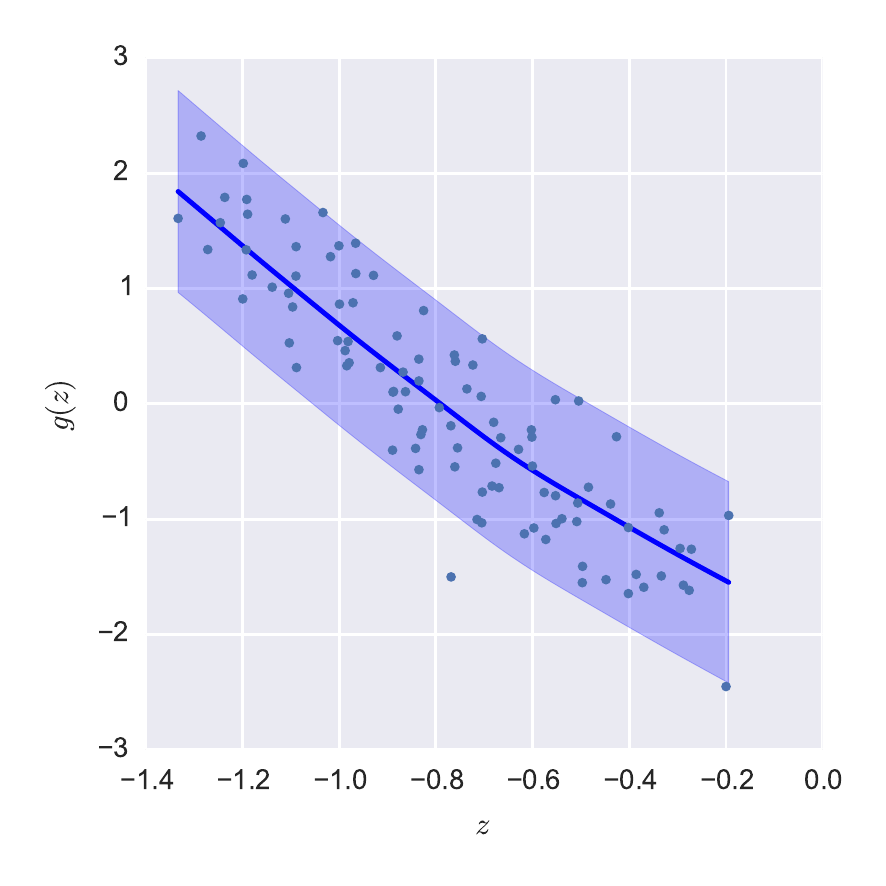}
    }
    \subfigure[] {
        \includegraphics[width=80mm]{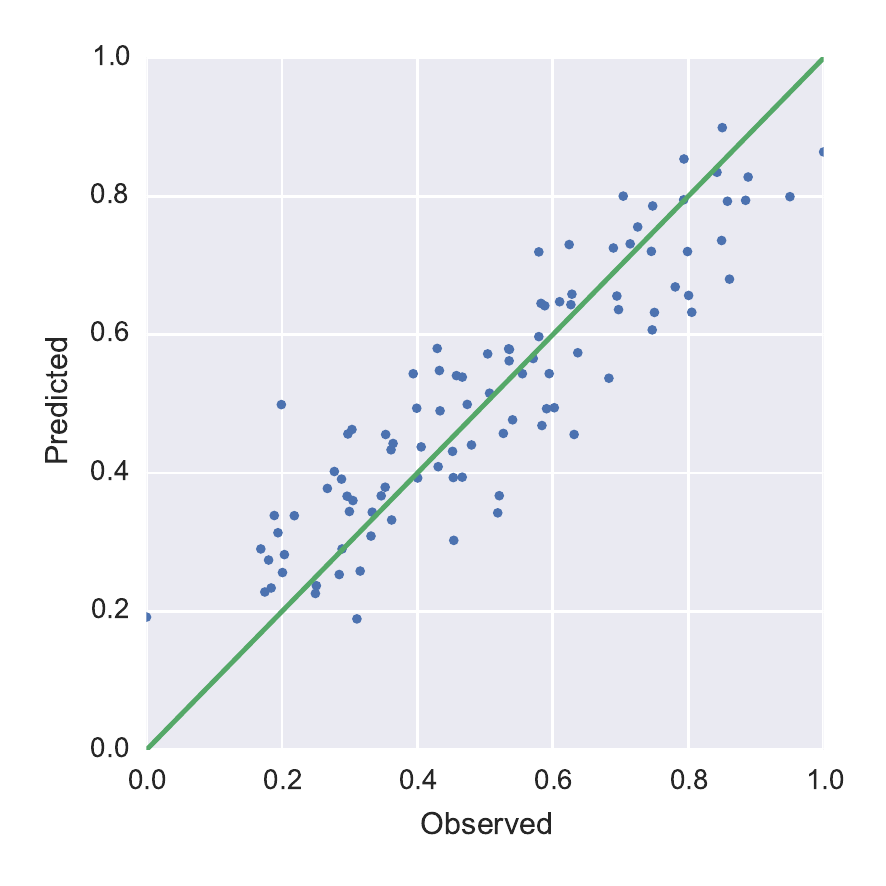}
    }
    \subfigure[] {
    	\includegraphics[width=80mm]{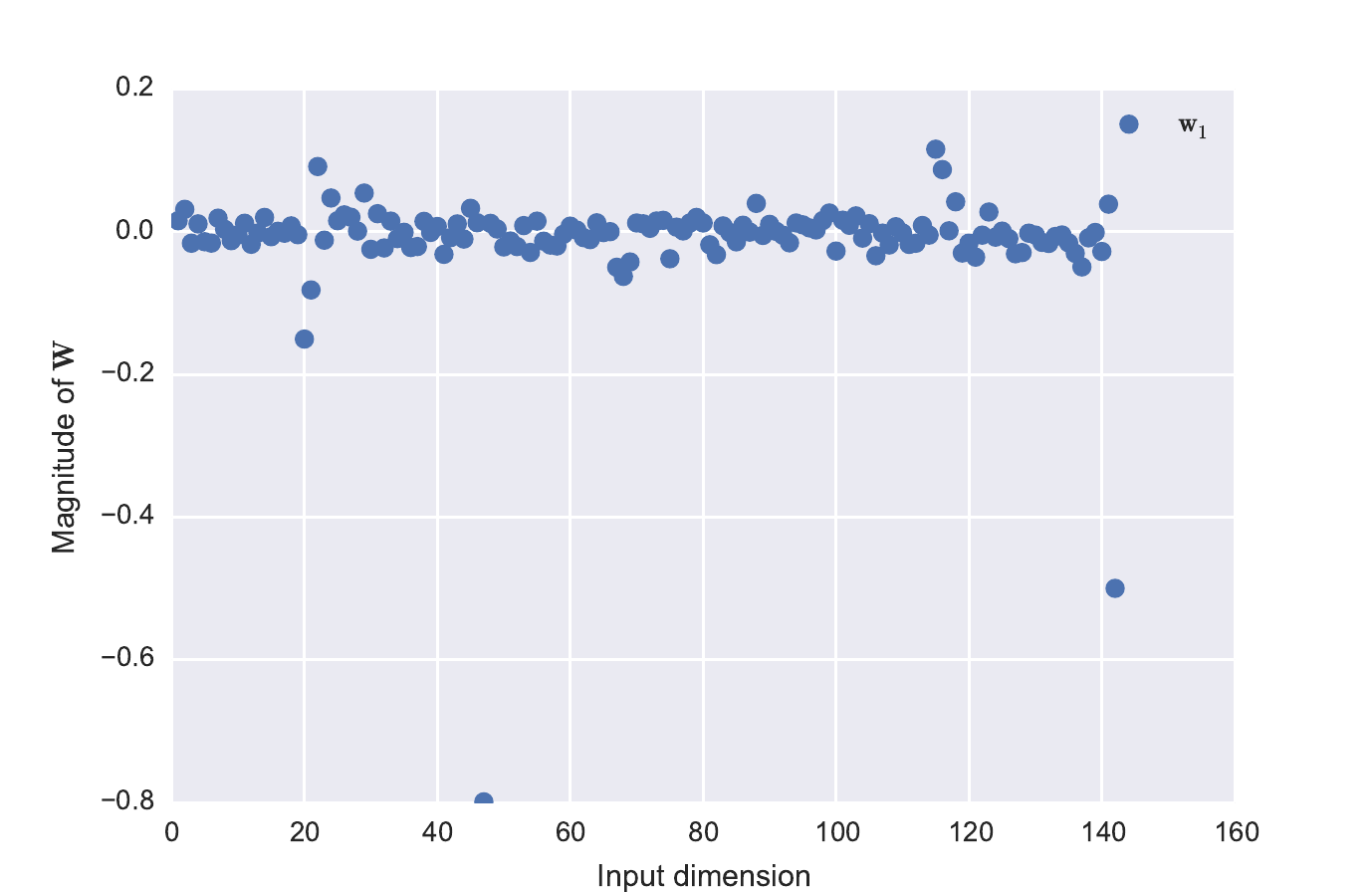}
    }
    \caption{
    One-dimensional granular crystal with gaps - amplitude of the soliton over the $20^{th}$ particle.
    The first plot shows the response surface in the active subspace. the second plot depicts the test observations vs model prediction plot. The final plot depicts the components of the projection matrix.
    }
    \label{fig:p_20_amp_gaps}
\end{figure*}

\begin{figure*}
    \centering
    \subfigure[] {
        \includegraphics[width=80mm]{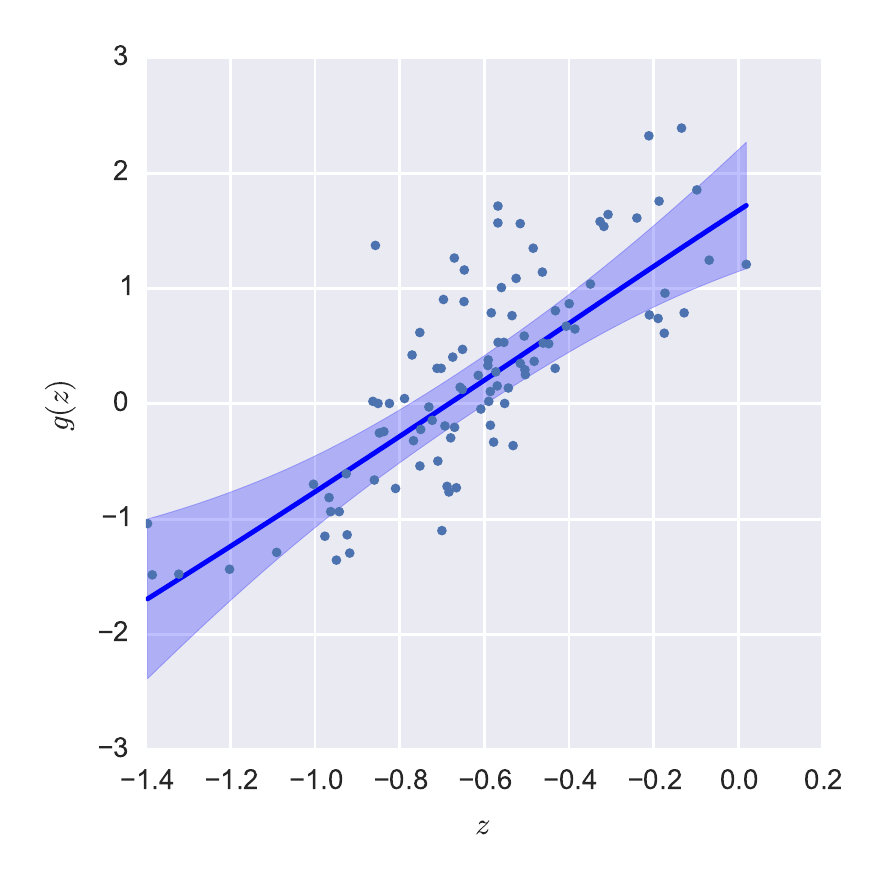}
    }
    \subfigure[] {
        \includegraphics[width=80mm]{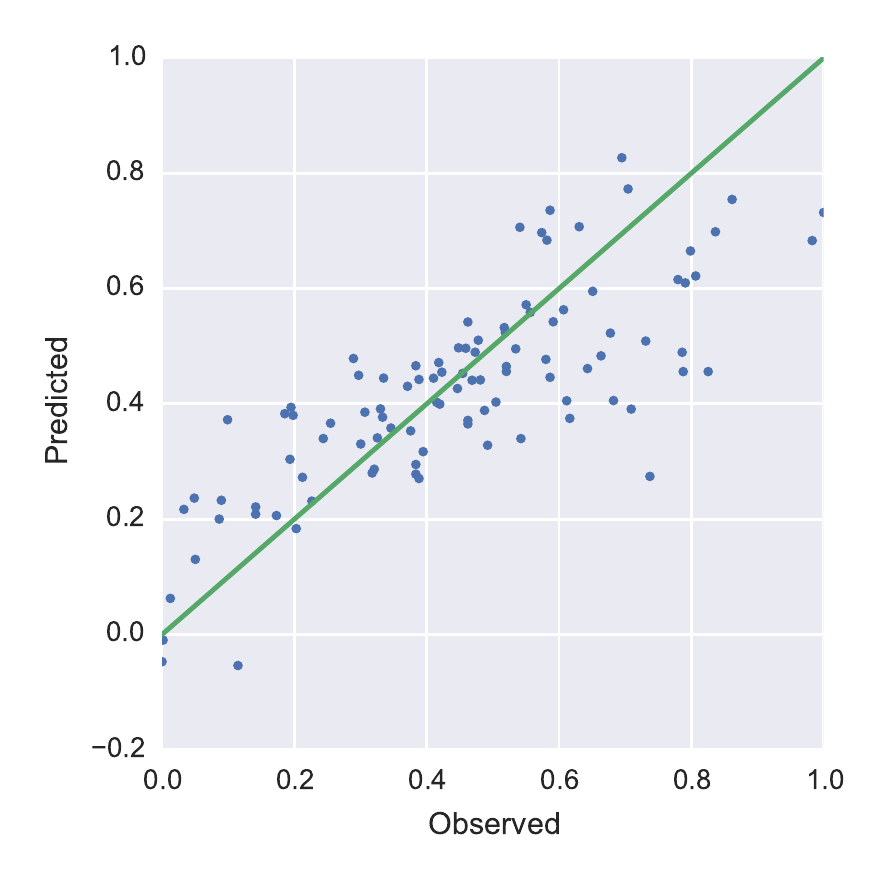}
    }
    \subfigure[] {
    	\includegraphics[width=80mm]{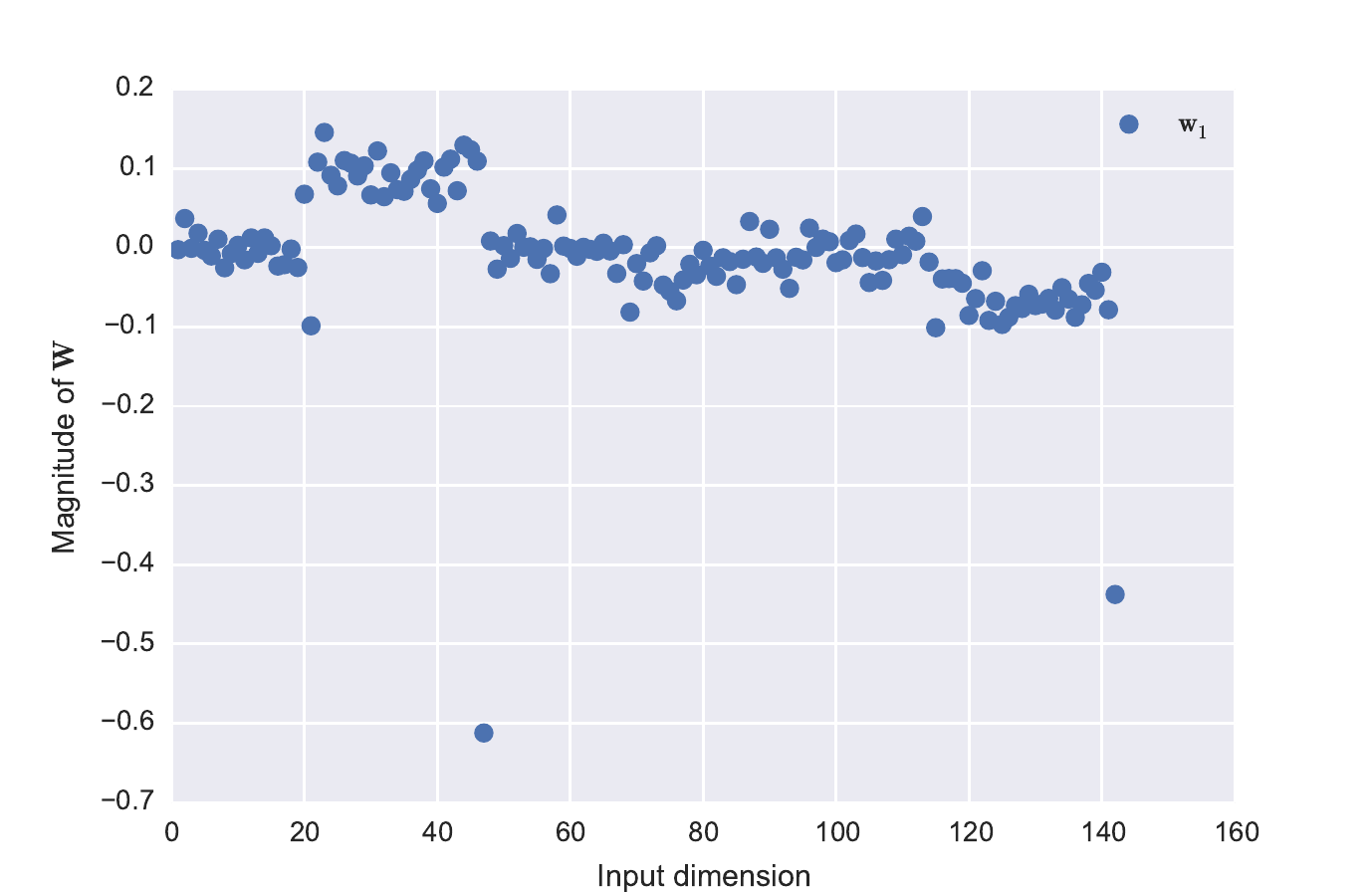}
    }
    \caption{
    One-dimensional granular crystal with gaps - time of flight of the soliton over the $20^{th}$ particle.
    The first plot shows the response surface in the active subspace. the second plot depicts the test observations vs model prediction plot. The final plot depicts the components of the projection matrix.
    }
    \label{fig:p_20_tof_gaps}
\end{figure*}

\begin{figure*}
    \centering
    \includegraphics[width=80mm]{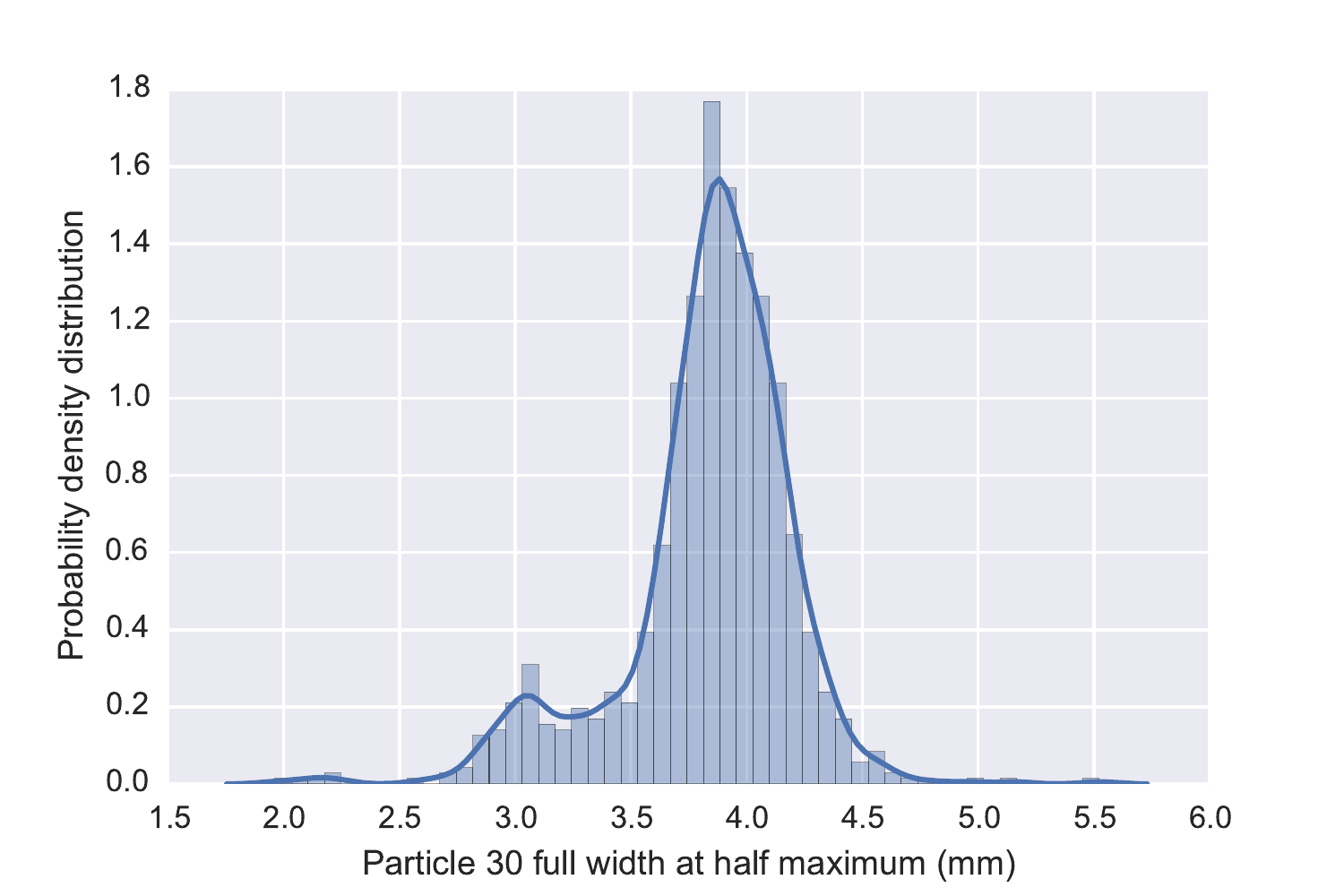}
    \caption{
    One-dimensional granular crystal with gaps - histogram of the observed full width outputs for the $30^{th}$ particle corresponding to the 
    gaps input case. 
    }
    \label{fig:hist_p30_obs_fw}
\end{figure*}

\begin{figure*}
    \centering
    \subfigure[] {
        \includegraphics[width=80mm]{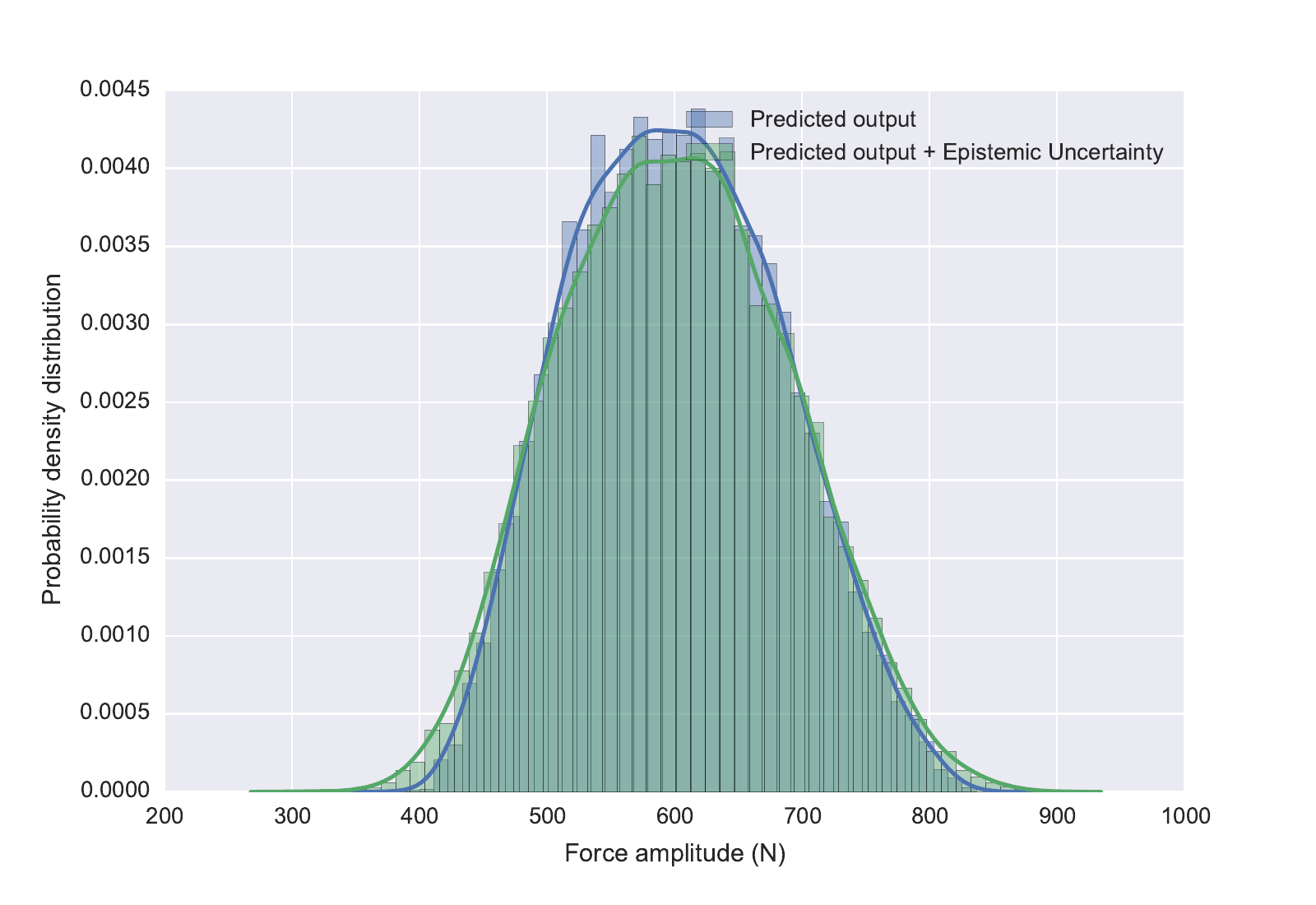}
    }
    \subfigure[] {
        \includegraphics[width=80mm]{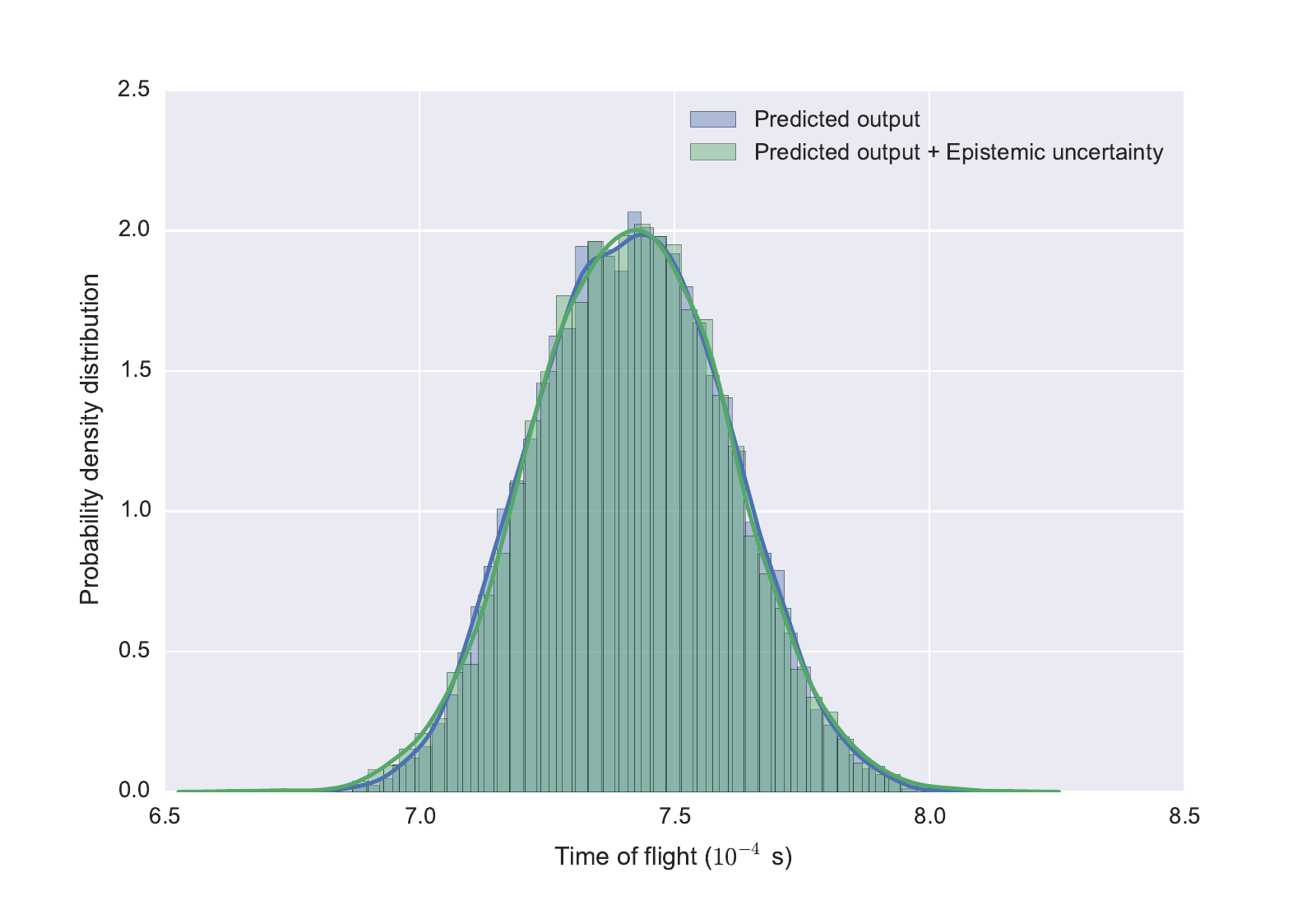}
    }
    \subfigure[] {
    	\includegraphics[width=80mm]{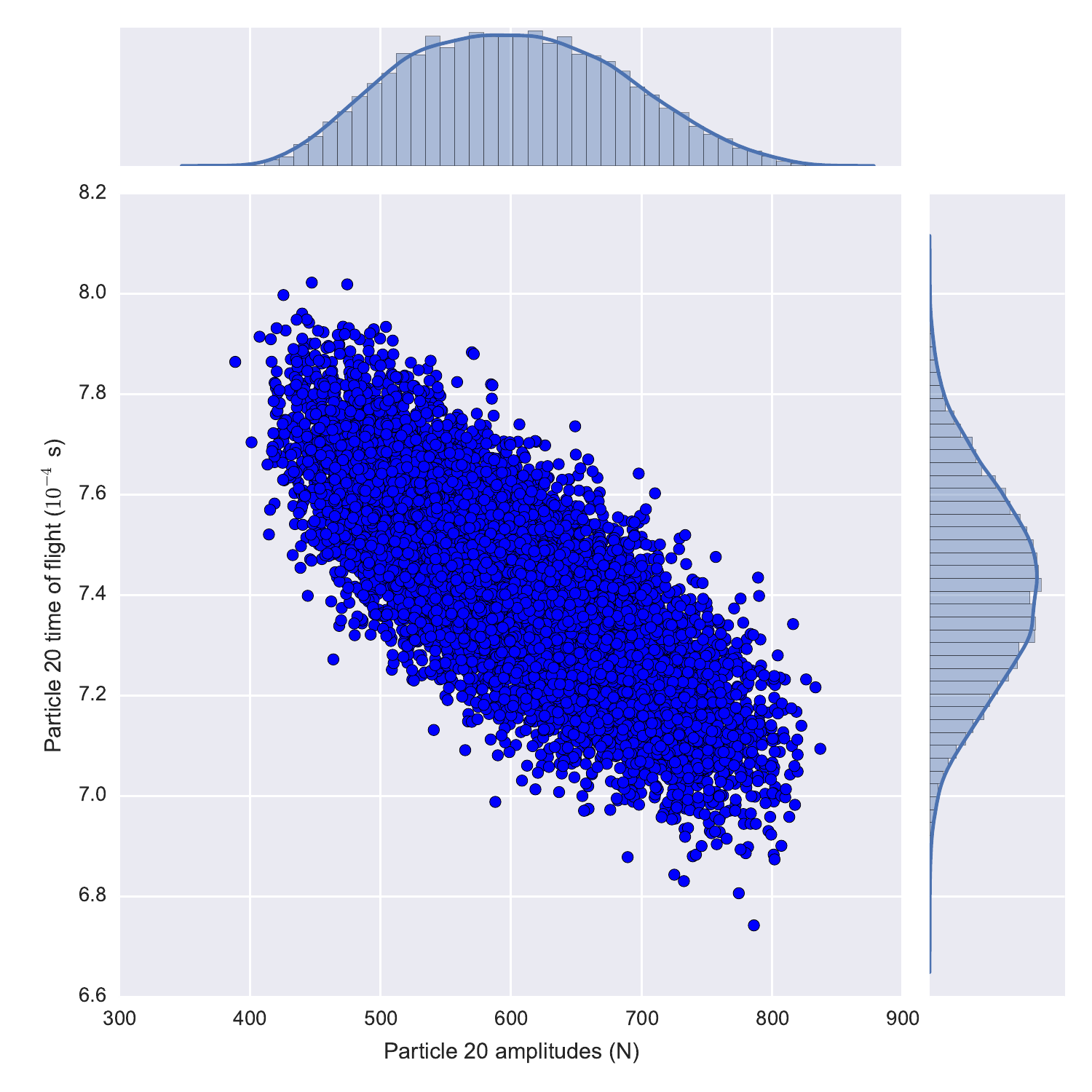}
    }
    \caption{One-dimensional granular crystal without gaps - propagating the uncertainty by assigning a uniform distribution to the inputs.
    (a) Marginal distribution of the amplitude of the soliton over the $20^{th}$ particle;  (b) Marginal distribution of the time of flight of the soliton over the $20^{th}$ particle; 
    (c) Joint distribution of the amplitude and time of flight of the soliton over the $20^{th}$ particle.
    For the plots corresponding to the marginal distributions the blue histogram and curve represents the mean predictions of the surrogate while the green histogram and curve represents the mean predictions added with Gaussian noise of variance equal to the variance of the GP surrogate. The difference between the blue and green curves is a measure of the associated epistemic uncertainty 
    in the output QoI.
       }
    \label{fig:p20_uq}
\end{figure*}

\begin{figure*}
    \centering
    \subfigure[] {
        \includegraphics[width=80mm]{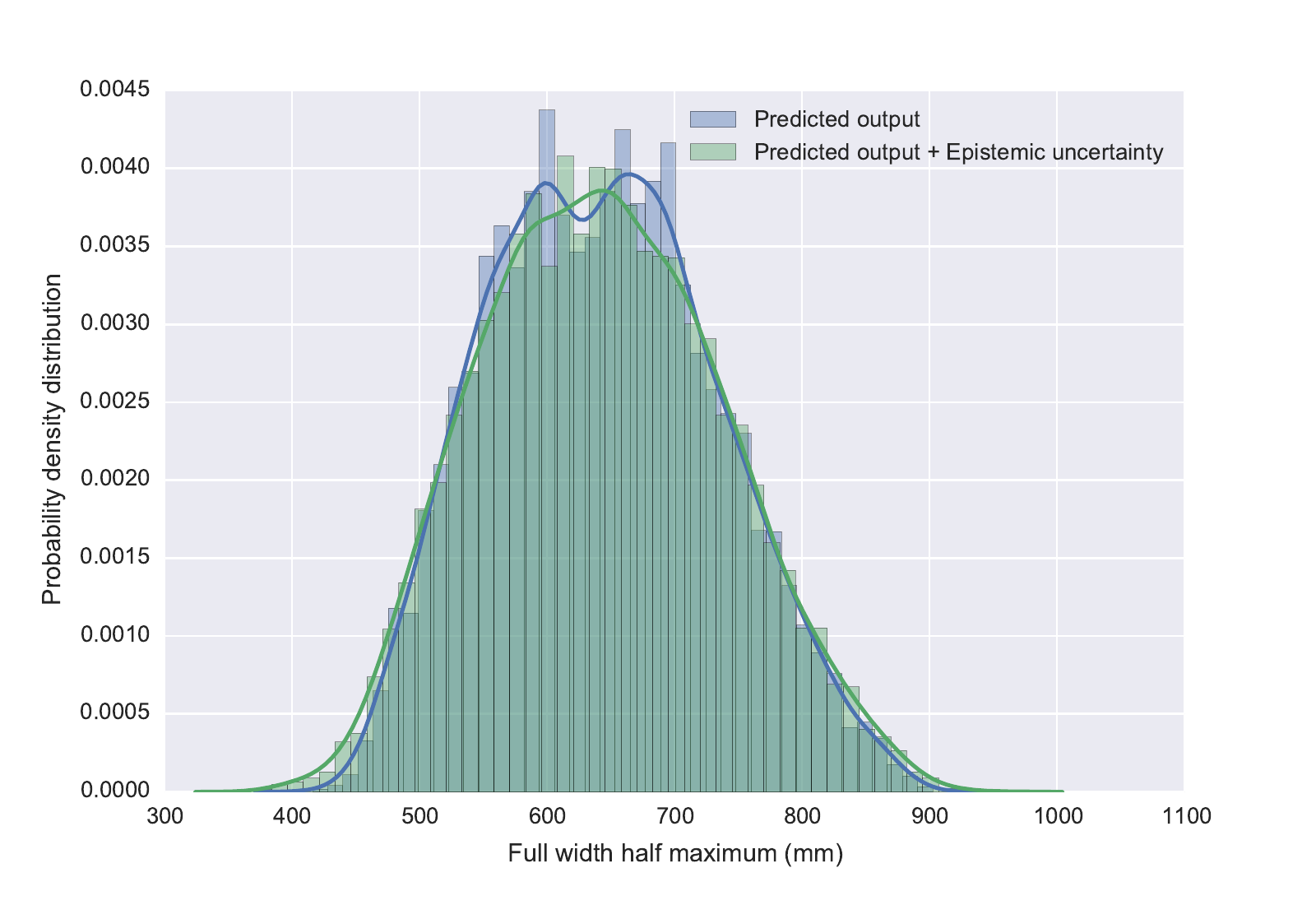}
    }
    \subfigure[] {
        \includegraphics[width=80mm]{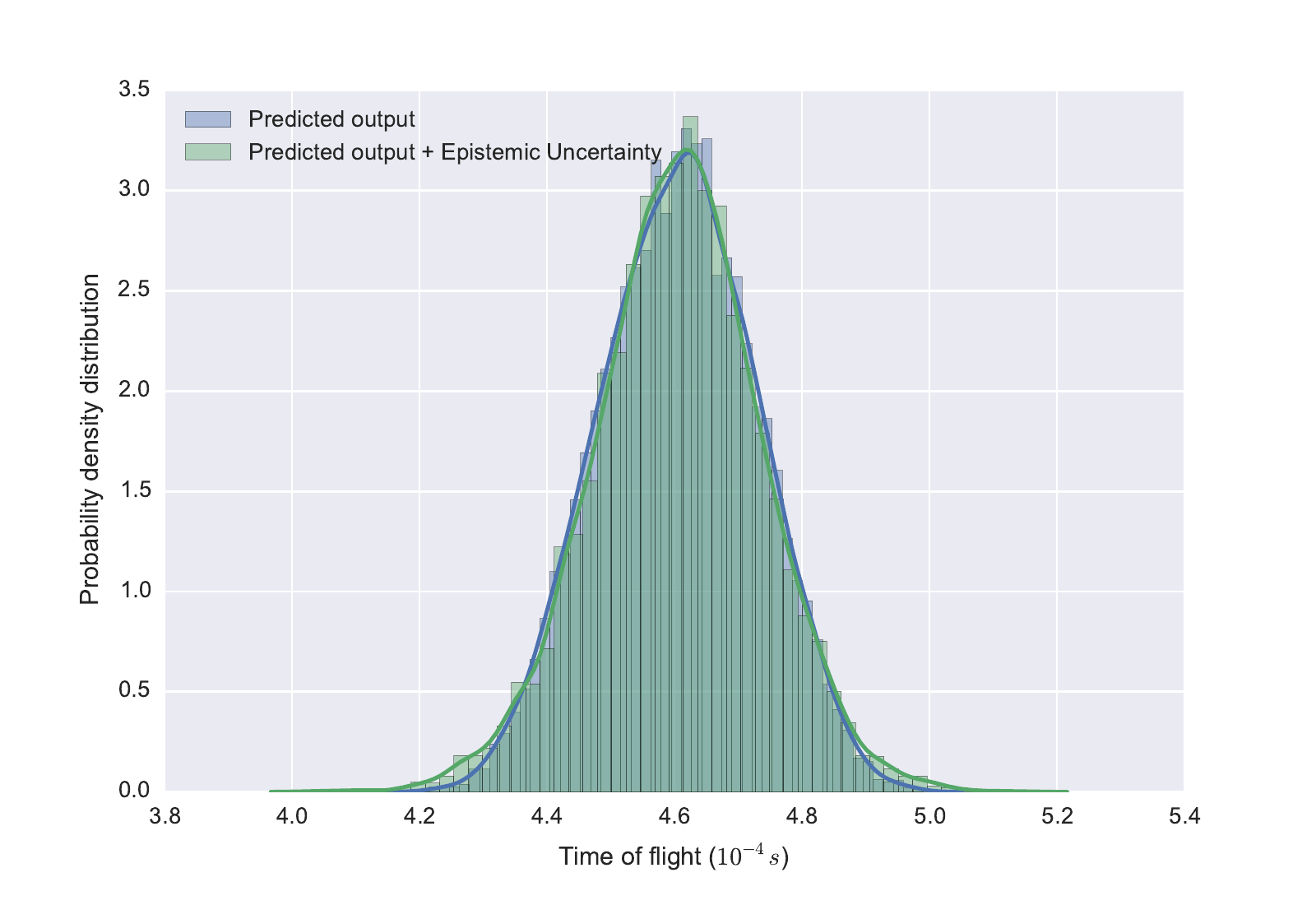}
    }
    \subfigure[] {
    	\includegraphics[width=80mm]{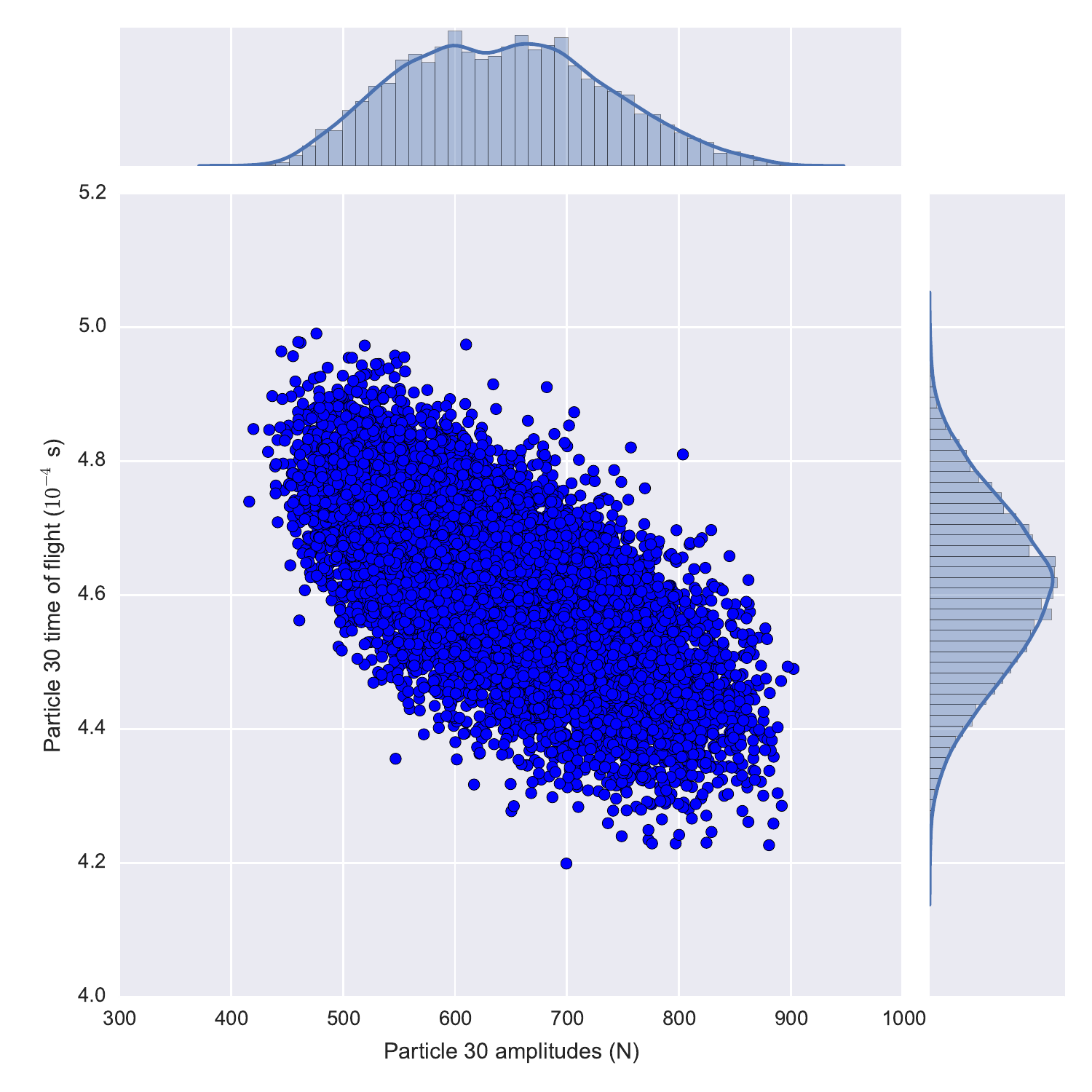}
    }
    \caption{One-dimensional granular crystal without gaps - propagating the uncertainty by assigning a uniform distribution to the inputs.
    (a) Marginal distribution of the amplitude of the soliton over the $30^{th}$ particle;  
    (b) Marginal distribution of the time of flight of the soliton over the $30^{th}$ particle; 
    (c) Joint distribution of the amplitude and time of flight of the soliton over the $30^{th}$ particle.
    For the plots corresponding to the marginal distributions the blue histogram and curve represents 
    the mean predictions of the surrogate while the green histogram and curve represents the 
    mean predictions added with Gaussian noise of variance equal to the variance of the 
    GP surrogate. The difference between the blue and green curves is a measure of the 
    associated epistemic uncertainty 
    in the output QoI.
       }
    \label{fig:p30_uq}
\end{figure*}

\begin{figure*}
    \centering
    \subfigure[] {
        \includegraphics[width=80mm]{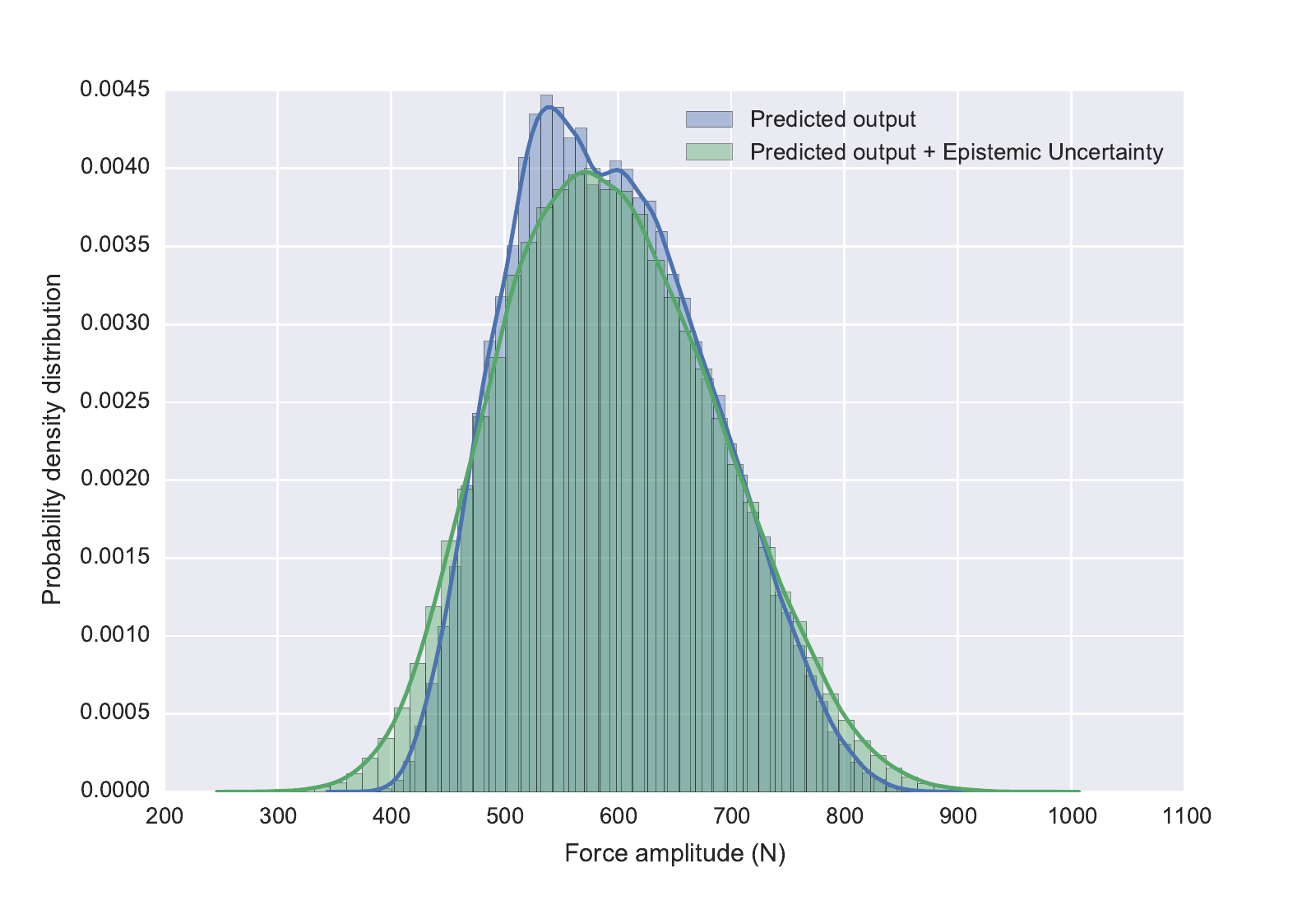}
    }
    \subfigure[] {
        \includegraphics[width=80mm]{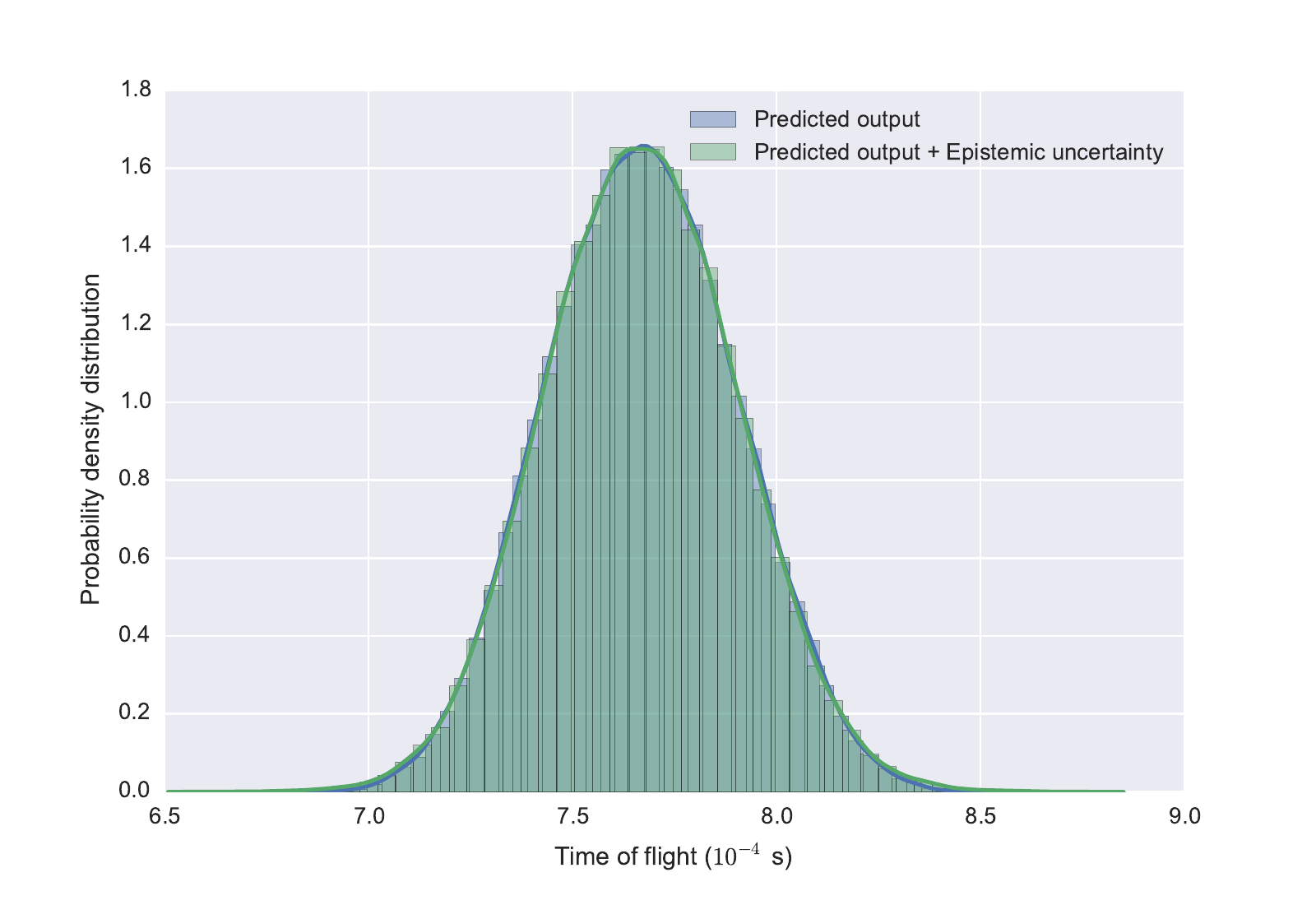}
    }
    \subfigure[] {
    	\includegraphics[width=80mm]{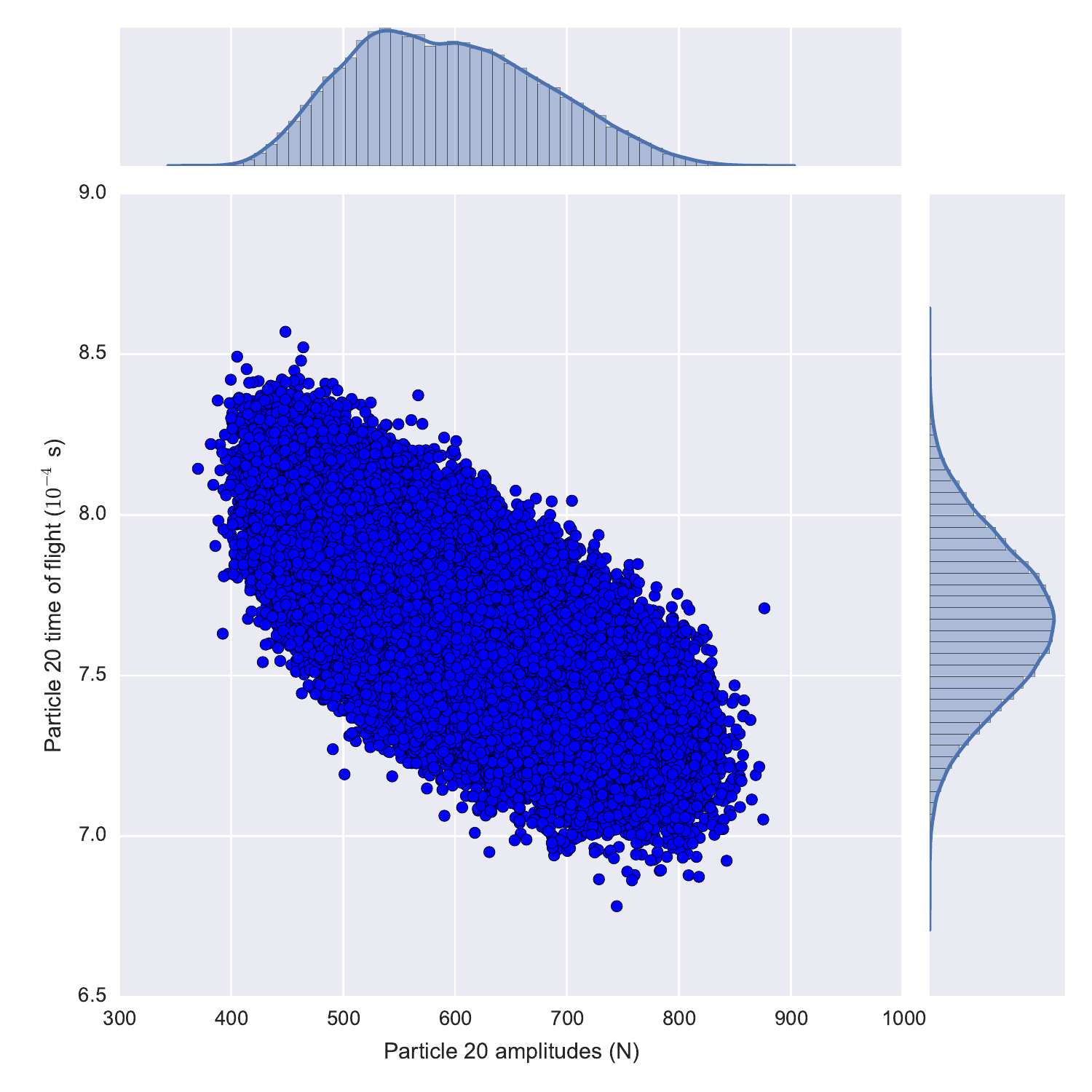}
    }
    \caption{One-dimensional granular crystal with gaps - propagating the uncertainty by assigning a uniform distribution to the inputs.
    (a) Marginal distribution of the amplitude of the soliton over the $20^{th}$ particle;  (b) Marginal distribution of the time of flight of the soliton over the $20^{th}$ particle; 
    (c) Joint distribution of the amplitude and time of flight of the soliton over the $20^{th}$ particle.
    For the plots corresponding to the marginal distributions the blue histogram and curve represents the mean predictions of the surrogate while the green histogram and curve represents the mean predictions added with Gaussian noise of variance equal to the variance of the GP surrogate. The difference between the blue and green curves is a measure of the associated epistemic uncertainty 
    in the output QoI.
       }
    \label{fig:p20_uq_gaps}
\end{figure*}

\begin{figure*}
    \centering
    \subfigure[] {
        \includegraphics[width=80mm]{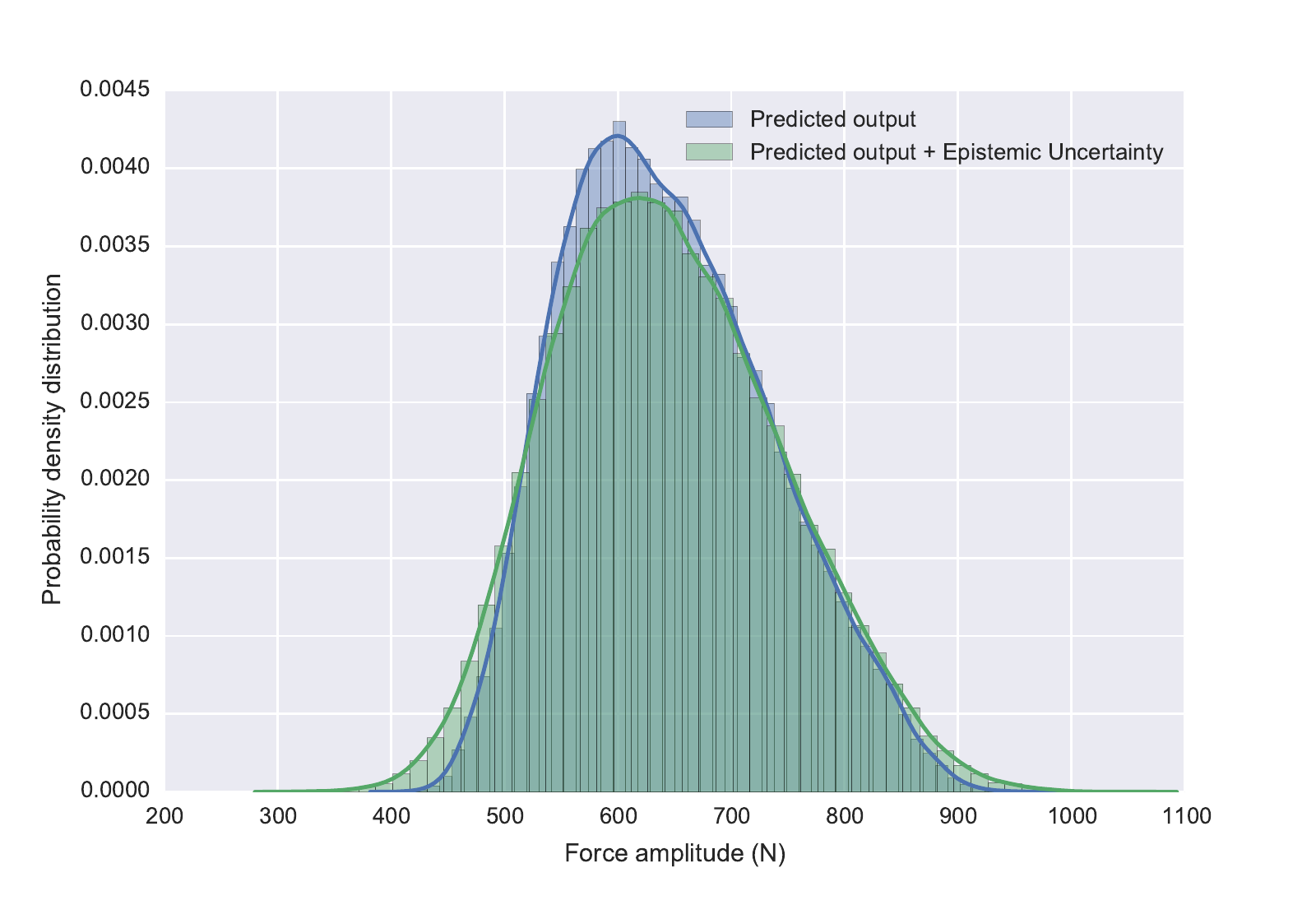}
    }
    \subfigure[] {
        \includegraphics[width=80mm]{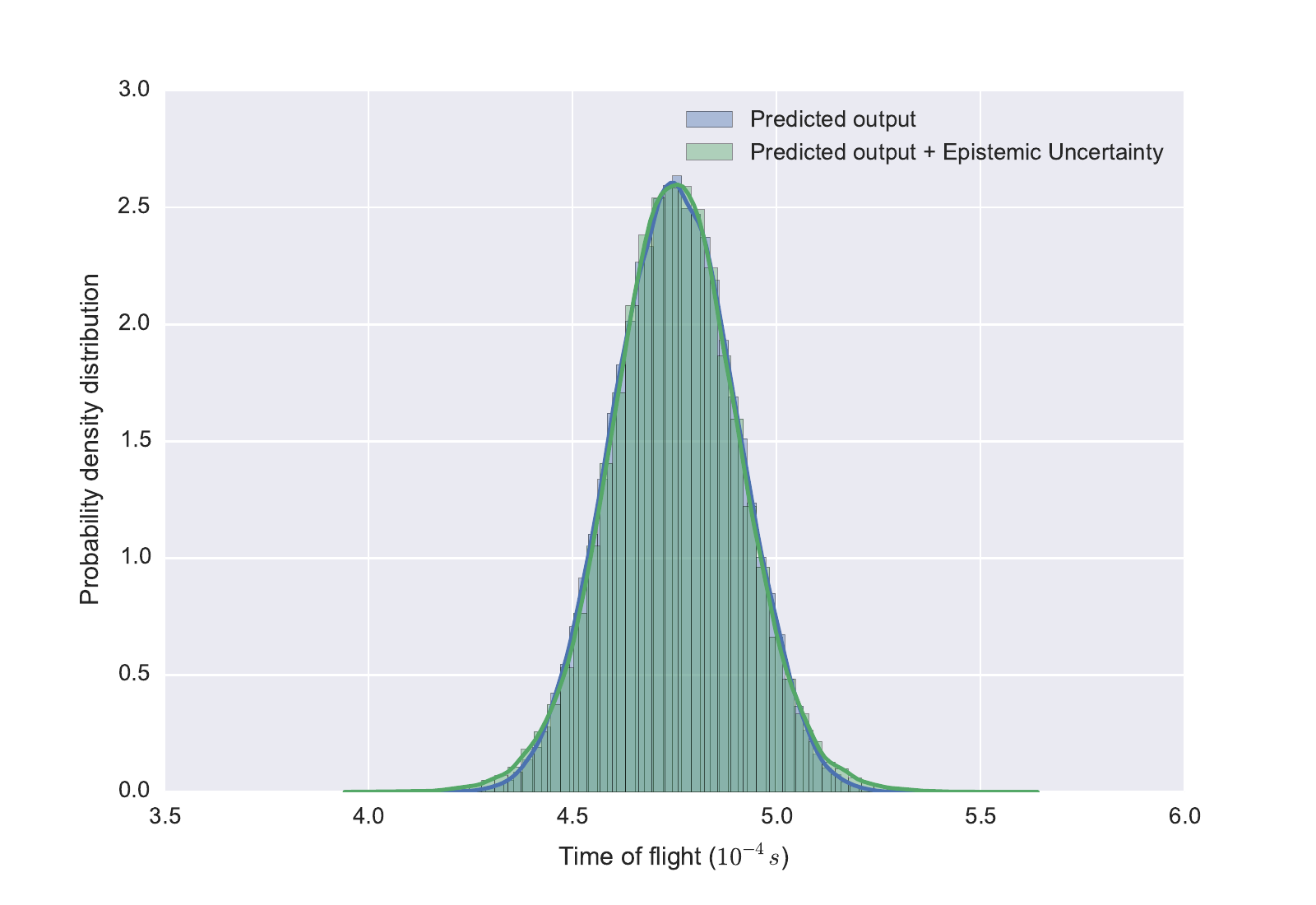}
    }
    \subfigure[] {
    	\includegraphics[width=80mm]{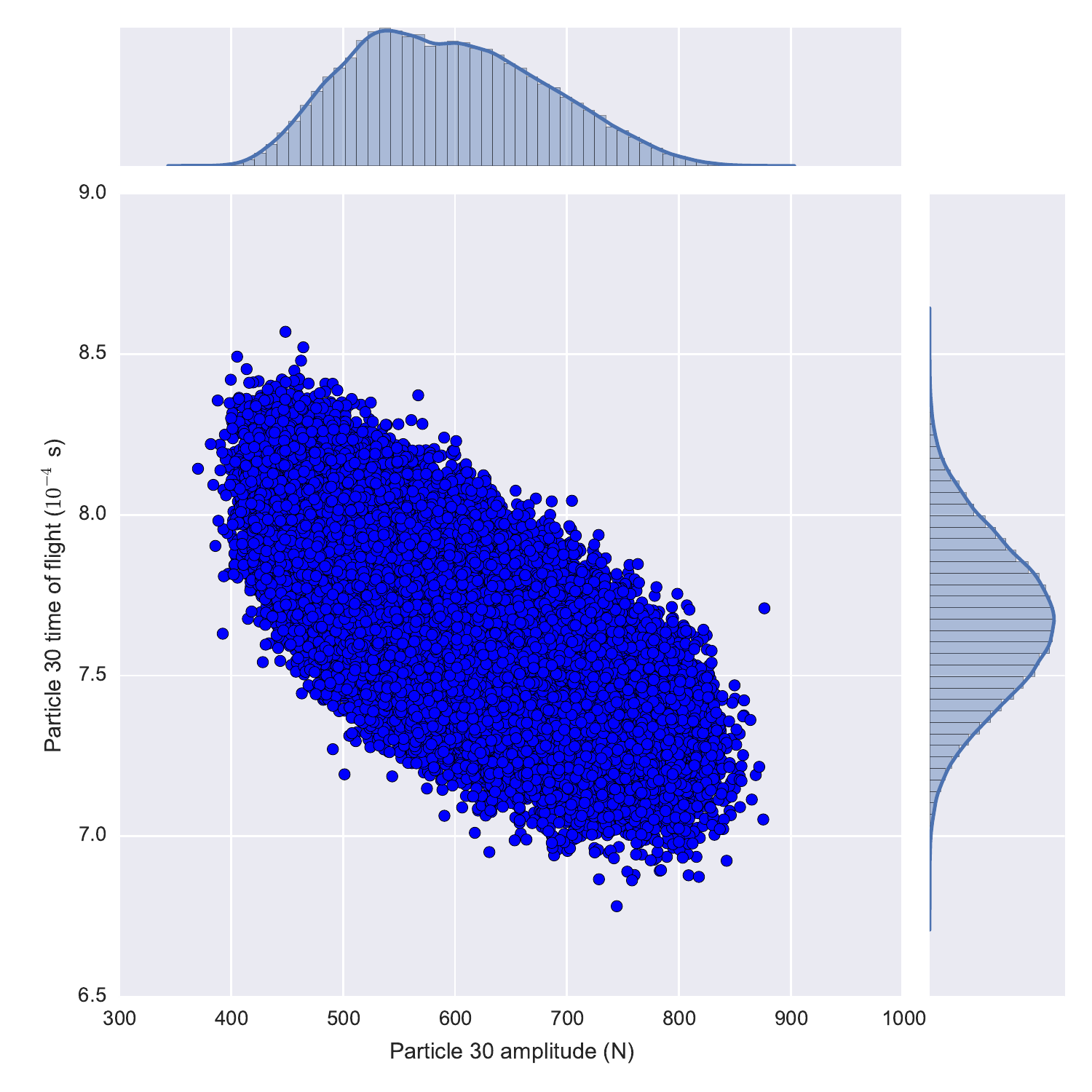}
    }
    \caption{One-dimensional granular crystal with gaps - propagating the uncertainty by assigning a uniform distribution to the inputs..
    (a) Marginal distribution of the amplitude of the soliton over the $30^{th}$ particle;  (b) Marginal distribution of the time of flight of the soliton over the $30^{th}$ particle; 
    (c) Joint distribution of the amplitude of the soliton over the $30^{th}$ particle and the time of flight of the soliton over the $30^{th}$ particles.
    For the plots corresponding to the marginal distributions the blue histogram and curve represents the mean predictions of the surrogate while the green histogram and curve represents the mean predictions added with Gaussian noise of variance equal to the variance of the GP surrogate. The difference between the blue and green curves is a measure of the associated epistemic uncertainty 
    in the output QoI.
       }
    \label{fig:p30_uq_gaps}
\end{figure*}


\subsubsection{Results}
The GP surrogate was trained using 1,000 data samples and validated using  
a further 100 samples. 
It is observed that most of the stochasticity of this high dimensional problem is exhibited 
on a one dimensional active subspace. We present plots of the AS representation 
of the link function, projection matrix and predictions vs observations plots for the 
cases noted above. Note that the underlying response surfaces obtained are 
approximately linear. There is very good agreement between the training data-set 
output predictions as compared to the actual training set outputs.
In all the cases that we just demonstrated, we observe the localized nature 
of the soliton.

We first discuss the case when the particles are in point contact i.e. the input design 
matrix is $\bX_1$.  
From plots corresponding to the soliton amplitude over particles 20 and 30 , shown in Fig. \ref{fig:p_20_amp} 
and \ref{fig:p_30_amp} we observe that the projection matrix has zero entries for most components 
except those corresponding  to the radius and Young's modulus of the respective particle 
under observation and the particle at the striker end. 
Likewise, from plots corresponding to the case of the time of flight outputs over the respective particles, 
in Fig. \ref{fig:p_20_tof} and Fig. \ref{fig:p_30_tof}, we observe that the entries of the projection matrix 
are approximately zero for components corresponding to the radii and Young's moduli of the particles 
before the particle under observation. We also note that the time of flight depends more on the particle 
radii than the particle Young's moduli, as evidenced by the larger weights assigned to the radii terms 
in the projection matrix. 
Finally, we present the plots corresponding to the full width of the soliton as it stands over particle 30
in Fig. \ref{fig:p_30_fwhm}.
The change in the BIC score from the 1st active dimension to the 2nd active dimension is insignificant 
and as such, suggests that we should be looking at a one dimensional active subspace. We note that 
the link function shows slight non-linearity. The large non-zero weights are associated with the radii terms 
in and around the $30^{th}$ particle. We believe that the remaining components of the projection 
matrix should have been closer to zero and the likely cause of this not being case would be that 
the optimization routine gets stuck in a local minima despite several restarts of the algorithm 
from random initial points. 

We now discuss the more challenging case when the particles are separated by gaps i.e. when the 
input design matrix is $\bX_2$. In Fig. \ref{fig:p_20_amp_gaps}, we show the plots corresponding to the case of 
soliton amplitude over the $20^{th}$ particle.  Looking at the projection matrix, we observe that it shows 
a similar trend to the corresponding matrix when we considered point contact inputs, i.e., the magnitude 
of most of the components of the projection matrix are approximately zero and significantly large weights 
are associated with the radius of the $20^{th}$ particle and the striker velocity. From Fig. \ref{fig:p_20_tof_gaps}
we observe that the components of the projection matrix, when looking at the time of flight of the soliton 
over the $20^{th}$ particle. We note that there are non-zero components of the projection matrix 
on a few terms corresponding to the radii and Young's moduli in and around the $20^{th}$ particle, with 
the weights being higher on the radii as opposed to the Young's moduli terms. 
Our method was unable to converge to an active subspace for the case of the full width output 
and as such we don't present it here. In Fig. \ref{fig:hist_p30_obs_fw}, we present the histogram of the 
observed full width output for 
the $30^{th}$ particle and find that it is bimodal. This suggests the presence 
of a discontinuity in the response which can be resolved, given sufficient 
data and clustering of the data into appropriate batches. 
Overall, we observe that the trends in the link function and the 
projection matrix for different outputs corresponding to the 
two different input cases are similar. It is not surprising that the results look significantly 
better when we consider the case of point contact between the particles as opposed to 
the case where we consider the case where gaps exist between the particles.  
This is because in the former case,  
the GP surrogate has to learn fewer parameters 
from the same limited quantity of data. Given the very dimensional nature of the 
associated optimization problem in the case of the gaps inputs, 
it is also likely that the optimization algorithm gets stuck in a local minima 
despite several restarts of the algorithm from random initial starting 
points. We do not present plots for the link function and projection matrices 
for the soliton properties corresponding to the $30^{th}$ particle in the gaps input 
case since they follow trends similar to what has been discussed so far. 

\subsubsection{Uncertainty Quantification}
Having built a cheap-to-evaluate response surface, we can tackle the uncertainty
propagation problem in an efficient manner. We assign a uniform distribution to the inputs. 
Then, we sample 10,000 observations of the input from this distribution and use
the surrogate to generate predictions for the time of flight and amplitude of the soliton as 
it passes over 
the respective particles. Finally, we plot the marginal and joint
distributions of some of the outputs as shown in Fig.
\ref{fig:p20_uq}, \ref{fig:p30_uq}, \ref{fig:p20_uq_gaps} and \ref{fig:p30_uq_gaps}.  For each of the marginal distribution plots, we also show the associated epistemic 
uncertainty. We observe that for a uniform distribution over the inputs, the outputs look approximately normal.

\section{Conclusions}
\label{sec:conclusion}
We have developed a gradient-free approach to active subspace (AS) discovery 
and exploitation suitable for dealing with noisy outputs.
We did so by developing  a novel Gaussian 
process regression model with built-in dimensionality reduction. Specifically we represented the AS 
as an orthogonal projection matrix that constitutes a hyper-parameter of the covariance function
to be estimated from the data by maximizing the likelihood.
Towards this end, we devised a two-step 
optimization procedure that ensures the orthogonality of the projection matrix 
by exploiting recent results on the description of the Stiefel 
manifolds. 
An addendum of the probabilistic approach is the ability to use the Bayesian
information criterion (BIC) to automatically select the dimensionality of the
AS. We validated our method using both synthetic examples with known AS and by
comparing directly to the classic gradient-based AS approach.
Finally, we used our method to study the effect of geometric and material
uncertainties in force waves propagated through granular crystals.

This work is a first step towards a fully Bayesian AS-based surrogate, a
persistent theme of our current research plans.
As argued in \cite{bilionis2015c}, Bayesian
surrogates should be capable of quantifying all the epistemic uncertainty induced by
limited data, since quantification of this epistemic uncertainty is the key to deriving
problem-specific information acquisition policies, i.e., rules for deciding
where to sample the model next in order to obtain the maximum amount of information towards a specific task.
A fully Bayesian treatment requires the specification of priors for
all the hyper-parameters of the covariance function and the derivation of
Markov Chain Monte Carlo (MCMC) schemes to sample from the posterior of the model.
The big challenge is the construction of proposals that force $\bW$ to 
remain on the Stiefel manifold, which could be achieved, for example, by modifying
the Riemann manifold  Hamiltonian MC of~\cite{RSSB:RSSB765}. 
Such approaches would open the way for more
robust AS dimensionality selection, e.g., by reversible-jump MC~\cite{green1995}
or by directly computing the model evidence.

Many physical models do not have an AS.
They may have, however, a non-linear low-dimensional manifold exhibiting
maximal response variability.
Assuming that this low-dimensional manifold is a Riemann manifold, i.e.,
locally isomorphic to a Eucledian space, a potential approach would be 
to consider mixtures of the model proposed in this work.
To this end, the results of~\cite{chen2015} on infinite mixtures of GP's
could be leveraged.
The latter is also a subject of on-going research.

\begin{acknowledgment}
Ilias Bilionis and Marcial Gonzalez acknowledge the startup support provided by
the School of Mechanical Engineering at Purdue University.
\end{acknowledgment}

\bibliography{bibliography}{}
\bibliographystyle{unsrt}

\end{document}